\documentclass[
 aip, 
 amsmath,amssymb,
preprint,%
author-numerical,%
]{revtex4-1}
\usepackage{amssymb}
\usepackage{graphicx}
\usepackage{amsmath}
\usepackage{subfigure}
\usepackage{amssymb}
\usepackage{xcolor}
\renewcommand{\textcolor}[2]{#2}
\usepackage{comment}
\usepackage[acronym]{glossaries}
\usepackage{soul}
\usepackage{hyperref}
\hypersetup{
    colorlinks=true,
    linkcolor=blue,
    filecolor=cyan,
    citecolor=blue,
    urlcolor=magenta,
    }
\makeglossaries
\usepackage[normalem]{ulem}   %

\newcommand{\eal}{{\em et al.} }
\def\avg#1{{\left<{#1}\right>}}
\def\pd#1#2{\frac{\partial #1}{\partial #2}}

\newcommand{\Rio}{{Ri}_0}
\newcommand{\Ro}{Ro}
\newcommand{\Ra}{Ra}

\newcommand{\Pran}{Pr}

\newcommand{\Ek}{Ek}

\draft 

\begin{document}

\title{Baroclinic wave dynamics in the Ekman-free rotating rectangular annulus with localized forced plume} 



\author{Shivam Swarnakar}
\email[]{shivamswarnakar72@gmail.com}
\altaffiliation[Presently at ]{ Borosil Renewables Limited, Hinjewadi Phase - I, Pune, Maharashtra, 411057}
\affiliation{Department of Mechanical Engineering, Indian Institute of Technology Bombay, Mumbai, 400076, India}

\author{Ayan Kumar Banerjee}
\email[]{ak\_banerjee@cb.amrita.edu}
\altaffiliation{Corresponding author}
\affiliation{Amrita School of Artificial Intelligence, Coimbatore, Amrita Vishwa Vidyapeetham, India
}

\author{Sridhar Balasubramanian}
\affiliation{Department of Mechanical Engineering, Indian Institute of Technology Bombay, Mumbai, 400076, India}

\author{Amitabh Bhattacharya}
\affiliation{Department of Applied mechanics, Indian Institute of Technology Delhi, New Delhi, 110016, India}

\noindent
Note: The following article has been published in
\textit{Physics of Fluids} \textbf{38}, 096614 (2026):
\url{https://doi.org/10.1063/5.0344131}

\medskip

\noindent
\copyright 2026 Shivam Swarnakar, Ayan Kumar Banerjee,
Sridhar Balasubramanian, and Amitabh Bhattacharya.
This article is distributed under a Creative Commons
Attribution NonCommercial 4.0 International
(CC BY-NC) License.
\url{https://creativecommons.org/licenses/by-nc/4.0/}

\vspace{0.3cm}

\begin{abstract}
We report numerical simulations of a rotating rectangular annulus that isolates 
the Ekman-free bulk of the cylindrical baroclinic annulus, subjected to 
bi-directional temperature gradients imposed by a uniformly cooled inner wall 
and a localized forced heated plume at the outer bottom. The finite-volume 
OpenFOAM solver is employed across combinations of source Richardson number 
$Ri_0 = 99, 4, 1$ and Rossby number $Ro = 0.3, 0.1, 0.07$. A non-dimensional scaling of the governing equations identifies 
geostrophic-hydrostatic balance as the leading-order bulk state, a result 
confirmed a posteriori by the $x$ and $z-$momentum budgets. Baroclinic waves of mode $m=2$ at $Ro=0.3$ transition to $m=3$ as $Ro$ decreases, consistent with the contraction of the Eady deformation radius $L_\rho = NH/f$;  Complex Empirical Orthogonal Function (CEOF) analysis characterizes the wave regime and detects a Hopf-bifurcated vacillating state \textcolor{blue}{(quasi-periodic amplitude modulation)} at $Ri_0 = 99,~Ro = 0.1$. The plume morphology, classified through the Morton length scale and source flux-balance parameter, transitions from weak, laterally-swept structures at 
$Ri_0 = 99$ to sustained columnar plumes traversing the full baroclinic depth 
at $Ri_0 \leq 4$. The plume entrainment coefficient $\Gamma(z)$ shows opposite rotational sensitivities at low and high $Ri_0$, which we organize through a local plume Rossby number $Ro_p = w/(2\Omega b)$. A mixing-length argument predicts a bulk turbulent heat flux $\avg{u'T'} \propto Ri_0^{-1/2}$, anticipating an 
order-of-magnitude enhancement from $Ri_0 = 99$ to $Ri_0 = 1$, in agreement 
with the simulations. A regime map in the $(Ri_0, Ro)$ plane reveals that, 
within the explored range, the plume-regime and wave-selection problems are 
approximately separable: $Ri_0$ sets the plume regime while $Ro$ selects the 
dominant baroclinic wave mode.
\end{abstract}


\maketitle 

\printglossaries
 
\section{Introduction}
\label{sec:Intro}

Convective circulations observed in the atmosphere are driven by a combination of a temperature gradient, between the equator and pole, and the earth's rotation. As the thermal gradient and rotation become more prominent, baroclinic waves (usually seen at the mid-latitude level) form that aid in the transport and exchange of mass and energy in the atmosphere and also affect the weather in the extratropical region (Zhang \eal \cite{zhang2021seasonal}).
The baroclinic waves are studied at the laboratory scale using the differentially heated rotating cylindrical annulus setup (Hide \cite{hide1958experimental}). It is commonly known as the baroclinic annulus setup. It consists of a cylindrical annulus on a rotating table, the outer and inner walls of the annulus are uniformly heated and cooled respectively. The baroclinic annulus setup, with working fluid as water, is studied extensively in the literature using experimental techniques (Read \eal \cite{read2014general}, Harlander \eal \cite{harlander2011piv}, Vincze \eal \cite{vincze2014experimental}, Von Larcher \eal \cite{von2005experiments}, Hide and Fowlis \cite{hide1965thermal}, Douglas \eal\cite{douglas1972investigation}, Pfeffer \eal \cite{pfeffer1968wave}, Hide \cite{hide1958experimental}, Read \cite{Read2020Baroclinic}, and Harlander \eal \cite{CRPHYS_2024__25_S3_379_0}) and numerical simulations (Vincze \eal \cite{vincze2014benchmarking}, Read \cite{read2003combined}, Williams \cite{williams1969numerical, williams1971baroclinic}, James \eal \cite{james1981combined}, Hignett \eal\cite{hignett1985comparison}, Ukaji \eal \cite{ukaji1989comparison}, and Meletti \eal \cite{Meletti2026SurfaceBoundary, Meletti2025HeatTransfer}). The baroclinic waves are formed in the region between the inner and outer walls of the annulus and propagate in the annulus. Its structure and the wave modes depend on the geometrical configuration of the annulus i.e. annulus gap $(R)$ and depth $(d)$. 

One of the major shortcomings of the baroclinic annulus setup is the presence of a unidirectional temperature gradient i.e., a horizontal temperature gradient between the inner and outer wall of the annulus. However, in a realistic atmosphere, both meridional (horizontal) and vertical temperature gradients are present. The baroclinic annulus lacks the vertical temperature gradient due to the imposition of uniform heating/cooling of the outer/inner wall. To overcome this limitation, several variants of the ``classical" baroclinic annulus configurations have been proposed and investigated using both experimental and numerical approaches. One class of configurations employs a localized heating region at the bottom outer periphery of the annulus together with uniform cooling of the inner cylindrical wall, thereby establishing a bi-directionally forced thermal gradient (Banerjee \eal \cite{banerjee2018experimental,banerjee2021investigation,banerjee2024axisymmetric}; Swarnakar \eal \cite{swarnakar2023numerical}). Another class introduces localized cooling near the top of the inner wall in combination with localized heating at the bottom outer periphery, thereby generating both radial and vertical thermal stratification (Scolan \eal \cite{scolan2017rotating}; Wright \eal \cite{wright2017regimes}; Ding \eal \cite{ding2026stratification}; Harlander \eal \cite{harlander2023new}; Rodda \eal \cite{Rodda2020AtmosphericLike}; Sukhanovskii \eal \cite{Sukhanovskii2023ShallowLayer}).

\textcolor{blue}{In our previous work (Swarnakar \textit{et al.} \cite{swarnakar2023numerical}), we numerically investigated convection dynamics in an annulus subjected to bi-directional thermal forcing. The flow was characterized by the Rayleigh and Taylor numbers,}
\begin{equation}
Ra=\frac{\alpha g\Delta T_x l_x^{3}}{\nu\kappa},
\qquad
Ta=\frac{4\Omega^{2}l_x^{4}}{\nu^{2}},
\end{equation}
\textcolor{blue}{which quantify the buoyancy forcing due to the imposed temperature difference, $\Delta T_x$, and the influence of the imposed rotation rate, $\Omega$, respectively. It's worth noting that here $Ra$ is the horizontal (baroclinic) Rayleigh number, constructed using the horizontal temperature contrast $\Delta T_x$ and the horizontal gap $l_x$ that drive the thermal wind. This is the standard measure for differentially side-heated (sloping-convection) systems, in which gravity is vertical while the destabilizing temperature gradient is horizontal. It should be distinguished from the vertical (Rayleigh--B\'enard) Rayleigh number, $Ra_z=\frac{\alpha g\,\Delta T_z\,l_z^3}{\nu\kappa}$
associated with the plume's vertical temperature contrast. 

The simulations were performed at $Ra=4.76\times10^{8}$ and three Taylor numbers, $Ta=6.5\times10^{8}$, $1.5\times10^{9}$, and $2.7\times10^{9}$.} We observed a mode 4 steady baroclinic wave in the annulus at the lowest $Ta$. As the $Ta$ number increases, the wave shifts to a transition regime and eventually moves to an irregular regime at the highest $Ta$ number, where the baroclinic wave disintegrates into small eddies. In addition, at low $Ta$, we observed distinct plume structures extending to the top of the domain. With an increase in $Ta$, the plume structure became narrower and shorter, eventually fragmenting into smaller eddies at the highest $Ta$ value. These large-scale structures play a crucial role in heat and mass transport within the annulus, and their breakdown significantly influences the transport phenomenon. Furthermore, at each $Ta$ value, we found that at the top and bottom boundaries of the annulus, the Ekman boundary layers are formed. These boundary layers control the rate of radially outward flow at the bottom and the radially inward flow at the top of the annulus. However, their effect on the mean flow is limited to the regions very close to the walls and does not affect the region in the bulk of the annulus where baroclinic waves are present (Swarnakar \eal \cite{swarnakar2023numerical}). Moreover, we found that for $Ta = 6.5 \times 10^8$, the radial heat transport through the Ekman layers and the bulk of the annulus is similar in magnitude. For $Ta > 6.5 \times 10^8$, the radial heat transport is mainly through the bulk region of the annulus, and the Ekman layers do not contribute much to the radial heat transport. In addition, resolving the Ekman boundary layers in a numerical simulation is computationally very expensive. Since our primary interest is in understanding the dynamics of baroclinic waves, which form in the annulus, and the columnar plumes that form near the heating zone, we believe that it is prudent to ignore the Ekman boundary layer and only focus on the bulk region, dominated by co-existence of wave and plume. 

 A distinguishing feature of these recently studied configurations~\cite{banerjee2018experimental,banerjee2021investigation, banerjee2024axisymmetric,swarnakar2023numerical,scolan2017rotating} is that the localized heating generates buoyant plume structures near the heating zone, which coexist with and can interact with the baroclinic waves present in the annular bulk. Despite the extensive literature on both baroclinic annulus dynamics~\cite{hide1965thermal,douglas1972investigation,pfeffer1968wave,hide1958experimental,williams1969numerical, williams1971baroclinic,james1981combined,hignett1985comparison, read2014general,harlander2011piv,vincze2014experimental,von2005experiments,vincze2014benchmarking,read2003combined, banerjee2018experimental,banerjee2021investigation, banerjee2024axisymmetric,swarnakar2023numerical,scolan2017rotating, Agaoglou2024Building,Harlander2022Probability} and rotating and/or non-rotating plumes~\cite{turner1986turbulent,woods2010turbulent,morton1956turbulent,turner1986turbulent,fischer1979mixing,kaye2008turbulent,wright1979two,helfrich1991experiments,bush1999generation} in isolation, the interaction between a buoyant plume and baroclinic waves in a rotating environment has received comparatively little attention. In configurations with bi-directionally forced temperature gradient, the localized bottom heating naturally generates plume structures that coexist with the annular baroclinic waves. Previous studies of such configurations (Scolan \eal \cite{scolan2017rotating}; Harlander \eal \cite{harlander2023new}; Banerjee \eal \cite{banerjee2018experimental,banerjee2021investigation}) have focused primarily on characterizing the baroclinic wave dynamics arising from natural convection, where the induced plume is relatively weak. The effect of a considerably stronger, forced plume on baroclinic wave structure, mode selection, and amplitude has not been systematically investigated. We hypothesize that a stronger plume could fundamentally alter the baroclinic wave dynamics in the annulus, a possibility that warrants careful investigation. Consequently, we aim to investigate a configuration that includes a forced plume and depicts a bulk region of the baroclinic annulus, simulating a boundary layer-free domain.

\textcolor{blue}{The mid-latitude atmosphere
is forced by spatially localised, banded diabatic heating —
latent-heat release in storm tracks, land–sea and orographic contrasts — superposed on a stable
vertical stratification. The localized forced plume is precisely an idealisation of such concentrated
convective forcing, and it restores the vertical temperature gradient absent from the classical Hide
annulus, making the configuration closer to the real atmosphere in this respect. The coupling between localized convective plumes and the baroclinic wave field is moreover of
direct relevance to atmospheric/oceanic energy pathways and the shaping of mesoscale–submesoscale
energy spectra, as recently emphasised by Ding \textit{et al}\cite{ding2026stratification}.}

\begin{figure}
\centering
\subfigure[]{%
\includegraphics[scale = 0.4]{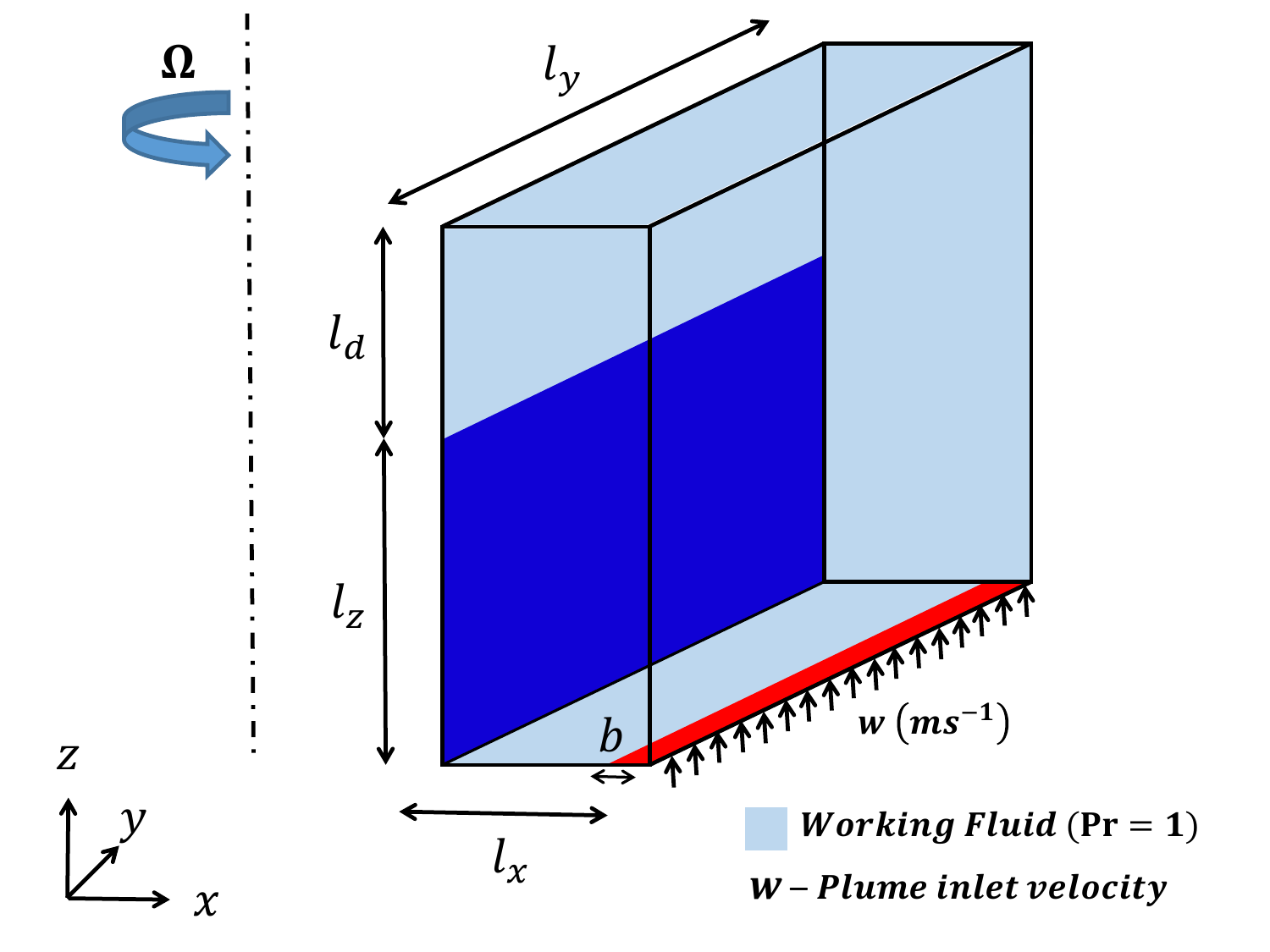}
\label{fig:TMean_w_1_2_Ta_low}}
\hspace{1cm}
\subfigure[]{%
\includegraphics[scale = 0.4]{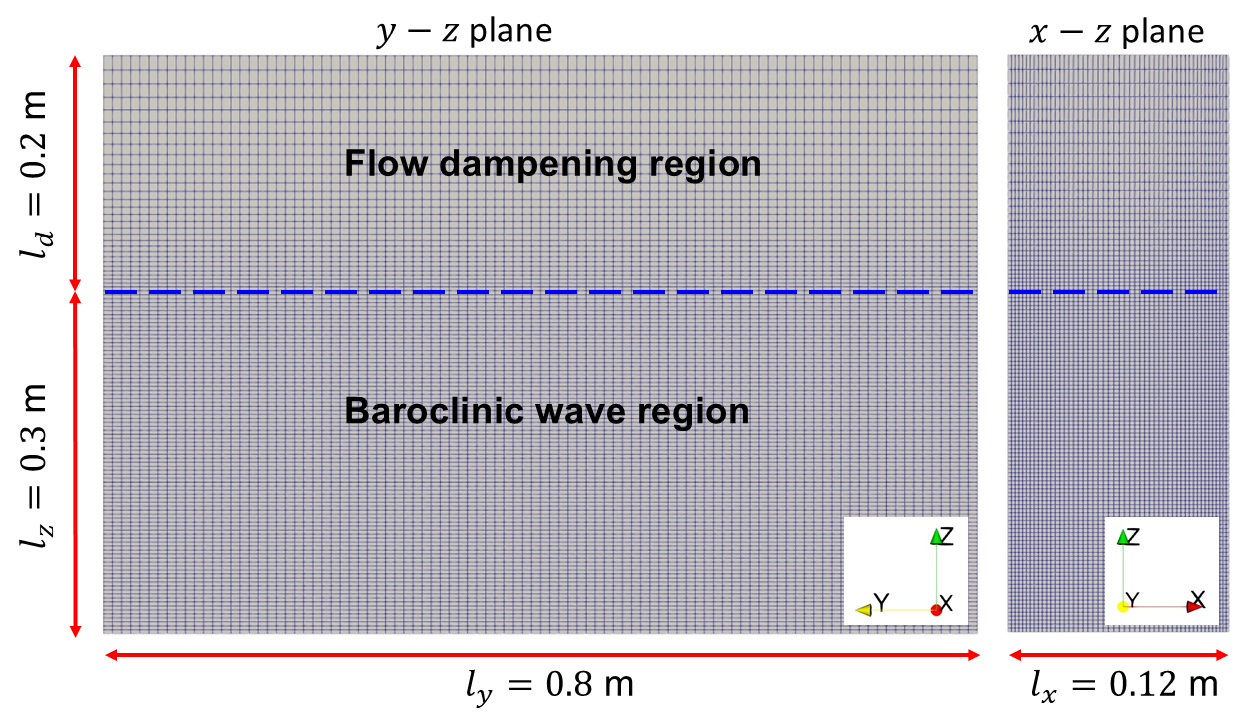}
\label{fig:TMean_w_1_2_Ta_mod}}
\hspace{1cm}
\\

  \caption{(a) Three-dimensional schematic of the rectangular configuration with bi-directional thermal forcing. The blue and red patches denote the cold wall and the inlet of the forced heated plume, respectively. A flow-damping region of height $l_d=0.20$ m occupies the upper section ($z>l_z$). The dimensions are $l_x=0.12$ m, $l_y=0.80$ m, and $l_z=0.30$ m. The top boundary is open to the atmosphere. Here, $\Omega$ is the rotation rate and $b$ is the width of the heated plume. \textcolor{blue}{(b) Structured \texttt{blockMesh} grid ($51\times96\times116$ in $x,y,z$) showing uniform cells in the baroclinic region ($0<z<l_z$), geometric stretching (expansion ratio 4) in the damping region ($z>l_z$), and local refinement around the bottom-edge heated strip of width $b$.}}
    \label{fig:schematic}
\end{figure}

In this work, we consider a three-dimensional rectangular section focusing on dynamics in the bulk region of the rotating cylindrical annulus as shown in Figure \ref{fig:schematic}. The bi-directional temperature gradients are implemented by a forced heated plume (with inlet velocity, $w$) present at the bottom outer periphery of the rectangular region and uniformly cooled inner wall. Such a configuration allows us to study the flow physics away from the Ekman layer, thereby saving computational time. The presence of a forced thermal plume introduces the source Richardson number, in addition to the Rayleigh and Taylor numbers, as a governing parameter of the flow,
\begin{equation}
Ri_{0}=\frac{\alpha g\,\Delta T_{p}\,b}{w^{2}},
\label{eq:Ri}
\end{equation}
which quantifies the ratio of buoyancy to the inlet momentum of the thermal plume. Since the present simulations exclude the Ekman boundary layers, the influence of rotation is more appropriately characterized by the flow Rossby number,
\begin{equation}
Ro=\frac{U_{y}}{2\Omega l_{x}},
\label{eq:Ro}
\end{equation}
where $U_y$ is the characteristic thermal-wind velocity 

\begin{equation}
U_y=\frac{\alpha g\,l_z\,\Delta T_x}{2\Omega l_x}.
\label{eq:RoT}
\end{equation}
and $l_x$ is the horizontal length scale. Unlike the Taylor number, which measures the relative importance of Coriolis and viscous forces, the Rossby number directly quantifies the ratio of inertial to Coriolis forces governing the bulk flow dynamics. Using Eqs.~(\ref{eq:Ro})--(\ref{eq:RoT})

\begin{equation}
Ro=\frac{\alpha g\,l_z\,\Delta T_x}{4\Omega^2l_x^2}.
\end{equation}

It is therefore exactly one quarter of the classical thermal Rossby number,
\begin{equation}
\Theta\equiv
\frac{\alpha g\,\Delta T_x\,l_z}{\Omega^2l_x^2}
=4Ro.
\end{equation}

Furthermore, it maps onto the convective Rossby number (with
$\kappa=\nu$, i.e.\ $Pr=1$) as

\begin{equation}
Ro_c\equiv
\left(\frac{Ra_z}{Pr\,Ta}\right)^{1/2}
\end{equation}

Table.~\ref{tab:Rossby} lists all five different non-dimensional numbers for the three rotation rates, so the readers can place our runs in any suitable rotating-convection parameter space \cite{ahlers2009heat, chong2020vortices, ding2026stratification, harlander2023new, scolan2017rotating}:
\newpage

\begin{table}[htbp]
\centering
\label{tab:Rossby}
\begin{tabular}{cccccc}
\hline
$\Omega$ (rad\,s$^{-1}$) & \qquad
$Ro$ & \qquad
$\Theta=4Ro$ & \qquad
$Ro_c$ & \qquad
$Ta$ & \qquad
$Ek=\nu/\Omega l_z^2$\\
\hline
0.89 & \qquad 0.27  & \qquad 1.07 & \qquad 0.2-- 0.7 & \qquad $6.5\times10^8$ & \qquad $1.3\times10^{-5}$\\
1.35 & \qquad 0.12  & \qquad 0.46 & \qquad 0.26--0.5 & \qquad $1.5\times10^9$ & \qquad $0.8\times10^{-5}$\\
1.81 & \qquad 0.065 & \qquad 0.26 & \qquad 0.14--0.36 & \qquad $2.7\times10^9$ & \qquad $0.6\times10^{-5}$\\
\hline
\end{tabular}
\caption{Comparison of different nondimensional parameters at three rotation rates.}
\end{table}

The nominal labels $Ro=0.3$, $0.1$, and $0.07$ are rounded values. Please note that $Ro_c$ spans a range of values at each $\Omega$, corresponding to the range of $\Delta T_z$ at each $\Omega$, and hence $Ra_z$.



A central question is whether the present rectangular configuration is dynamically equivalent to the bulk region of the cylindrical annulus, and in particular whether the removal of the Ekman layers alters the mode-selection physics. This issue was addressed experimentally in our companion study Banerjee \eal \cite{banerjee2018experimental} and numerically in the axisymmetric simulations of Banerjee\cite{banerjee2024axisymmetric}. Here we summarize the scaling argument underlying the present reduction and make the connection explicit in Sec. \ref{compdom}.


\textcolor{blue}{Two independent estimates justify neglecting the Ekman layers. First, the Ekman layer thickness
$h_E=(\nu/\Omega)^{1/2}=\sqrt{\Ek}\,l_z\approx1.0\text{--}0.7$~mm, so $h_E/l_z\approx2.5\text{--}3.6\times10^{-3}$
and the layers occupy less than $0.4\%$ of the fluid depth; their direct
viscous contribution to the vertically integrated bulk balance is
correspondingly small. Second, the Ekman pumping suction velocity scales as
$w_E/U_g\sim\sqrt{\Ek}/\Ro\approx0.03\text{--}0.04$ across our parameter range,
so the secondary meridional cell it drives is an order of magnitude
weaker than the primary geostrophic zonal circulation, and much weaker in the
low-$\Ek$, high-$\Ro$ corner.
The two estimates are consistent with the well-known Ekman-negligible branch
of the Eady problem~\cite{harlander2023new}, where baroclinic instability is
significantly altered by Ekman friction only when $h_E/H\gtrsim0.04$; the
present ratio $h_E/H\approx3\times10^{-3}$ is more than an order of magnitude
below this threshold, and the associated Ekman spin-down time
$\tau_E=\Omega^{-1}\Ek^{-1/2}\sim225\text{--}310$~s is $20\text{--}50$ times the
advective/baroclinic timescale $\tau_{\mathrm{adv}}=l_x/U_g=(\mathit{Ro}\,\Omega)^{-1}$, so friction acts on
a slow envelope relative to the wave dynamics. As empirical corroboration, the bulk mean flow predicted by the present model reproduces the sloping isotherms, the prograde upper and retrograde lower zonal jets, and the peak mean azimuthal velocity $|\langle v\rangle|$, observed in our Ekman-resolving cylindrical simulations (Swarnakar \textit{et al.}~\cite{swarnakar2023numerical}) to within $\pm15\%$.}

Furthermore, the Ekman boundary layers contribute only an $\mathcal{O}(\Ek^{1/2})$ ageostrophic mass flux that closes the weak secondary meridional circulation. Additionally, the Ekman-driven secondary streamfunction is smaller than the primary zonal circulation by a factor of $\mathcal{O}(\Ek^{1/2})$. For $\Ek = \nu/\Omega l_z^2 \approx \mathcal{O}(10^{-5})$, the resulting correction is only $\mathcal{O}(\Ek^{1/2}) \approx \mathcal{O}(10^{-3})$, indicating that the leading-order bulk dynamics are largely insensitive to whether the Ekman layers are explicitly resolved, provided the focus is on the quasi-steady baroclinic flow rather than spin-up transients or heat transport through the boundary layers themselves. This interpretation is consistent with the measurements reported in Banerjee \eal \cite{banerjee2018experimental}. The temperature-standard-deviation profiles in Figure~4 of Banerjee \eal \cite{banerjee2018experimental} show that the dominant thermal variance is concentrated in the baroclinic bulk, while the Ekman-layer contribution remains confined to thin regions near the top and bottom boundaries ($z\lesssim1$\,cm). The same conclusion follows from classical linear baroclinic-instability theory. On an $f$-plane, the most unstable Eady wavelength is determined by the bulk stratification and rotation according to        
\begin{equation}
\lambda_{\max}
=
2\pi \frac{NH}{f}\,
|k_{\mathrm{Eady}}|^{-1},
\end{equation}
where $k_{\mathrm{Eady}}\approx1.61$ is the corresponding non-dimensional wavenumber, $N$ is the Brunt--V\"ais\"al\"a frequency, $H$ is the fluid depth and the Coriolis parameter$f=2\Omega$. Evidently $\lambda_{\max}$ does not depend on the Ekman layers themselves. Removing the Ekman layers therefore does not modify the predicted dominant wavelength or the intrinsic baroclinic mode-selection mechanism. Their principal role is instead to provide a weak viscous damping of order $\mathcal{O}(\Ek^{1/2})$ that only slightly affects the wave growth rate and equilibration amplitude.

\textcolor{blue}{To summarize the above discussions, removing the Ekman layers, therefore, modifies neither the leading-order geostrophic balance nor the linear-stability characteristics
of the bulk. Consequently, the present idealization neglects Ekman pumping
and boundary-layer heat transport; the results should be interpreted as
describing the quasi-steady bulk wave--plume dynamics and bulk heat transport
rather than spin-up transients or boundary-layer heat-flux budgets.}

Sec. \ref{sec:num_methd} discusses the numerical methodology which includes the description of the computational domain, boundary condition, governing equations, and OpenFOAM solver. Sec. \ref{sec:rnd} discusses the baroclinic wave dynamics and forced heated plume dynamics using the mean temperature and velocity contours. The budgets of the $x$ and $z-$direction momentum equations are discussed to better comprehend the system. Moreover, the heat transport in the annulus is discussed in this section. Significant conclusions are discussed in Sec. \ref{sec:con}.

\section{\label{sec:num_methd}Numerical methodology}

In this section, we discuss the geometrical configuration of the domain, meshing, boundary conditions, governing equations, solver description, and stability criteria used in this work.

\subsection{\label{compdom} Computational domain, boundary conditions, and meshing}

The computational domain consists of a three-dimensional rectangular annulus as shown in Figure \ref{fig:schematic}. It is located at a distance of $x = 0.075$ m from the rotational axis which is centered at coordinates (0, 0, 0). The reference frame is rotating about the axis at a rate of $\Omega$ rad/s. The annulus is partitioned into two distinct regions: the baroclinic wave region $(0<z<l_z)$ and the flow-dampening region $(z > l_z)$. The baroclinic zone has a height of $l_z = 0.30$ m and extends $l_y = 0.80$ m in the azimuthal ($y$) direction. The annular gap width is $l_x = 0.12$ m, while the flow-dampening region of height $l_d = 0.20$ m is incorporated above the baroclinic zone. To facilitate the development of the baroclinic wave in the y-direction, we kept boundary walls perpendicular to $y-$direction as periodic. Furthermore, the height $(l_z)$ of the baroclinic region was selected to ensure the proper spreading of the forced plume in the $z-$direction. In Figure \ref{fig:schematic}, the blue patch depicts the uniformly cooled inner wall. The forced heated plume, characterized by inlet velocity $(w)$, serves as a heat source and is illustrated by the red patch in Figure \ref{fig:schematic}. This arrangement of heating and cooling manifests bi-directional temperature gradients in the rectangular annulus. \textcolor{blue}{The width of the heated \emph{source strip}, denoted $b$, measures $5$~mm; this sets the plume source size, while the plume width itself grows with height by entrainment and is not independently prescribed. The value of $b$ is chosen to match the experimental configuration of Banerjee \emph{et al.} \cite{banerjee2018experimental} for direct comparison.} The wave-mode selection is set by $l_y$, not $b$, provided $b\!\ll\!l_y$ (which is well-satisfied: $b/l_y = 0.006$). The top of the rectangular annulus remains open to the atmosphere, enabling the flow to exit the domain and thereby ensuring continuity.

\begin{table*}
    \centering
    \begin{tabular}{c c c c}
\hline 
\textbf{Parameter} & \textbf{Symbol} & \textbf{Variation} & \textbf{Units} \\
\hline
Frame rotation rate & $\Omega$ & 0.89, 1.35, 1.81 & rad $s^{-1}$\\
Plume inlet velocity & $w$ & 0.0012, 0.006, 0.012 & $m/s$ \\
Density & $\rho$ & 998.2 & kg $m^{-3}$\\
Kinematic viscosity & $\nu$ & $1.004\times10^{-6}$ & $m^{2}$ $s^{-1}$\\
Thermal diffusivity & $\kappa$ & $1.004\times10^{-6}$ & $m^{2}$ $s^{-1}$\\
Coeff. of thermal expansion & $\alpha$  & 0.000207 & $K^{-1}$\\
Rectangular annulus gap  & $\textit l_x$ & 0.12 & $m$ \\
Azimuthal length & $\textit l_y$ & 0.8 & $m$ \\
Baroclinic zone height  & $\textit l_z$ & 0.3 & $m$ \\
Flow-dampening region  & $\textit l_d$ & 0.2 & $m$ \\
Plume width & $b$ & 0.005 & $m$ \\
Initial ambient fluid temperature & $T_i$ & 293 & $K$ \\
Plume temperature at source & $T_p$ & 307  & $K$ \\ 
Cold wall temperature & $T_{c}$ & 287  & $K$ \\
Horizontal temperature gradient & $\Delta T_x = T_p - T_{c}$ & 20 & $K$ \\ 
Plume temperature gradient & $\Delta T_p = T_p - T_i$ & 14 & $K$ \\
Prandtl number & $Pr$ & 1& \\
Rayleigh number &$Ra = \frac{\alpha g \Delta T_x l_{x}^{3}}{\nu \kappa}$ & $ 7 \times 10^{7}$ \\
Characteristic zonal velocity & $U_{y} = \frac{\alpha g l_z \Delta T_x}{2\Omega l_x}$ \\
Rossby number & $Ro = \frac{U_{y}}{2 \Omega l_x}$ & 0.3, 0.1, 0.07 \\ 
Source Richardson number & $Ri_{0} = \frac{\alpha g \Delta T_p b}{w^2}$ & 99, 4, 1 \\

\hline
     \end{tabular}
    \caption{Properties and parameters used in the present study.}
    \label{tab:ParaTable}
\end{table*}

The fluid properties at $293\,\mathrm{K}$ are
$\nu = 1.004\times10^{-6}$\,m$^2$/s and
$\rho = 998.2$\,kg/m$^3$. \textcolor{blue}{The thermal diffusivity is set to $\kappa=\nu=1.004\times10^{-6}\,\mathrm{m^2\,s^{-1}}$, giving $Pr=1$. This is a deliberate idealisation rather than the use of a physical fluid: the kinematic viscosity of water is retained to preserve the characteristic length and velocity scales of the experimental apparatus of Banerjee \textit{et al.}\cite{banerjee2018experimental} for direct comparison, while only $\kappa$ is increased to reduce computational cost. A water-like fluid ($Pr\approx7$) would produce a thermal boundary layer thinner by a factor of $\sqrt{7}\approx2.6$, requiring substantially finer near-source spatial resolution and smaller time steps. The approximation primarily affects the thermal boundary layer, whereas the bulk baroclinic and geostrophic dynamics depend only weakly on $Pr$ for $Pr\gtrsim \mathcal{O}(1)$. In particular, the most-unstable Eady mode is inviscid and independent of $Pr$, consistent with the agreement in wave-mode selection between the present simulations and the $Pr\approx7$ cylindrical results reported in Table~\ref{tab:comparison_table}.}

We have incorporated cyclic boundary conditions for the velocity, temperature, and pressure fields at the front and back patches of the rectangular annulus (normal to the y-direction). This arrangement enables the flow leaving the front patch of the rectangular annulus to re-enter through the back patch, facilitating the proper development of baroclinic waves. \textcolor{blue}{At the heating zone a Dirichlet condition fixes the full velocity vector to the inlet value $(u,v,w)=(0,0,w)$ (purely vertical injection at speed $w$) and the temperature to $T=307$~K.}, while at the cold wall the temperature and velocity boundary conditions are set as fixed and free slip with no penetration $\left( \frac{\partial \mathbf{u}}{\partial \mathbf{n}} \quad \text{and} \quad \mathbf{u.n} = 0 \right)$ respectively. \textcolor{blue}{For the top boundary, the OpenFOAM \texttt{inletOutlet} boundary condition (Weller et al., \cite{weller1998tensorial}) is employed for the velocity field. This condition is preferred over a fixed-pressure outlet because the latter can permit local inflow through the open top whenever the vertical velocity reverses, thereby contaminating the upper bulk and the damping layer. The \texttt{inletOutlet} condition instead applies a zero-gradient condition during outflow ($\avg w>0$) and prescribes the specified inlet value during inflow ($\avg w<0$), thereby suppressing spurious recirculation while preserving global mass continuity.} Pressure and temperature boundary conditions are set to zero and adiabatic $(\frac{\partial T}{\partial n} = 0)$ respectively. For the remaining boundaries, the velocity is subjected to free slip with no penetration, while the temperature is set to adiabatic conditions. 

The computational domain is discretized using a structured grid. The grid points in the $x$, $y$, and $z$ directions are 51, 96, and 116 respectively. A structured grid is generated using the \texttt{`blockMesh'} utility in OpenFOAM. In the baroclinic wave region $(0<z<l_z)$, a uniform structured grid is employed, while in the flow-dampening region $(z > l_z)$ a non-uniform structured grid is utilized. The non-uniform grid incorporates stretching in the $z-$direction, with a grid expansion ratio set to 4, with an aim to numerically diffuse the forced plume. \textcolor{blue}{The baseline mesh ($51\times96\times116$) resolves the dynamically relevant bulk scale --- the deformation radius $L_\rho\sim \mathcal{O}(l_x)$ --- with $\sim50$ cells across the gap, and, because the Ekman layers are removed, there is no thin viscous boundary layer to resolve. The plume is buoyancy-dominated over essentially the whole depth ($z/L_M>5$ for all $Ri_0$), so the near-source jet region need not be finely resolved; we note that the source width $b=5$~mm is spanned by $\sim2$ cells in $x$, so the immediate near-inlet plume is only marginally resolved, which affects the absolute $\Gamma(z)$ near $z=0$ but not the bulk-integrated diagnostics. To confirm grid independence we repeated the representative case $Ri_0=4,\Ro=0.1$ on coarse and fine meshes (please refer Sec. \ref{appendix}).}

\subsection{\label{goveqtn} Governing equations, solver description, stability criteria, and flow dampening}

In this investigation, the working fluid is assumed to be incompressible $(\rho = const)$, and the Boussinesq approximation is utilized to model the buoyancy-driven flow.  Here, fluid flow is governed by a system of continuity (Eq. \ref{eqn:continuity}), momentum (Eq. \ref{eqn:momentum}), and temperature equations (Eq. \ref{eqn:Tequation}). These equations are solved over time to determine the pressure ($p(\mathbf{x},t)$), velocity $(\mathbf{u}(\mathbf{x},t))$, and temperature ($T(\mathbf{x},t)$) fields.

\begin{eqnarray}
\label{eqn:continuity}
\nabla\cdot \mathbf{u} &=&0 \\
\label{eqn:momentum}
\frac{\partial \mathbf{u}}{\partial t}+\mathbf{u}\cdot\nabla\mathbf{u} +\mathbf{f}\times \mathbf{u} &=&-\nabla p -\mathbf{g\cdot h}\nabla \rho _{k} + \nu \nabla^2 \mathbf{u}  \\
\label{eqn:Tequation}
\dfrac{\partial T}{\partial t} + \mathbf{u}.\nabla T  &=& \kappa \nabla^2 T 
\end{eqnarray} 

Here, the Coriolis parameter is defined as $\mathbf{f}=f\mathbf{k}$, with $ f = 2 \Omega$. In this expression, $\Omega$ represents the frame rotation rate, and its direction is anti-clockwise (into the plane of the paper). In the momentum equation (Eq. \ref{eqn:momentum}), the pressure $(p)$ represents the fluctuation from the hydrostatic pressure $(p_{hyd} = \rho \mathbf{g}\cdot\mathbf{h})$, divided by the density $(\rho)$. The Coriolis acceleration term $(\mathbf{f} \times \mathbf{u})$ is included to account for the effect of frame rotation on fluid flow. Additionally, $\rho_k = 1 - \alpha(T -T_{ref})$, where $\alpha$ is the coefficient of thermal expansion, $T$ is the temperature, $T_{ref}$ is the reference temperature, and $\mathbf{h}$ represents the position vector. The fluid properties and other relevant parameters are detailed in Table \ref{tab:ParaTable}. An open-source CFD toolkit OpenFOAM (v2006) is employed to solve the governing equations utilizing finite volume methods. These equations are discretized on a structured grid. \textcolor{blue}{To solve the discretized governing equations, the Pressure Implicit with Splitting of Operators (PISO) algorithm, originally developed by Issa \cite{issa1986solution}, was employed for pressure--velocity coupling, using two pressure-correction steps (\texttt{nCorrectors} = 2) and one non-orthogonal correction (\texttt{nNonOrthogonalCorrectors} = 1). Being a non-iterative within-step scheme, PISO uses no steady-state under-relaxation (all under-relaxation factors set to unity); per-step linear-solver tolerances were $10^{-7}$ (pressure, GAMG/PCG) and $10^{-8}$ (velocity/temperature, smoothSolver). Statistical stationarity was confirmed by the vanishing unsteady budget term ($U_z\approx0$, Figure \ref{fig:z_mom_bdgt}) and by stationary time-averaged statistics accumulated over the final $5000$~s.} The solution of the governing equations is first-order accurate in time and second-order accurate in space. The flow in the rectangular annulus is considered laminar and hence turbulence models are not used. The stability of the numerical simulation is ensured by the Courant–Friedrichs–Lewy (CFL) condition. The Courant number, $Co$, is maintained below 0.1 in all simulations to ensure numerical stability. In each simulation, the time step $(\delta t)$ was progressively reduced with increasing frame rotation rate $(\Omega)$ to satisfy the numerical stability criterion, while the maximum Courant number was continuously monitored using the runtime diagnostics available in OpenFOAM. The observed maximum Courant numbers were $\mathrm{Co}_{\max}\approx 0.08$ for $\Omega = 0.89$\,rad/s with $\delta t = 0.01$\,s, $\mathrm{Co}_{\max}\approx 0.07$ for $\Omega = 1.35$\,rad/s with $\delta t = 0.005$\,s, and $\mathrm{Co}_{\max}\approx 0.06$ for $\Omega = 1.81$\,rad/s with $\delta t = 0.0025$\,s, all of which remain comfortably below the imposed bound of $\mathrm{Co}<0.1$. The chosen time steps are also significantly smaller than the characteristic time scale associated with the slowest baroclinic-wave oscillations in the system. Estimating the wave period as $T_w = 2\pi/c|k| \sim
2\pi/(\mathcal{O}(10^{-3}\text{-}10^{-4})\times 2\pi m/0.8) \approx 267\text{-}4000\,\mathrm{s}$, the corresponding number of time steps per wave period is $T_w/\delta t = \mathcal{O}(10^4\text{-}10^6)$, indicating that the temporal resolution is substantially higher than that required to accurately resolve the baroclinic-wave dynamics. The phase-speed estimates use the time-lag $\tau_{\mathrm{lag}}$ between the
real and imaginary PCs of the CEOF, which is determined to within one sample
($\delta\tau = 8.3$\,s at our diagnostic sampling rate). The fractional uncertainty is $\delta c/c \sim \delta\tau/\tau_{\mathrm{lag}} \approx 4\%$. Each simulation ran for $\sim 8000$\,s of physical time, equivalent to $\sim$1300--2300 rotation periods. Statistical steady state is reached after $\sim$3000\,s; statistics are computed over the final 5000\,s.

\textcolor{blue}{The forced plume rising from the heating zone is absorbed in the flow-dampening region ($z>l_z$) before it reaches the open top, so that numerical reflection of its upward momentum flux at the open boundary cannot excite spurious oscillations in the baroclinic zone. The absorbing layer applies a smooth cosine ramp}

\begin{equation}
f(z)=\tfrac12\,k_d\!\left[1-\cos\!\left(
\frac{\pi\,|z-z_1|}{z_2-z_1}
\right)\right],
\qquad
z_1=0.30\ \mathrm{m}\le z\le z_2=0.50\ \mathrm{m}.
\label{eq:damping_function}
\end{equation}

\textcolor{blue}{which is the standard form for an absorbing layer in geophysical-fluid simulations and is reflection-free at the interface because $\partial f/\partial z=0$ at $z=z_1$, eliminating any impedance mismatch. For $Ri_0 = 99$ and $4$, the dampening coefficient is 1, while for the case of $Ri_0 = 1$, it is 3. The dampening must counteract the
plume momentum that would otherwise reflect off the open top boundary.  The plume vertical momentum flux at the inlet scales as $M_0 = w^2 b \propto 1/\Rio$ (at fixed $\Delta T_p$), so for $\Rio=1$ the
plume carries $\sim$100$\times$ more vertical momentum than at $\Rio=99$. A dampening coefficient $k_d=1$ was insufficient at $\Rio=1$, wherein residual
oscillations of the bulk fluid persisted with $\sim$5\% of the inlet velocity amplitude, contaminating the CEOF. A dampening coefficient $k_d=3$ reduced these oscillations to $<0.5\%$.}

\textcolor{blue}{Flow dampening is achieved through a reduction in the instantaneous velocity $(\mathbf{u})$ within the dampening region. This is accomplished by subtracting the instantaneous velocity vector $(\mathbf{u})$ from the product of the mean velocity vector and the cosine function $(\mathbf{\overline{u}}*f(z))$ where $\mathbf{\overline{u}}$ is the time-averaged of $\mathbf{u}$. Damping is enforced on the velocity field only; the temperature field is not damped. All data from $z>l_z$ are excluded from the analysis, so the contours of Figures~\ref{fig:mean_plot_u}--%
\ref{fig:mean_plot_w} (plotted for $z<l_z$) are unaffected. Subsequently, the data from the flow-dampening region is not included in the preparation of the results, discussed in Sec. \ref{sec:rnd}. The implementation modifies the acoustic-damping utility of OpenFOAM (v2006).
}

The modified OpenFOAM (v2006) solver has been validated against the experiments of Banerjee \eal \cite{banerjee2018experimental} and existing numerical literature (see Figure 4 and Figure 15 of Swarnakar \eal \cite{swarnakar2023numerical}). In particular, the validation includes comparison of the mean zonal velocity profiles with the experimental and numerical results reported in \cite{banerjee2018experimental, banerjee2021investigation}, yielding relative errors of less than $8\%$ in the peak $\langle v\rangle$ and less than $10\%$ in the dominant wave mode number. 

\subsection{\label{justification_rect_domain} Justification of the rectangular approximation}

There are several reasons for considering rotating three-dimensional rectangular section over the rotating cylindrical annulus in this study. These are: 

\begin{enumerate}
\renewcommand{\labelenumi}{(\roman{enumi})}
    \item The annulus
gap--to--mean--radius ratio in the Swarnakar \eal\cite{swarnakar2023numerical} is
\(\delta = (r_o-r_i)/\tfrac12(r_o+r_i)
       = 0.12/0.135 \approx 0.89,\) corresponding $r_i=0.075$ m and $r_o=0.195$ m. Therefore, the gap is comparable to the radius and the curvature is not formally
small.  However, the dynamical question is not whether curvature is
geometrically small but whether it modifies the \emph{baroclinic--wave
mechanism} we wish to study.  In the bulk region, the relevant length scale
for the baroclinic mode is the deformation radius
\(L_\rho = NH/f \sim \mathcal{O}(l_x)\),
which is set by stratification and rotation, not by the mean radius. Here $N = \sqrt{-\frac{g}{\rho_0}\frac{d\rho}{dz}}$ is the Brunt-Väisälä frequency (measure of stratification).
Curvature enters only through the \textcolor{blue}{curvature parameter }
\(\beta_{\mathrm{cyl}} = -f/r\),
whose magnitude relative to the leading Coriolis term is
\(\beta_{\mathrm{cyl}} l_y/f \approx l_y/r_m \approx 0.59\), where $r_m = (r_o + r_i)/2$. This correction would matter for a single wave that travels around the full
cylinder, but in our reduced rectangular section with periodic side walls each
mode samples only the bulk dynamics, not the curvature-induced phase
modulation.  This is precisely the same level of approximation that underlies
the classical Eady--type analyses of mid-latitude baroclinic instability on a
$\beta$-plane. \textcolor{blue}{We emphasise that $\beta_{\mathrm{cyl}}$ is a curvature metric term and is \emph{not} the planetary $\beta=\mathrm{d}f/\mathrm{d}y$ of the $\beta$-plane; it produces a phase modulation only for a wave that circumnavigates the full cylinder, which the periodic rectangular section excludes.}

\begin{table*}
\centering
\small
\setlength{\tabcolsep}{6pt}
\renewcommand{\arraystretch}{1.2}

\begin{tabular}{c c c}
\hline
\textbf{Quantity} & \textbf{Cylindrical \cite{swarnakar2023numerical}} & \textbf{Rectangular (present work)} \\
\hline

Slope of bulk isotherms 
& inclined 
& inclined \\

Sign of zonal jet asymmetry 
& \begin{tabular}[c]{@{}c@{}}prograde upper / \\ retrograde lower\end{tabular}
& \begin{tabular}[c]{@{}c@{}}prograde upper/ \\ retrograde lower\end{tabular} \\

Magnitude of $\langle v\rangle$ peak ($m\,s^{-1}$) 
& $\sim 0.02$ 
& $\sim 0.02$ (within $\pm 15\%$) \\

Geostrophic balance in bulk 
& yes 
& yes \\

\hline
\end{tabular}

\caption{Comparison between the cylindrical annulus results of Swarnakar \eal\ \cite{swarnakar2023numerical} and the present rectangular model.}
\label{tab:comparison_table}

\end{table*}

\item In Sec. \ref{sec:Rec_mean_field} we qualitatively compared the mean flow structures with the cylindrical annulus simulations of Swarnakar \eal\ \cite{swarnakar2023numerical}. A more quantitative comparison at comparable forcing conditions ($Ta = 6.5\times10^8$ in the cylindrical case and $\Ro = 0.3$ with $\Rio = 4$ in the present rectangular configuration) is summarized in Table \ref{tab:comparison_table}. Most of the large-scale dynamical signatures, including the inclined bulk isotherms, the upper prograde and lower retrograde zonal jets, and the magnitude of the mean azimuthal velocity, remain in close agreement between the two configurations. 

The dominant wave mode is the primary quantity directly influenced by curvature. In a cylindrical annulus. the integer mode
number $m$ is fixed by $2\pi r_m/\lambda$, whereas in our rectangular section the spectrum is set by $l_y/\lambda$.  The two are simply related: if our geometry sustained the same wavelengths as the cylindrical case, the rectangular mode would be $m_{\mathrm{rect}} = (l_y/2\pi r_m)\, m_{\mathrm{cyl}}
\approx 0.94\, m_{\mathrm{cyl}}$. Hence a cylindrical mode 4 maps to a rectangular
mode of $\approx 4$ at the same wavelength--so the modes 2-3 we observe
correspond to physically longer baroclinic waves than those of
Swarnakar \eal\cite{swarnakar2023numerical}, an effect we attribute to the strong forcing of the plume
(see Sec.~\ref{sec:Rec_plume}).

\end{enumerate}

\section{\label{sec:rnd}Results and discussions}

In this section, we look into the analysis and discussion of the flow physics for different parameters used in our study. As illustrated in Table \ref{tab:ParaTable}, the source Richardson number values are $Ri_0 =$ 99, 4, and 1, and the Rossby number values are $Ro = $ 0.3, 0.1, and 0.07. The largest $Ri_0$ value is based on the maximum vertical velocity $(w)$ observed in simulations of Swarnakar \eal \cite{swarnakar2023numerical}. Other values of $Ri_0$ number are selected such that the significant effect of increased plume inlet velocity could be observed on the flow dynamics of the annulus. Here, the Rayleigh number, $Ra = \frac{\alpha g \Delta T_x l_x^{3}}{\nu \kappa }$, value is  $7 \times 10^7$. and it is determined using the annulus gap $(l_x)$ and the temperature gradient along the annulus gap $(\Delta T_x)$.

\subsection{\label{sec:Rec_mean_field}Analysis of mean temperature and velocity field}

In this section, we examine plots of the mean temperature $\avg{T}$ and velocity fields $\avg{\mathbf{u}}$ in an $x-z$ plane for a range of  $Ri_0 = 99, 4, 1$ and $Ro = 0.3, 0.1, 0.07$. Here, the operator $\avg{\cdot}$ denotes averaging over time and in the $y$-direction. 

\begin{figure}
\centering
\subfigure[]{%
\includegraphics[scale = 0.42]{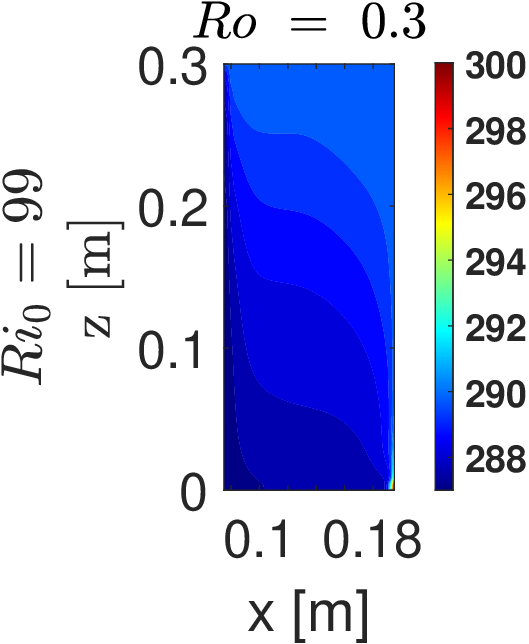}
\label{fig:TMean_w_1_2_Ta_low}}
\hspace{1cm}
\subfigure[]{%
\includegraphics[scale = 0.42]{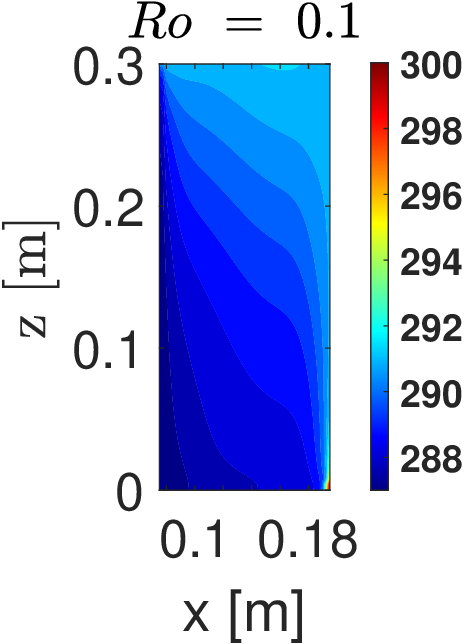}
\label{fig:TMean_w_1_2_Ta_mod}}
\hspace{1cm}
\subfigure[]{%
\includegraphics[scale = 0.42]{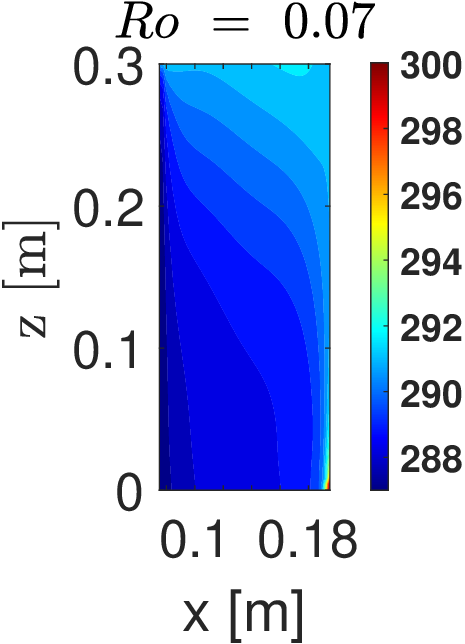}
\label{fig:TMean_w_1_2_Ta_high}}
\hspace{1cm}
\\
\subfigure[]{%
\includegraphics[scale = 0.4]{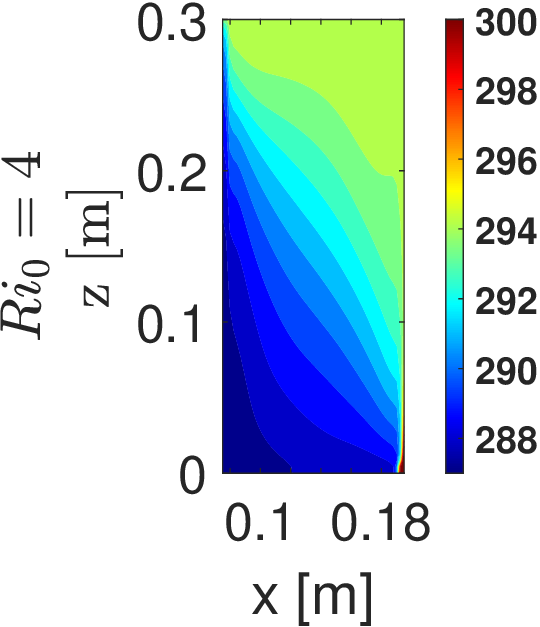}
\label{fig:TMean_w_6_Ta_low}}
\hspace{1cm}
\subfigure[]{%
\includegraphics[scale = 0.4]{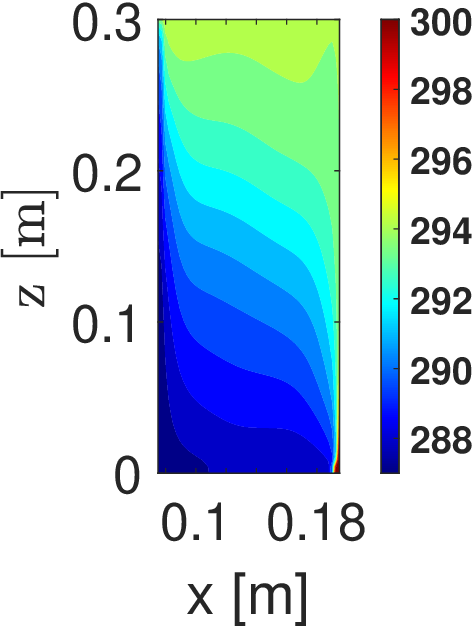}
\label{fig:TMean_w_6_Ta_mod}}
\hspace{1cm}
\subfigure[]{%
\includegraphics[scale = 0.4]{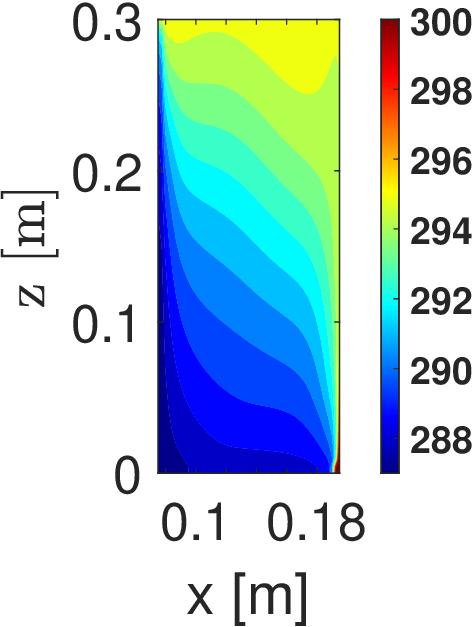}
\label{fig:TMean_w_6_Ta_high}}
\\
\subfigure[]{%
\includegraphics[scale = 0.4]{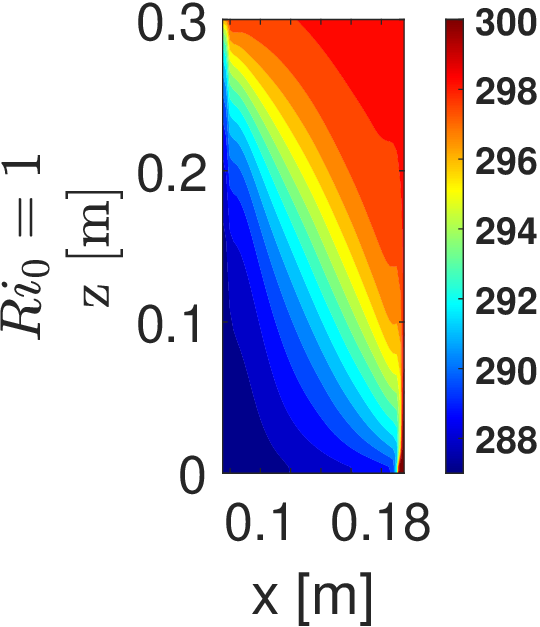}
\label{fig:TMean_w_12_Ta_low}}
\hspace{1cm}
\subfigure[]{%
\includegraphics[scale = 0.4]{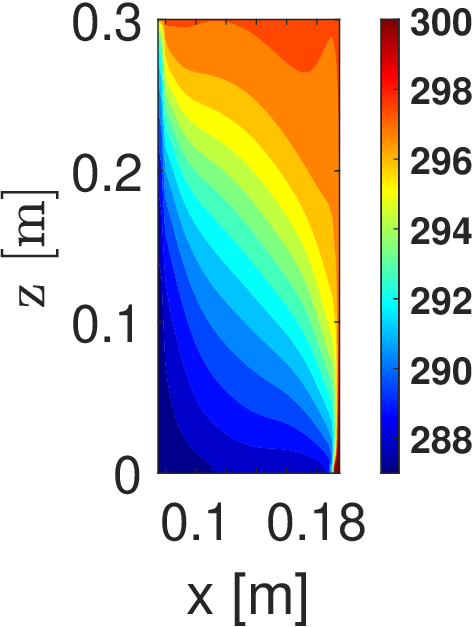}
\label{fig:TMean_w_12_Ta_mod}}
\hspace{1cm}
\subfigure[]{%
\includegraphics[scale = 0.4]{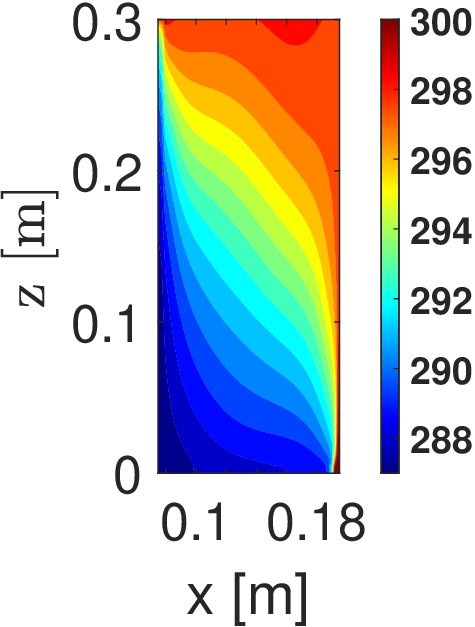}
\label{fig:TMean_w_12_Ta_high}}

\caption{Isocontours of the time and y-direction averaged temperature $\avg{T}$ \textcolor{blue}{for the rectangular baroclinic annulus,} for a range of $Ri_0$ and $Ro$. Here, $Ri_0=$ 99 and $Ro =$ 0.3, 0.1, 0.07 for [(a), (b), (c)], $Ri_0=$ 4 and $Ro =$ 0.3, 0.1, 0.07 for [(d), (e), (f)], and $Ri_0=$ 1 and $Ro =$ 0.3, 0.1, 0.07 for [(g), (h), (i)] respectively. Unit is in $K$. \textcolor{blue}{The axes are not drawn to scale. The horizontal-to-vertical aspect ratio is exaggerated for improved visualization of the flow field.}}

\label{fig:mean_plot_T}
\end{figure}

The structure of the mean fields in Figures~\ref{fig:mean_plot_T}--%
\ref{fig:mean_plot_w} is most readily understood by casting the
governing equations Eqs.~\eqref{eqn:continuity}-\eqref{eqn:Tequation} in non-dimensional form.  Using the
scales
\begin{equation}
x^{*} = \frac{x}{l_x}, \quad
z^{*} = \frac{z}{l_z}, \quad
u^{*} = \frac{u}{U_y}, \quad
w^{*} = \frac{w}{U_y}, \quad
t^{*} = \frac{t}{l_x/U_y}, \quad
p^{*} = \frac{p}{f\,U_y\,l_x}, \quad
\theta^{*} = \frac{T-T_{\mathrm{ref}}}{\Delta T_{x}},
\label{eq:nondim_vars}
\end{equation}
the $y$-averaged horizontal and vertical momentum
equations become
\begin{align}
\Ro\!\left(\frac{\partial u^{*}}{\partial t^{*}}
  + u^{*}\frac{\partial u^{*}}{\partial x^{*}}
  + w^{*}\frac{\partial u^{*}}{\partial z^{*}}\right)
  - v^{*}
  &= -\frac{\partial p^{*}}{\partial x^{*}}
     + \Ek\,\nabla^{*2} u^{*},
\label{eq:nondim_xmom}\\[4pt]
\Ro\!\left(\frac{\partial w^{*}}{\partial t^{*}}
  + u^{*}\frac{\partial w^{*}}{\partial x^{*}}
  + w^{*}\frac{\partial w^{*}}{\partial z^{*}}\right)
  &= -\frac{\partial p^{*}}{\partial z^{*}}
     + \frac{\Ra\,\Ek^{2}}{\Pran\,\Ro}\,\theta^{*}
     + \Ek\,\nabla^{*2} w^{*},
\label{eq:nondim_zmom}
\end{align}
while the temperature equation reads
\begin{equation}
\frac{\partial \theta^{*}}{\partial t^{*}}
+\mathbf{u}^{*}\!\cdot\!\nabla^{*}\theta^{*}
=\frac{1}{\mathrm{Pe}}\,\nabla^{*2}\theta^{*},
\qquad
\mathrm{Pe} = \frac{\Ro\,\Pran}{\Ek}.
\label{eq:nondim_theta}
\end{equation}
For the present parameter range, $\Ro = 0.07$--$0.30$,
$\Ek = \mathcal{O}(10^{-5})$, $\Ra\,\Ek^{2}/(\Pran\,\Ro) = l_{x}/l_{z}
\approx 0.4 = \mathcal{O}(1)$, and $\mathrm{Pe} = \mathcal{O}(10^{3}\text{--}10^{4})$.
The leading-order balance ($\Ro\to 0$, $\Ek\to 0$) is therefore
\begin{equation}
v^{*} = \frac{\partial p^{*}}{\partial x^{*}}
\qquad\text{(geostrophic)},
\qquad\qquad
\frac{\partial p^{*}}{\partial z^{*}}
   = \frac{l_{x}}{l_{z}}\,\theta^{*}
\qquad\text{(hydrostatic)},
\label{eq:leading_balance}
\end{equation}
and cross-differentiating eliminates the pressure to give the
thermal-wind relation \textcolor{blue}{(see, e.g., Vallis, \emph{Atmospheric and Oceanic Fluid Dynamics}, 2017\cite{vallis2017})}
\begin{equation}
\frac{\partial v^{*}}{\partial z^{*}}
= \frac{l_{x}}{l_{z}}\,\frac{\partial \theta^{*}}{\partial x^{*}},
\qquad\text{equivalently}\qquad
\frac{\partial \avg{v}}{\partial z}
= \frac{g\alpha}{f}\,\frac{\partial \avg{T}}{\partial x}.
\label{eq:thermal_wind}
\end{equation}
Three features of the mean fields follow directly.

(i) The zonal jet $\avg{v}$ is the geostrophic flow (Figure~\ref{fig:mean_plot_v}).
At leading order $\avg{v}$ is determined entirely by the horizontal pressure gradient through Eq.~\eqref{eq:leading_balance}, and its vertical shear is set by the horizontal temperature gradient through the thermal-wind relation Eq.~\eqref{eq:thermal_wind}. This explains why $\langle v \rangle$ attains its largest values in regions where $\partial \langle T \rangle/\partial x$ is strongest, namely near the sidewalls, and within the bulk only for low-$Ri_0$ cases that sustain a pronounced interior temperature gradient. It also explains why, for a fixed $Ri_0$, $\langle v \rangle$ decreases with decreasing $Ro$: increasing the rotation rate weakens the thermal-wind shear, which scales as $1/f$ for a given $\partial \langle T \rangle/\partial x$, while $\partial\avg{T}/\partial x$ itself also becomes weaker.

(ii) The slanted isotherms of
Figure~\ref{fig:mean_plot_T} are the thermal-wind balance made visible. Because $\mathrm{Pe}\gg 1$, the mean temperature field is governed by advection rather than diffusion away from thin wall layers, so the steady isotherm pattern is set by the balance between the imposed bidirectional forcing and the advecting circulation.  The slope of the bulk isotherms is a direct manifestation of the thermal-wind balance,
$\partial\avg{v}/\partial z \propto \partial\avg{T}/\partial x$: a sheared zonal jet and slanted isotherms are two views of the same geostrophic--hydrostatic state. When the inlet buoyancy flux is weak (e.g., $Ri_0 = 99$ at low $Ro$), the bulk radial temperature gradient, $\partial \langle T \rangle/\partial x$, is correspondingly small. As a result, the thermal-wind shear weakens, and the bulk isotherms become nearly vertical.

\begin{figure}
\centering
\subfigure[]{%
\includegraphics[scale = 0.42]{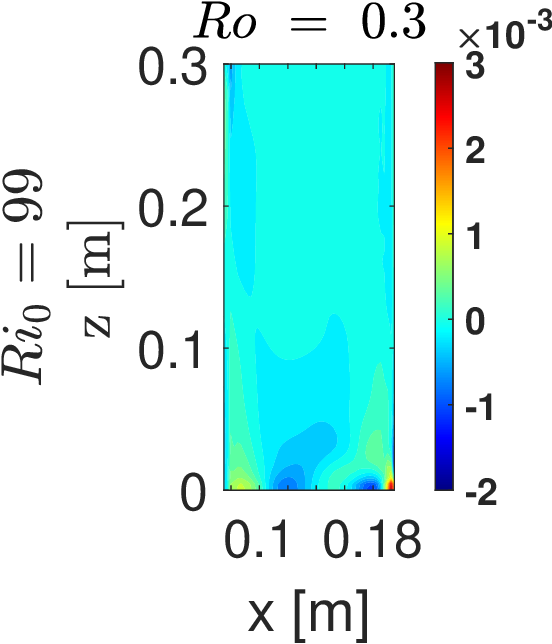}
\label{fig:UMean_w_1_2_Ta_low}}
\hspace{1cm}
\subfigure[]{%
\includegraphics[scale = 0.42]{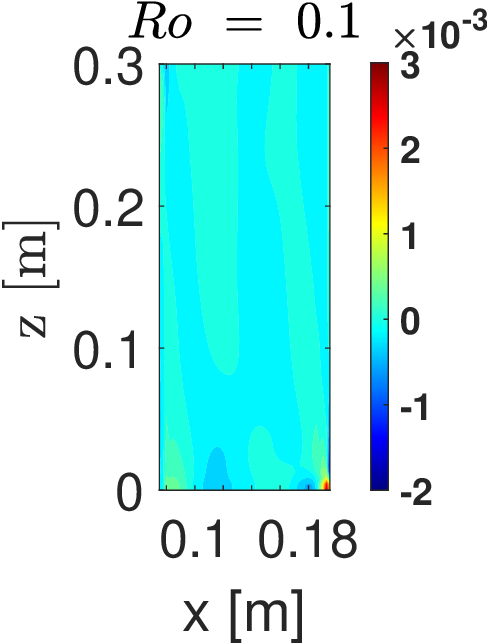}
\label{fig:UMean_w_1_2_Ta_mod}}
\hspace{1cm}
\subfigure[]{%
\includegraphics[scale = 0.42]{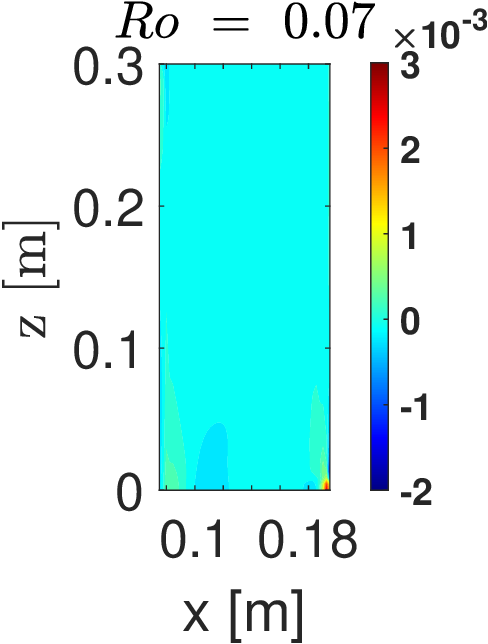}
\label{fig:UMean_w_1_2_Ta_high}}
\hspace{1cm}
\\
\subfigure[]{%
\includegraphics[scale = 0.4]{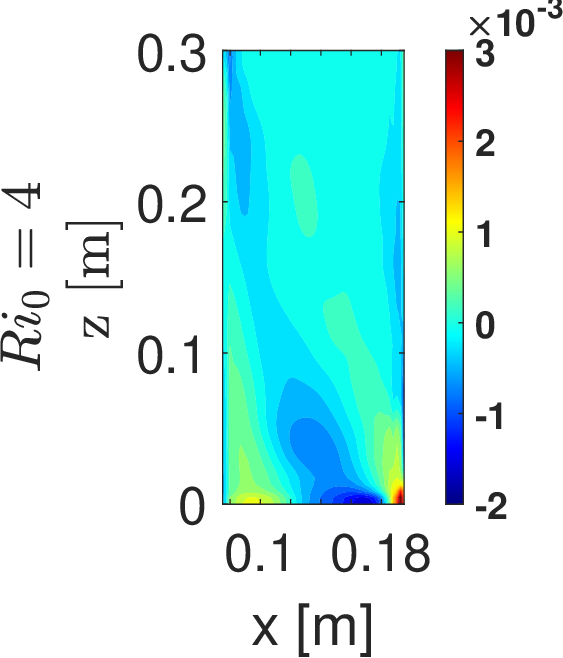}
\label{fig:UMean_w_6_Ta_low}}
\hspace{1cm}
\subfigure[]{%
\includegraphics[scale = 0.4]{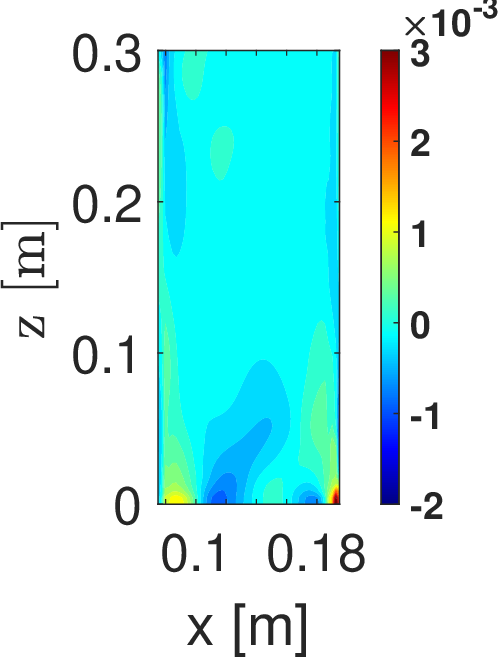}
\label{fig:UMean_w_6_Ta_mod}}
\hspace{1cm}
\subfigure[]{%
\includegraphics[scale = 0.4]{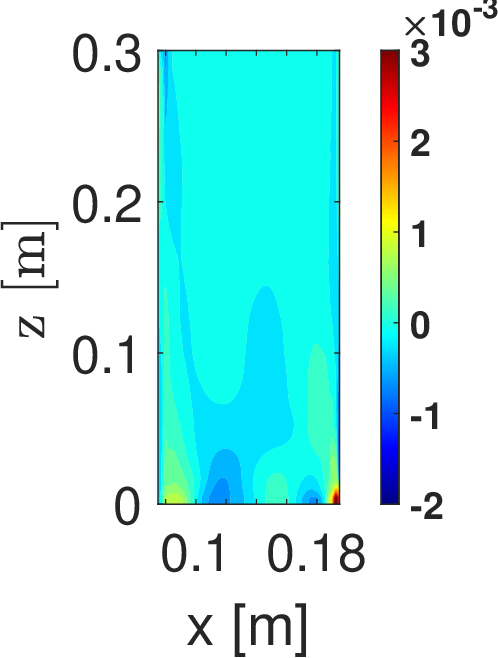}
\label{fig:UMean_w_6_Ta_high}}
\\
\subfigure[]{%
\includegraphics[scale = 0.4]{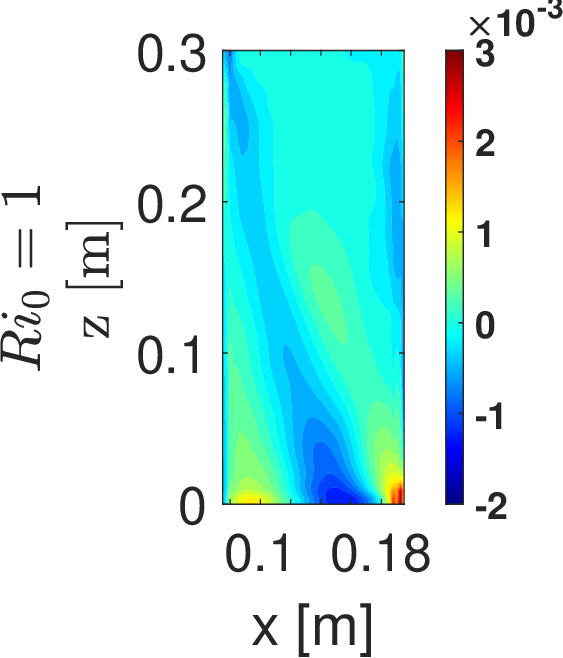}
\label{fig:UMean_w_12_Ta_low}}
\hspace{1cm}
\subfigure[]{%
\includegraphics[scale = 0.4]{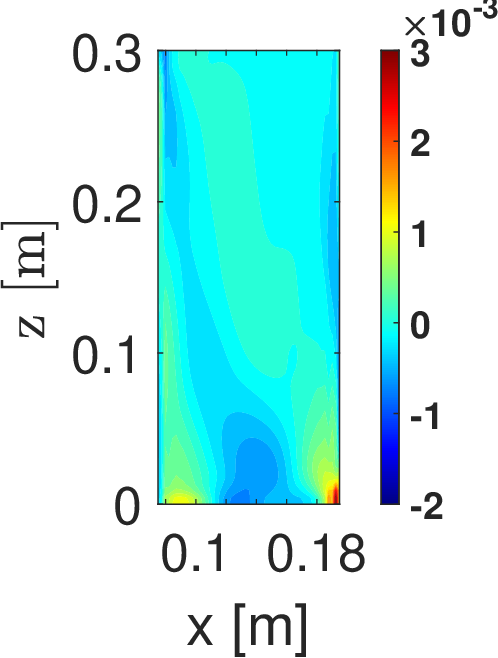}
\label{fig:UMean_w_12_Ta_mod}}
\hspace{1cm}
\subfigure[]{%
\includegraphics[scale = 0.4]{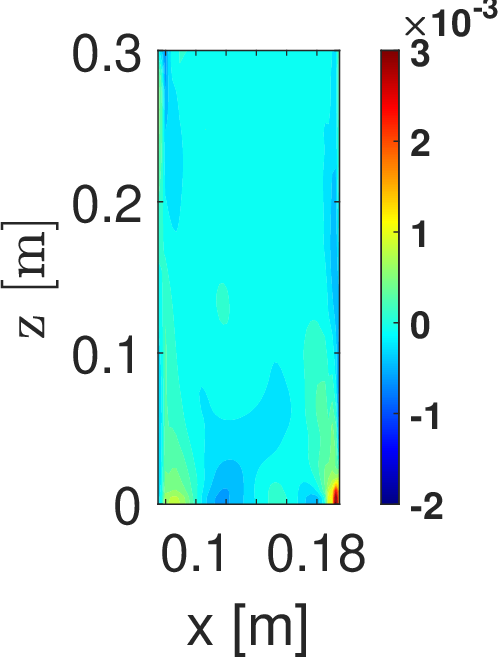}
\label{fig:UMean_w_12_Ta_high}}

\caption{Isocontours of the time and y-direction averaged $x$ component of velocity $\avg{u}$ \textcolor{blue}{for the rectangular baroclinic annulus,} for a range of $Ri_0$ and $Ro$. Here, $Ri_0=$ 99 and $Ro =$ 0.3, 0.1, 0.07 for [(a), (b), (c)], $Ri_0=$ 4 and $Ro =$ 0.3, 0.1, 0.07 for [(d), (e), (f)], and $Ri_0=$ 1 and $Ro =$ 0.3, 0.1, 0.07 for [(g), (h), (i)] respectively. Unit is in $m/s$. \textcolor{blue}{The axes are not drawn to scale. The horizontal-to-vertical aspect ratio is exaggerated for improved visualization of the flow field.}}

\label{fig:mean_plot_u}
\end{figure}

\begin{figure}
\centering
\subfigure[]{%
\includegraphics[scale = 0.42]{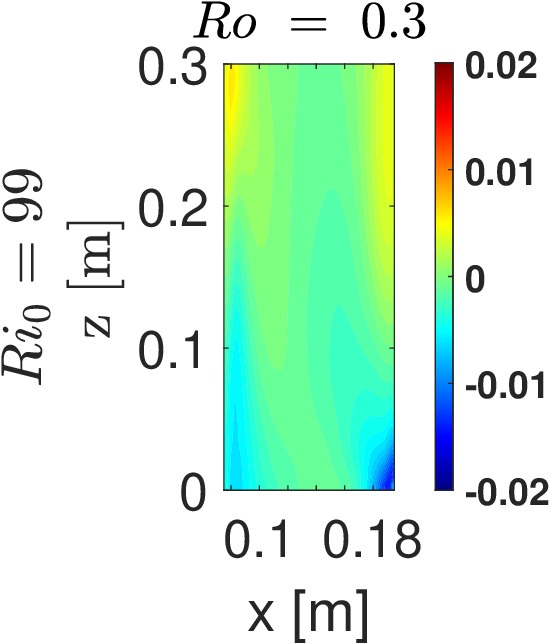}
\label{fig:VMean_w_1_2_Ta_low}}
\hspace{1cm}
\subfigure[]{%
\includegraphics[scale = 0.42]{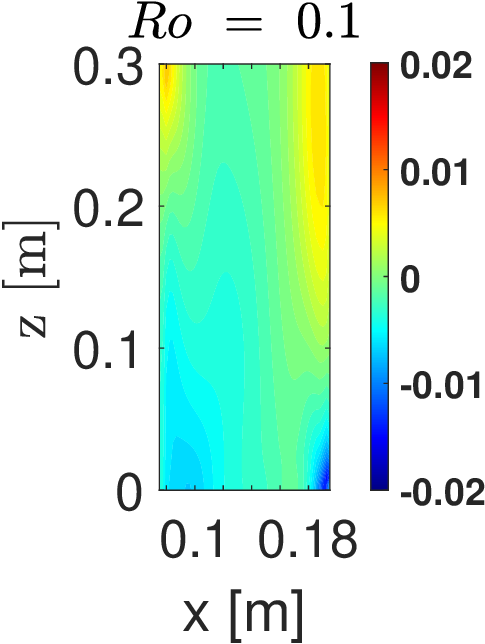}
\label{fig:VMean_w_1_2_Ta_mod}}
\hspace{1cm}
\subfigure[]{%
\includegraphics[scale = 0.42]{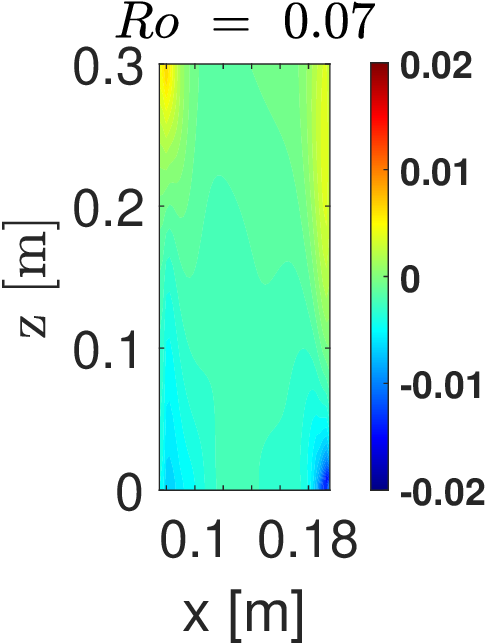}
\label{fig:VMean_w_1_2_Ta_high}}
\hspace{1cm}
\\
\subfigure[]{%
\includegraphics[scale = 0.4]{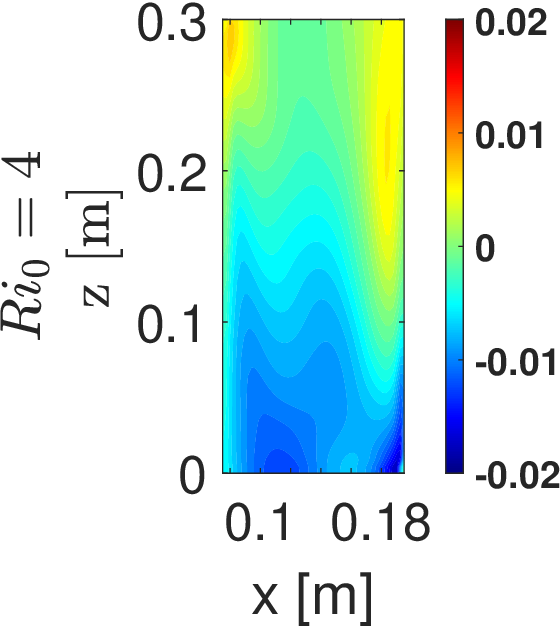}
\label{fig:VMean_w_6_Ta_low}}
\hspace{1cm}
\subfigure[]{%
\includegraphics[scale = 0.4]{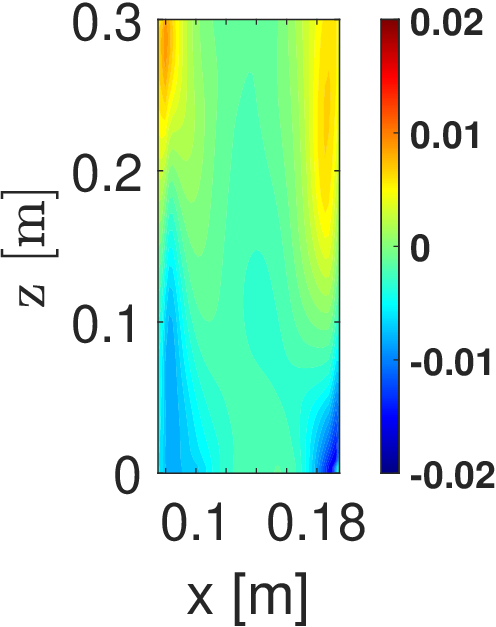}
\label{fig:VMean_w_6_Ta_mod}}
\hspace{1cm}
\subfigure[]{%
\includegraphics[scale = 0.4]{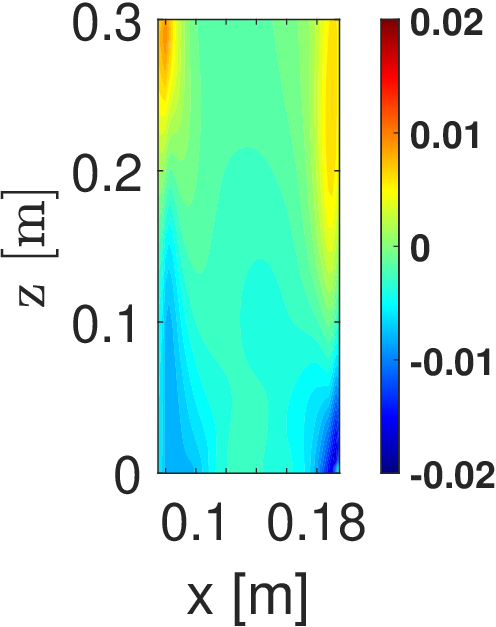}
\label{fig:VMean_w_6_Ta_high}}
\\
\subfigure[]{%
\includegraphics[scale = 0.4]{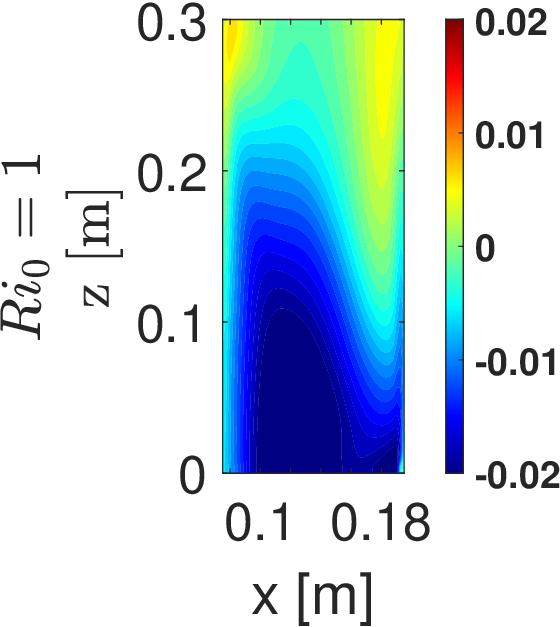}
\label{fig:VMean_w_12_Ta_low}}
\hspace{1cm}
\subfigure[]{%
\includegraphics[scale = 0.4]{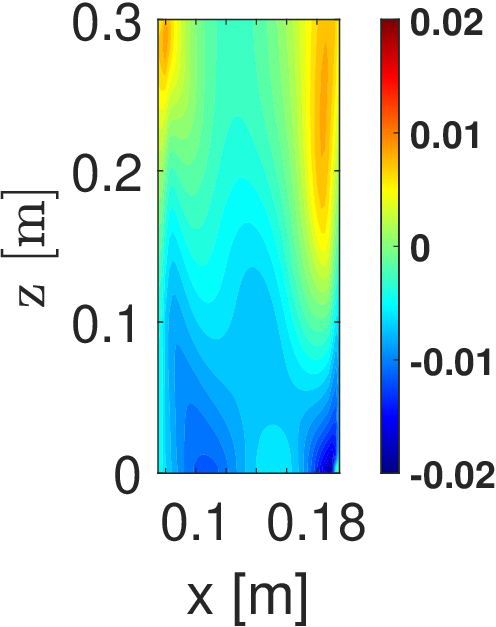}
\label{fig:VMean_w_12_Ta_mod}}
\hspace{1cm}
\subfigure[]{%
\includegraphics[scale = 0.4]{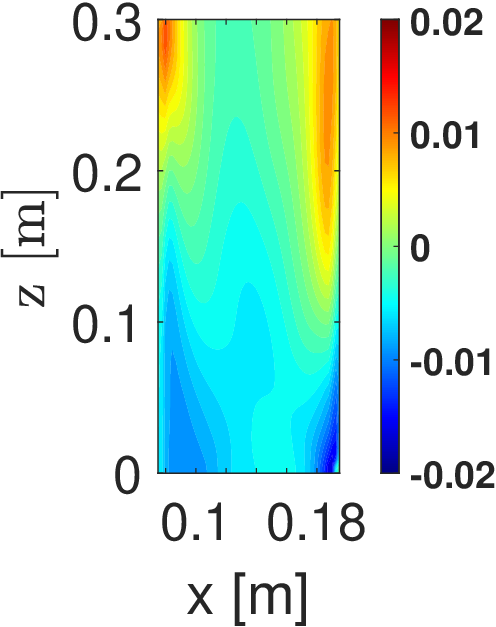}
\label{fig:VMean_w_12_Ta_high}}

\caption{Isocontours of the time and y-direction averaged $y$ component of velocity $\avg{v}$ \textcolor{blue}{for the rectangular baroclinic annulus,} for a range of $Ri_0$ and $Ro$. Here, $Ri_0=$ 99 and $Ro =$ 0.3, 0.1, 0.07 for [(a), (b), (c)], $Ri_0=$ 4 and $Ro =$ 0.3, 0.1, 0.07 for [(d), (e), (f)], and $Ri_0=$ 1 and $Ro =$ 0.3, 0.1, 0.07 for [(g), (h), (i)] respectively. Unit is in $m/s$. \textcolor{blue}{The axes are not drawn to scale. The horizontal-to-vertical aspect ratio is exaggerated for improved visualization of the flow field.}}

\label{fig:mean_plot_v}
\end{figure}

\begin{figure}
\centering
\subfigure[]{%
\includegraphics[scale = 0.42]{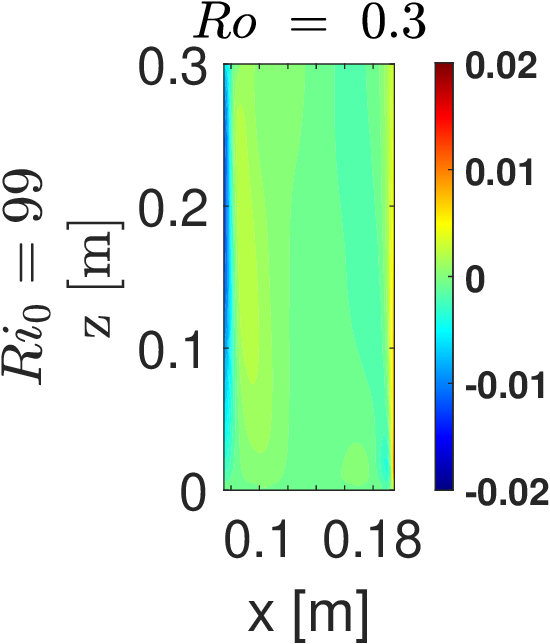}
\label{fig:WMean_w_1_2_Ta_low}}
\hspace{1cm}
\subfigure[]{%
\includegraphics[scale = 0.42]{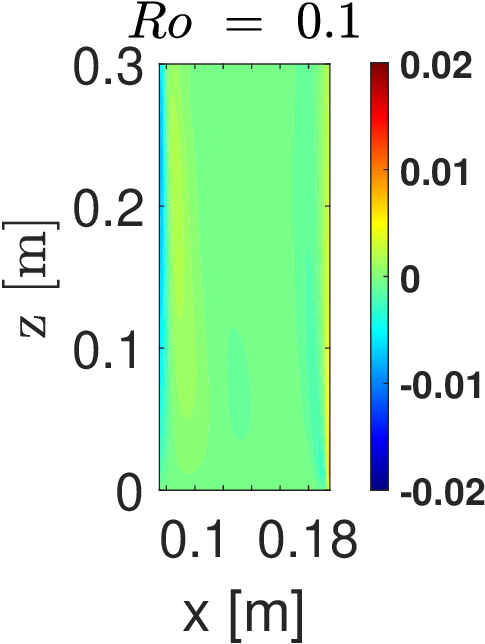}
\label{fig:WMean_w_1_2_Ta_mod}}
\hspace{1cm}
\subfigure[]{%
\includegraphics[scale = 0.42]{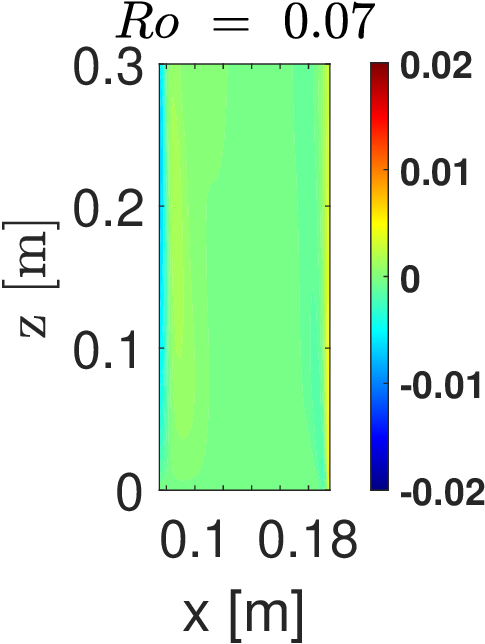}
\label{fig:WMean_w_1_2_Ta_high}}
\hspace{1cm}
\\
\subfigure[]{%
\includegraphics[scale = 0.4]{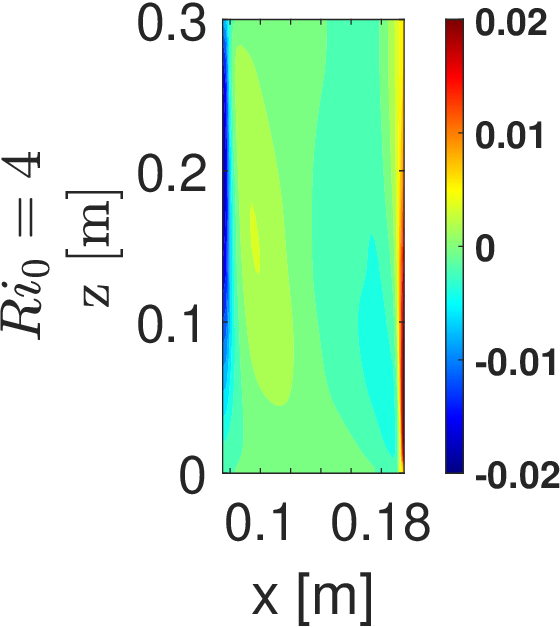}
\label{fig:WMean_w_6_Ta_low}}
\hspace{1cm}
\subfigure[]{%
\includegraphics[scale = 0.4]{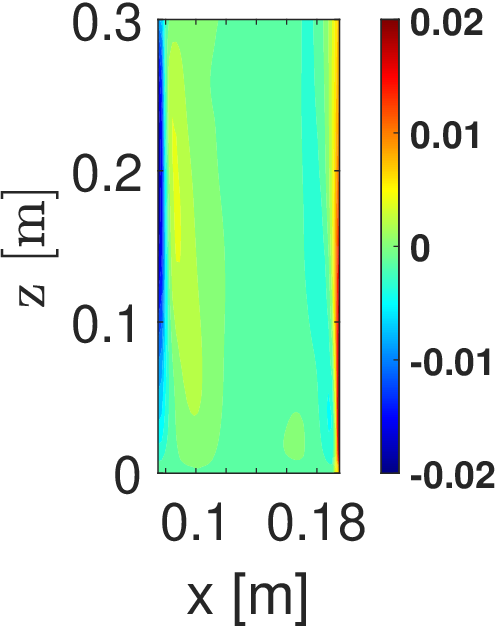}
\label{fig:WMean_w_6_Ta_mod}}
\hspace{1cm}
\subfigure[]{%
\includegraphics[scale = 0.4]{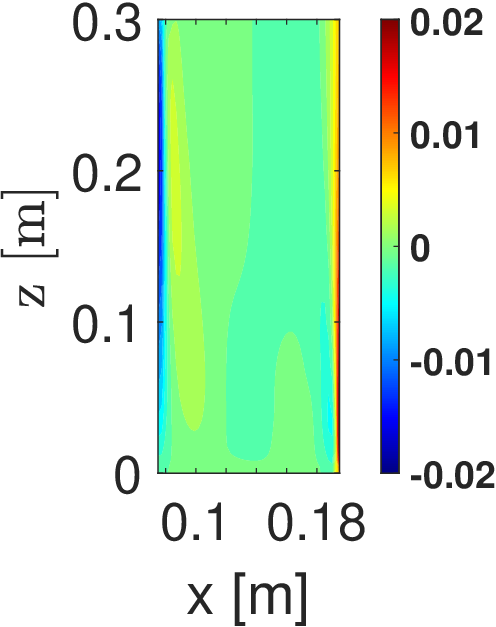}
\label{fig:WMean_w_6_Ta_high}}
\\
\subfigure[]{%
\includegraphics[scale = 0.4]{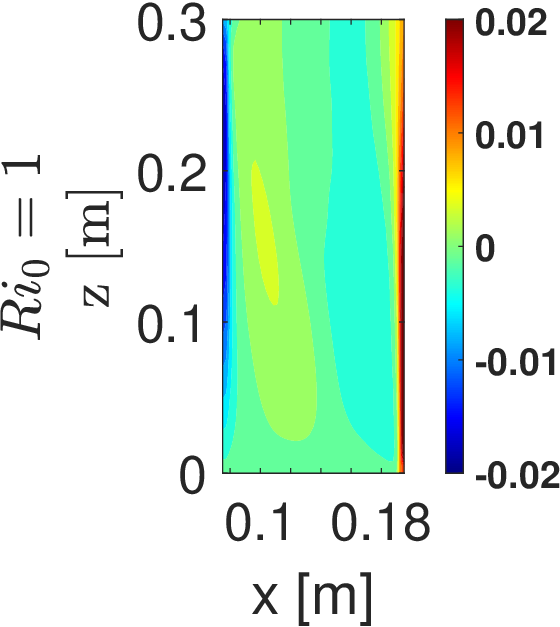}
\label{fig:WMean_w_12_Ta_low}}
\hspace{1cm}
\subfigure[]{%
\includegraphics[scale = 0.4]{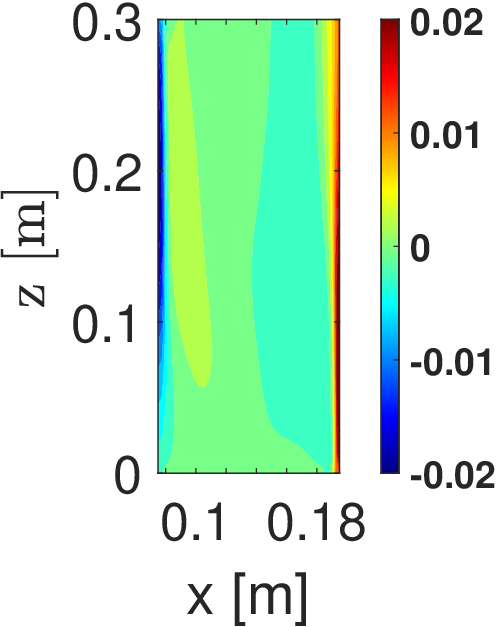}
\label{fig:WMean_w_12_Ta_mod}}
\hspace{1cm}
\subfigure[]{%
\includegraphics[scale = 0.4]{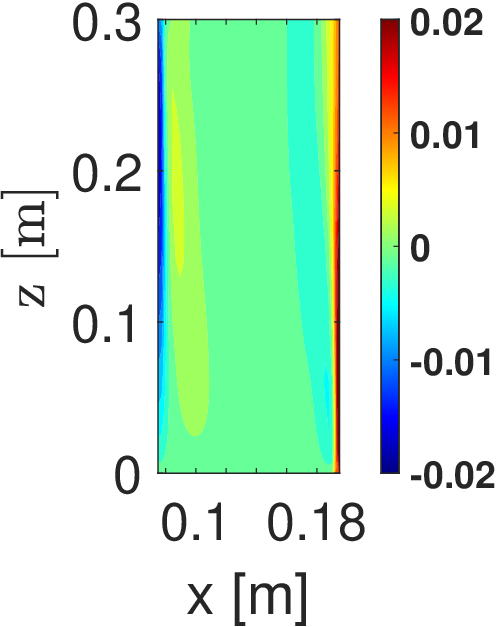}
\label{fig:WMean_w_12_Ta_high}}

\caption{Isocontours of the time and y-direction averaged $z$ component of velocity $\avg{w}$ for a range of $Ri_0$ and $Ro$. Here, $Ri_0=$ 99 and $Ro =$ 0.3, 0.1, 0.07 for [(a), (b), (c)], $Ri_0=$ 4 and $Ro =$ 0.3, 0.1, 0.07 for [(d), (e), (f)], and $Ri_0=$ 1 and $Ro =$ 0.3, 0.1, 0.07 for [(g), (h), (i)] respectively. Unit is in $m/s$. \textcolor{blue}{The axes are not drawn to scale. The horizontal-to-vertical aspect ratio is exaggerated for improved visualization of the flow field.}}

\label{fig:mean_plot_w}
\end{figure}

(iii) The secondary circulation $\avg{u}$, $\avg{w}$ is ageostrophic and hence weak in the bulk (Figures~\ref{fig:mean_plot_u} and~\ref{fig:mean_plot_w}). The leading-order balance Eq.~\eqref{eq:leading_balance} contains neither $u^{*}$ nor $w^{*}$: a purely geostrophic, hydrostatic state
supports only a zonal flow.  The meridional and vertical velocities appear only at next order, $\mathcal{O}(\Ro)$, where the inertial terms on the left of Eqs.~\eqref{eq:nondim_xmom}--\eqref{eq:nondim_zmom} enter. The bulk values of $\avg{u}$ and $\avg{w}$ are, therefore, smaller than the geostrophic $\avg{v}$ by a factor $\mathcal{O}(\Ro)\sim 0.1$--$0.3$, Consequently, $\langle w \rangle$ is negligible throughout the bulk, while $\langle u \rangle$ is significant only under weak rotation ($Ro=0.3$) and diminishes as the rotation rate increases (i.e., as $Ro$ decreases). In contrast, strong values of $\langle u \rangle$ and $\langle w \rangle$ are confined to the sidewalls and the localized heating region, where lateral plume entrainment and the buoyancy-driven updraft give rise to strong ageostrophic motions. The strength of this near-wall secondary circulation is governed not by the bulk thermal-wind scaling but by the inlet forcing, parameterized by the source Richardson number $Ri_{0}$; this is taken up in
Sec.~\ref{sec:Rec_plume} in connection with Figure~\ref{fig:YZ_vector_plot}.

The iso-contours of $\avg{T}$ are shown for all cases in Figure~\ref{fig:mean_plot_T}.  For every case, a high temperature is seen in the forced plume at the right wall and a low temperature in the cold descending flow at the left wall.  In the bulk the isotherms are slanted; as established above, this slant is the visible signature of the thermal-wind balance, Eq.~\eqref{eq:thermal_wind}, and reflects the joint presence of stable vertical stratification and a horizontal temperature gradient. Banerjee~\cite{banerjee2024axisymmetric} demonstrated that rotation amplifies wall temperature anomalies, whose non-local pressure perturbation field induces a bulk radial temperature perturbation gradient through geostrophic and thermal-wind balance, producing isotherm tilts that scale as $(Ta/Ra)^{1/4}$. The slanted isotherms, coupled with the frame rotation, can give rise to lateral mixing of temperature via baroclinic waves.  We observe a larger magnitude of temperature gradients in the bulk at lower $Ri_{0}$, owing to the higher buoyancy flux delivered at the larger inlet velocities.  For any given $Ri_{0}$, increasing the frame rotation rate (decreasing $Ro$) increases the detrainment height of the hot plume, so the isotherms spread further into the bulk at lower $Ro$.  For $Ri_{0}=1,4$ the slant of the iso-contour lines declines as $Ro$
decreases; for $Ri_{0}=99$ the weak inlet buoyancy flux leaves only a small bulk $\partial\avg{T}/\partial x$, so by the thermal-wind relation the isotherms there approach the vertical.

The iso-contours of the mean meridional velocity $\avg{u}$ are shown in Figure~\ref{fig:mean_plot_u}.  Large positive values of $\avg{u}$ appear at the left and right bottom corners of the domain, with strong negative $\avg{u}$ regions adjacent to them, indicating convergence and divergence zones near the left and right corners respectively.  Not surprisingly, there is no mean meridional fluid
transport from the cold to the hot side at the bottom, owing to the free-slip boundary condition imposed on $\mathbf{u}$ at $z=0$.  As
anticipated from the scaling analysis above, $\avg{u}$ is the ageostrophic component of the circulation and is therefore weak in the bulk: the appreciable bulk values seen for $Ri_{0}=1,4$ at the low rotation rate $Ro=0.3$ -- representing entrainment into the secondary updraft/downdraft on the sides of the annulus -- diminish as $Ro$ is reduced, consistent with the $\mathcal{O}(Ro)$ scaling of the secondary circulation.  The overall strength of $\avg{u}$ also increases as $Ri_{0}$ decreases, since the stronger plume drives more horizontal inflow to satisfy continuity.

The iso-contours of the mean zonal velocity $\avg{v}$ are shown in
Figure~\ref{fig:mean_plot_v}.  A positive (negative) sign of $\avg{v}$ indicates zonal flow in the anti-clockwise (clockwise)
direction.  As established above, $\avg{v}$ is the leading-order geostrophic flow, Eq.~\eqref{eq:leading_balance}, and its vertical
shear is fixed by the horizontal temperature gradient through the thermal-wind relation, Eq.~\eqref{eq:thermal_wind}.  The features of
Figure~\ref{fig:mean_plot_v} follow directly: the zonal flow is strongest near the side walls and, for the low-$Ri_{0}$ cases at $Ro=0.3$, in the bulk -- precisely where $\partial\avg{T}/\partial x$
is largest; it is weaker at high $Ri_{0}$, where the bulk temperature gradient is small; and for a fixed $Ri_{0}$ it weakens as $Ro$ decreases, both because the thermal-wind shear scales as $1/f$ and because $\partial\avg{T}/\partial x$ itself diminishes.


The iso-contours of the mean vertical velocity $\avg{w}$
(Figure~\ref{fig:mean_plot_w}) show down-welling flow along the cold wall and an upward-moving plume over the heating zone.  The strength
of the upward-moving plume increases as $Ri_{0}$ decreases, owing to the larger inlet momentum at lower $Ri_{0}$.  A recirculating downward (upward) flow forms next to the forced plume (down-welling flow); these recirculations produce the alternating negative and positive $u$ velocity near the bottom of the rectangular annulus seen in Figure~\ref{fig:mean_plot_u}. In the bulk, $\avg{w}$, similar to $\avg{u}$,  remains negligible over the range of $Ri_{0}$ and $Ro$ values considered, as it is an $\mathcal{O}(Ro)$ ageostrophic quantity. As a result, the buoyant plume remains attached to the outer wall, exhibiting little lateral spreading into the interior.

The scaling argument of items (i)-(iii) can be made fully quantitative
by solving the bulk mean-temperature problem in closed form and applying
the thermal-wind relation. Because $\mathrm{Pe}\gg 1$ in the bulk, the
steady mean-temperature field is set there by a balance between the
imposed bidirectional forcing and the very weak secondary circulation. In the bulk the mean-temperature equation
\(
\avg{\mathbf{u}}\!\cdot\!\nabla\avg{T}+\nabla\!\cdot\!\avg{\mathbf{u}'T'}
=\kappa\nabla^2\avg{T}
\)
is, to leading order, diffusively controlled: the mean advection
$\avg{\mathbf u}$ is weak in the bulk and the eddy flux divergence is
subdominant there (Figure \ref{fig:x_enrgyflux_plot}). \textcolor{blue}{Here $\langle \mathbf{u}'T' \rangle = \langle \mathbf{u}T \rangle
- \langle \mathbf{u} \rangle \langle T \rangle$
denotes the turbulent heat flux in the $x$-direction, with $\langle\cdot\rangle$ denoting the average over time and the $y$-direction.} Therefore, treating the bulk-region mean temperature as harmonic at leading order
gives
\begin{equation}
\nabla^{2}\theta = 0,
\qquad 0\le x\le l_{x},\ \ 0\le z\le L,
\label{eq:laplace_T}
\end{equation}
with $L\equiv l_z+l_d$, $\theta(x,z)=\avg{T}(x,z)-T_{c}$. The
boundary conditions encode the geometry of the experiment:
\begin{alignat}{2}
\theta(0,z) &= 0
  &&\quad(\text{cold inner wall, full height}),
\label{eq:bc_inner}\\
\theta(l_x,z) &= h(z)
  &&\quad(\text{heated outer edge}),
\label{eq:bc_outer}\\
\partial_z\theta &= 0\quad\text{at}\quad z=0,\,L
  &&\quad(\text{adiabatic bottom and top}).
\label{eq:bc_topbot}
\end{alignat}
The outer-edge condition $h(z)$ is \emph{piecewise}: the differential
heating is confined to the baroclinic zone and the outer wall is adiabatic
in the damping zone,
\begin{equation}
h(z)=
\begin{cases}
  \Delta T_x\,g(z), & 0\le z\le l_z,\\[4pt]
  0,                & l_z < z \le L,
\end{cases}
\label{eq:hz}
\end{equation}
where the one-parameter profile
\begin{equation}
g(z)=e^{-z/h_{tr}},\qquad g(0)=1,
\label{eq:gz}
\end{equation}
describes the decay of the outer-edge heating signature with height as the
plume detrains; the detrainment height scale $h_{tr}$ is set by the plume physics, which closes the model.

\begin{figure}[!ht]
\centering
\includegraphics[width=1.05\textwidth]{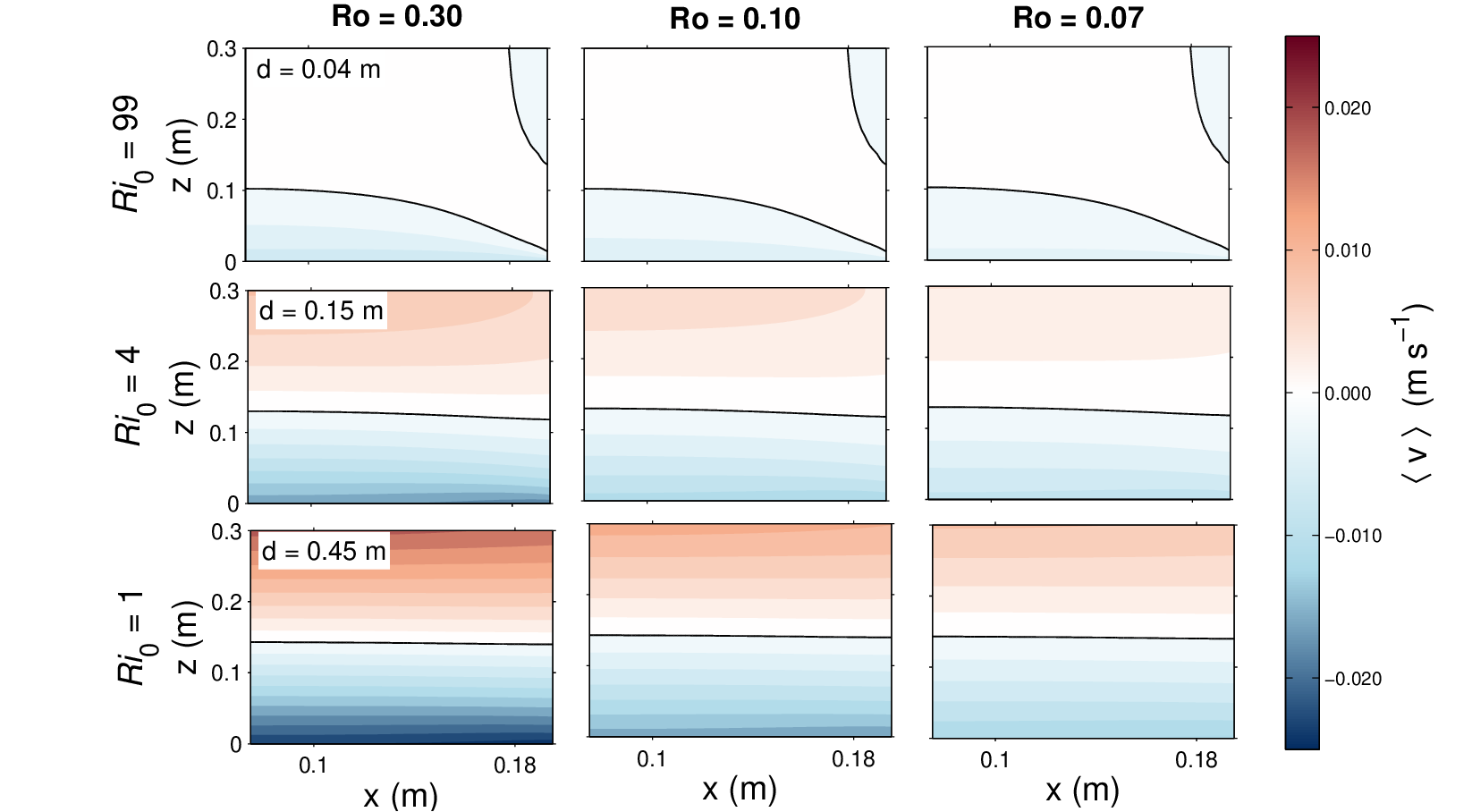}
\caption{Analytical mean zonal velocity $\avg{v}(x,z)$ obtained from the
closed-form thermal-wind solution
Eqs.~\eqref{eq:v_series}--\eqref{eq:v0_closure}, for the same nine
$(Ri_{0},Ro)$ combinations as the simulations of
Figure~\ref{fig:mean_plot_v}. The detrainment height scale is set to
$h_{tr}=0.04$, $0.15$, and $0.45$~m for $Ri_{0}=99,4,1$ respectively, in
accordance with the plume reach observed in
Figure~\ref{fig:YZ_vector_plot}. The dashed line marks the zero contour
separating the lower retrograde and upper prograde branches. Colour
scale: blue (red) indicates clockwise (anti-clockwise) zonal flow.
The figure should be compared directly with
Figure~\ref{fig:mean_plot_v}. \textcolor{blue}{Please note the physical domain aspect ratio is $l_x:l_z=0.12:0.30$; the panels are stretched horizontally by a factor $2.5$ for legibility and are \emph{not} drawn to equal $x$--$z$ scale.}}
\label{fig:v_analytical}
\end{figure}

Expanding $\theta$ in the natural cosine basis $\cos(k_{n}z)$ with
$k_{n}=n\pi/L$, the boundary condition (Eq.~\eqref{eq:laplace_T}) yields
\begin{equation}
\theta(x,z) = \frac{G_{0}\,x}{l_{x}}
+ \sum_{n=1}^{\infty}\frac{G_{n}\sinh(k_{n}x)}{\sinh(k_{n}l_{x})}\,
\cos(k_{n}z),
\label{eq:T_series}
\end{equation}
where the Fourier coefficients are
\begin{equation}
G_{0} = \frac{\Delta T_{x}h_{tr}}{L}
        \!\left(1-e^{-l_{z}/h_{tr}}\right),
\qquad
 G_n=\frac{2\Delta T_x}{L}\cdot
\frac{h_{tr}^{-1}\!\left(1-e^{-l_z/h_{tr}}\cos k_n l_z\right)
     +k_n\,e^{-l_z/h_{tr}}\sin k_n l_z}{h_{tr}^{-2}+k_n^2},
\qquad n\ge1       
\label{eq:Gn}
\end{equation}

The first term in Eq.~\eqref{eq:T_series} represents the vertically averaged ($n=0$) component of the imposed thermal forcing and produces a horizontally varying background temperature field analogous to the classical baroclinic-annulus gradient. Its magnitude $G_0$, however, is itself determined by the vertical profile of the localized heating. The higher Fourier modes ($n\ge1 $) represent departures from this vertical mean and carry the explicit vertical structure associated with the localized heating. Thus, the localized forcing contributes both to the mean horizontal gradient through $G_0$ and to the genuinely bidirectional $(x,z)$ - dependent structure through the higher modes. Inserting Eq.~\eqref{eq:T_series} into the thermal-wind relation
Eq.~\eqref{eq:thermal_wind} and integrating in $z$ from the bottom yields
the mean zonal jet,
\begin{equation}
\avg{v}(x,z) = \avg{v}_{0}(x)
+ \frac{g\alpha}{f}\!\left[
\frac{G_{0}\,z}{l_{x}}
+ \sum_{n\ge 1}\frac{G_{n}\cosh(k_{n}x)}{\sinh(k_{n}l_{x})}\,
\sin(k_{n}z)
\right].
\label{eq:v_series}
\end{equation}

  The reference profile $\avg{v}_{0}(x)$ is not determined by the thermal-wind relation, which constrains only the vertical shear of the mean zonal flow. To fix this barotropic component, we impose zero depth-integrated zonal transport at each $x$, $\int_{0}^{L}\!\avg{v}\,dz=0$. This condition removes any depth-uniform zonal component not determined by thermal-wind balance and is consistent with the absence of an externally imposed net azimuthal forcing under the adopted idealized boundary conditions. Using
$\int_{0}^{L}\!\sin(k_{n}z)\,dz=[1-(-1)^{n}]/k_{n}$, which vanishes
for even $n$ and equals $2/k_{n}$ for odd $n$,
\begin{equation}
\avg{v}_{0}(x) = -\frac{g\alpha}{f\,L}
\!\left[
\frac{G_{0}\,L^{2}}{2\,l_{x}}
+ \!\!\sum_{n\,\text{odd}}\!
\frac{2\,G_{n}\cosh(k_{n}x)}{k_{n}\sinh(k_{n}l_{x})}
\right].
\label{eq:v0_closure}
\end{equation}

Eqs.~\eqref{eq:v_series}--\eqref{eq:v0_closure} together give the
mean zonal jet in closed form, with the detrainment scale $h_{tr}$ as the
only physical input not contained in the governing parameters. The
associated geostrophic pressure follows from Eq.~\eqref{eq:leading_balance}
as $\avg{p}(x,z)=\avg{p}(0,z)+f\!\int_{0}^{x}\avg{v}\,dx'$, and one
verifies directly that
$\partial_{z}\!\left(\partial_{x}\avg{p}\right)
= f\,\partial_{z}\avg{v}
= g\alpha\,\partial_{x}\theta
= \partial_{x}\!\left(\partial_{z}\avg{p}\right)$,
so the analytical field is exactly consistent with \emph{both} the
geostrophic and hydrostatic balances of
Eq.~\eqref{eq:leading_balance}.

Figure~\ref{fig:v_analytical} shows the analytical $\avg{v}(x,z)$ for
all nine cases. The closed-form solution reproduces every qualitative
feature of the simulated $\avg{v}$ in Figure~\ref{fig:mean_plot_v}: the
sign reversal between the upper prograde ($\avg{v}>0$) and lower
retrograde ($\avg{v}<0$) branches of the jet, with the zero contour
sitting close to mid-depth; the concentration of $|\avg{v}|$ near the
outer wall, where $\partial\theta/\partial x$ is largest; the weakening of $|\avg{v}|$ as $Ro$ decreases at fixed $Ri_{0}$, through
the $1/f$ prefactor of the thermal wind; and the strengthening of
$|\avg{v}|$ as $Ri_{0}$ decreases at fixed $Ro$, through the growth of
$d$ and hence of the bulk $\partial\theta/\partial x$. The peak
magnitudes agree quantitatively as well: the model gives
$|\avg{v}|_{\max}\sim 4$--$9\times 10^{-3}$~m\,s$^{-1}$ for $Ri_{0}=99$
and $|\avg{v}|_{\max} \sim 1.2$--$2.4\times 10^{-2}$~m\,s$^{-1}$ for $Ri_{0}=1$, which matches the colourbar range of Figure~\ref{fig:mean_plot_v} to within
$\sim 20\%$ -- a level of agreement that is striking given that $h_{tr}$ is
the \emph{only} parameter beyond the Boussinesq constants and the
governing dimensionless groups. The closed-form prediction thus
places the qualitative thermal-wind argument of items (i)--(iii) on
an explicitly predictive footing and identifies the detrainment scale
$h_{tr}$ as the single physical quantity that the bulk mean state requires
the plume physics to provide.

\begin{figure}[t]
\centering
\subfigure[]{%
\includegraphics[scale = 0.36]{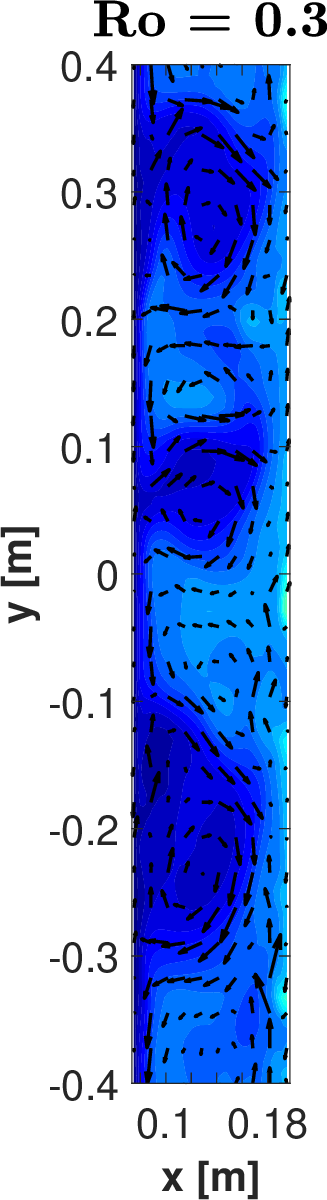}
\label{fig:Vect_plot_w_low_Ta_low}}
\subfigure[]{%
\includegraphics[scale = 0.36]{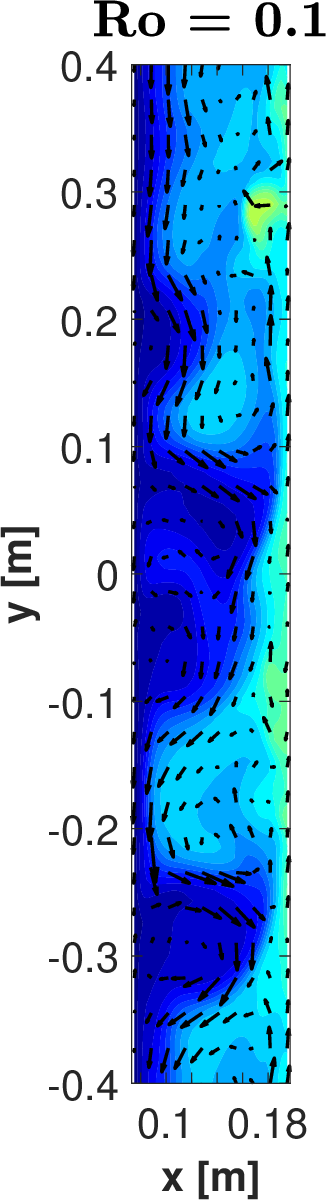}
\label{fig:Vect_plot_w_low_Ta_mod}}
\subfigure[]{%
\includegraphics[scale = 0.36]{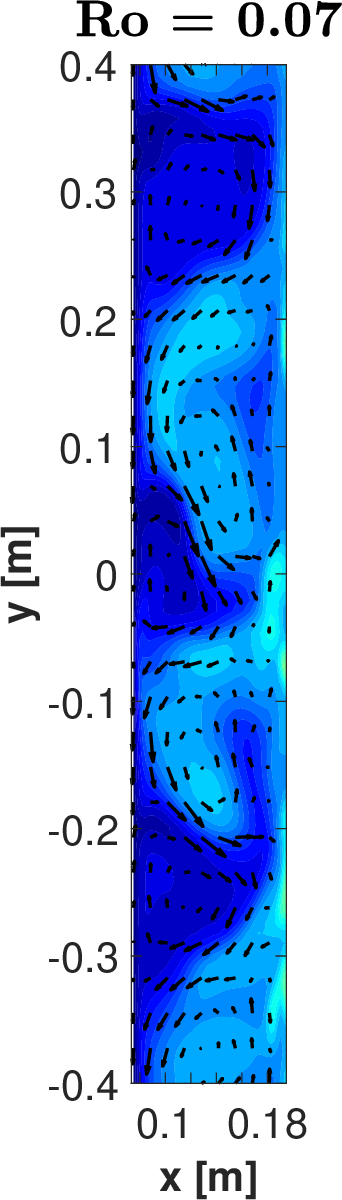}
\label{fig:Vect_plot_w_low_Ta_high}}
\subfigure[]{%
\includegraphics[scale = 0.36]{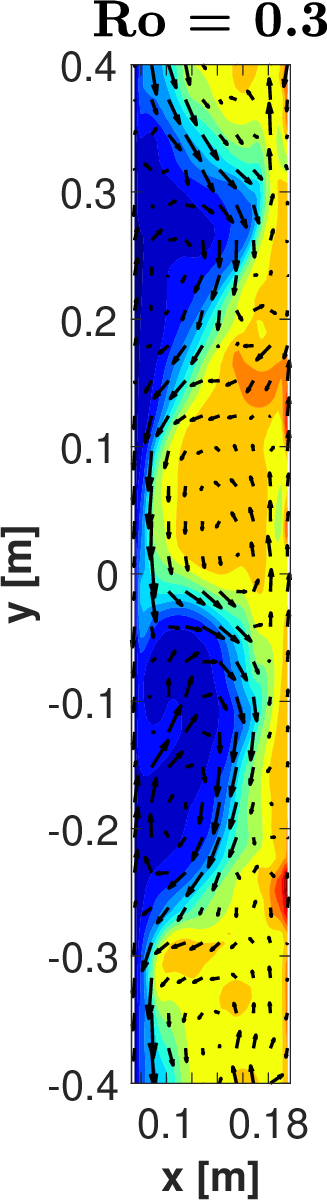}
\label{fig:Vect_plot_w_mod_Ta_low}}
\subfigure[]{%
\includegraphics[scale = 0.36]{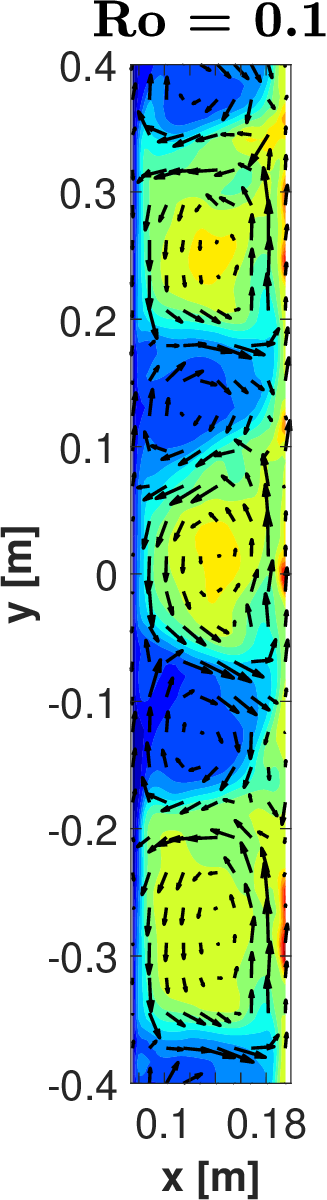}
\label{fig:Vect_plot_w_mod_Ta_mod}}
\subfigure[]{%
\includegraphics[scale = 0.36]{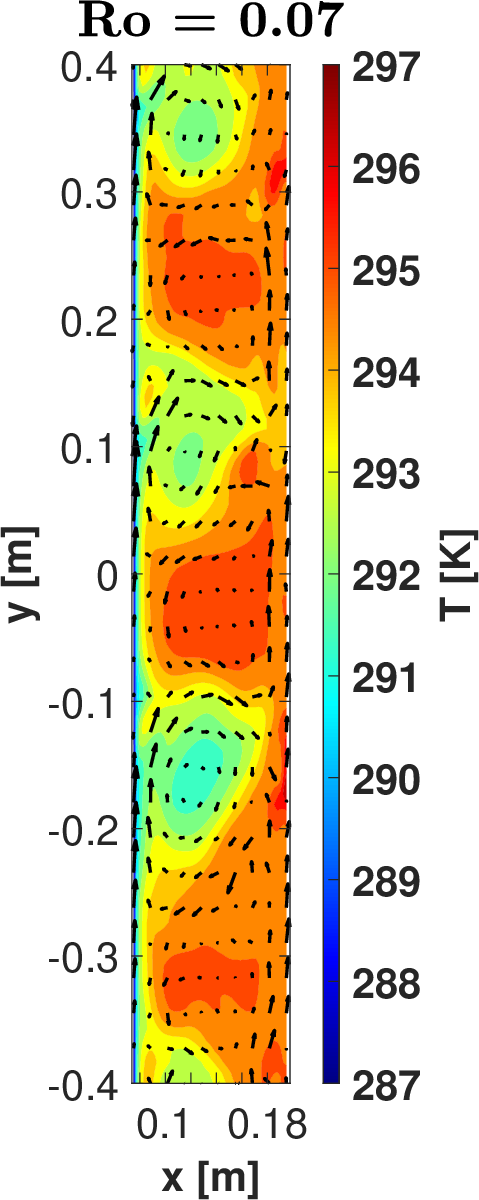}
\label{fig:Vect_plot_w_mod_Ta_high}}
\\
\subfigure[]{%
\includegraphics[scale = 0.36]{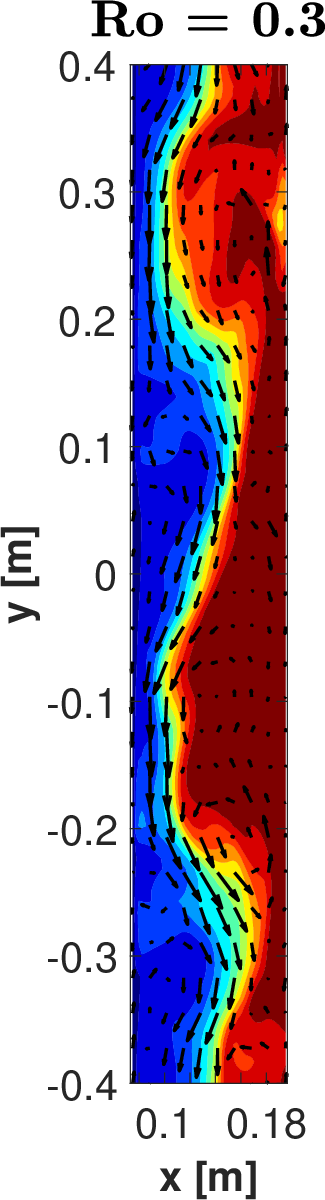}
\label{fig:Vect_plot_w_high_Ta_low}}
\subfigure[]{%
\includegraphics[scale = 0.36]{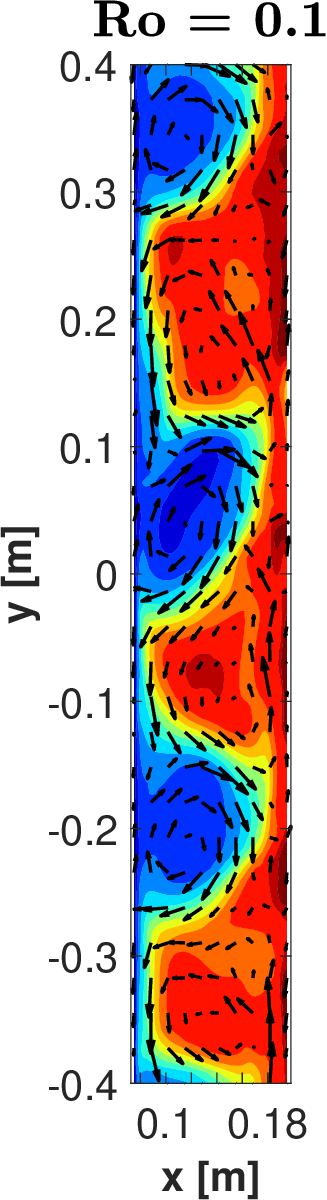}
\label{fig:Vect_plot_w_high_Ta_mod}}
\subfigure[]{%
\includegraphics[scale = 0.36]{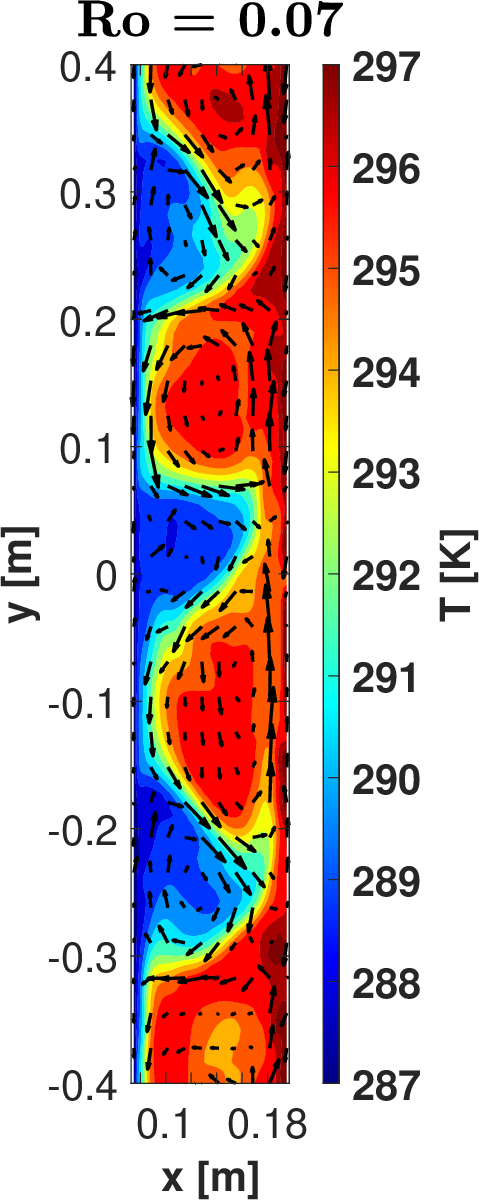}
\label{fig:Vect_plot_w_high_Ta_high}}

\caption{Instantaneous velocity vector superimposed on the temperature contour at $z=$ 0.15 m from the bottom of the rectangular annulus. Here, [(a), (b), (c)] are for $Ri_0=$ 99 and $Ro =$ 0.3, 0.1, 0.07, [(d), (e), (f)] are for $Ri_0=$ 4 and $Ro =$ 0.3, 0.1, 0.07, and [(g), (h), (i)] are for $Ri_0=$ 1 and $Ro =$ 0.3, 0.1, 0.07 respectively.}

\label{fig:Vectr_plot}
\end{figure}

Overall, the contours of $\avg{T}$, $\avg{u}$, $\avg{v}$, and $\avg{w}$ are consistent with the respective contours for the rotating cylindrical annulus
configuration with bi-directional temperature gradients reported by Swarnakar \eal \cite{swarnakar2023numerical}.  For instance, the
slanting isotherms in the bulk region in the rectangular baroclinic annulus (Figure~\ref{fig:mean_plot_T}) are also observed in Figure~5(c) for the cylindrical baroclinic annulus\cite{swarnakar2023numerical}.  Similarly, the organization of positive and negative zonal/vertical velocity regions in the rectangular baroclinic annulus
(Figures~\ref{fig:mean_plot_v} and~\ref{fig:mean_plot_w}) are also
observed in Figures~5(g)--5(i) and Figures~5(j)--5(l) respectively for the cylindrical baroclinic annulus\cite{swarnakar2023numerical}.  The major difference in mean
flow pattern between the two setups is that the cylindrical baroclinic annulus shows the presence of Ekman boundary layers at the
top and bottom boundaries (Figures~5(d)--5(f)), which are absent in the rectangular baroclinic annulus (Figure~\ref{fig:mean_plot_u}).
These results corroborate our earlier argument that the Ekman boundary layers are not required to sustain the mean flow statistics observed in the bulk region of
the cylindrical baroclinic annulus\cite{swarnakar2023numerical}.

\subsection{\label{sec:Rec_wave}Analysis of baroclinic wave structures}

In this section, we characterize the structure of baroclinic waves in the rectangular baroclinic annulus with imposed bi-directional temperature gradients. Figure \ref{fig:Vectr_plot} shows the instantaneous velocity vector plot superimposed on the isocontours of instantaneous temperature field at $z=$ 0.15 m, for various combinations of rotation rate and inlet velocity, namely $Ri_0=$ 99, 4, 1 and $Ro=$ 0.3, 0.1, 0.07. 

In Figure \ref{fig:Vectr_plot}, instantaneous velocity vector plots help to visualize the zonal jet, akin to the jet stream observed in the atmosphere, which emerges between the periphery of the cold and hot fluids. The zonal jet is a narrow region characterized by significantly higher flow velocity compared to its surroundings. Its formation is attributed to the local horizontal temperature gradient existing between the boundaries of the cold and hot fluids, along with the influence of frame rotation. The strength of the zonal jet depends upon the local horizontal temperature gradient. Additionally, the temperature contours depicted in Figure \ref{fig:Vectr_plot} indicate the presence of baroclinic waves within the rectangular annulus.

In the discussion below, we will refer to the number of lobes in the baroclinic waves with the integer $``m"$.
Baroclinic wave with two lobes $(m =2)$ is observed at $Ri_0 =4$, $Ro = 0.3$ (Figure \ref{fig:Vect_plot_w_mod_Ta_low}), and at  $Ri_0 =1$, $Ro = 0.3$ (Figure \ref{fig:Vect_plot_w_high_Ta_low}). In contrast, for other combinations of $Ri_0$ and $Ro$ values, baroclinic waves with three lobes $(m = 3)$ is observed (see Figure \ref{fig:Vectr_plot}). The Eady deformation radius $L_\rho = NH/f$ decreases with rotation rate (i.e. with decreasing $\Ro$) at fixed stratification.  Hence the most-unstable
wavelength $\lambda_{\max} \propto L_\rho$ also decreases with decreasing
$\Ro$, and the wave mode $m = l_y/\lambda_{\max}$ increases with
decreasing $\Ro$.  This is exactly the $m=2 \to m=3$ transition we observe. These waves persist across the entire depth of the baroclinic region $(z < l_z)$ and exhibit greater intensity near the bottom of the annulus, where the horizontal temperature gradient is more pronounced. \textcolor{blue}{At $Ri_0=99$ and $Ro=0.3$ the leading CEOF mode is $m=3$ (Figure~\ref{fig:PC_w_1.2_Ta_low}); the three lobes are of low amplitude in the instantaneous snapshot of Figure \ref{fig:Vect_plot_w_low_Ta_low}. As $\Ro$ is reduced to $0.1$, the wave develops an irregular, vacillating lobe pattern (Figure \ref{fig:Vect_plot_w_low_Ta_mod}). Further reduction of $Ro$ to 0.07 results in a baroclinic wave with relatively regular wave lobes, as shown in Figure \ref{fig:Vect_plot_w_low_Ta_high}. }
At $Ri_0 = 4$ and $Ro = 0.3$, a two-mode baroclinic wave is observed (Figure \ref{fig:Vect_plot_w_mod_Ta_low}), which evolves into a well-defined three-mode baroclinic wave as $Ro$ decreases to 0.1 (Figure \ref{fig:Vect_plot_w_mod_Ta_mod}) and 0.07 (Figure \ref{fig:Vect_plot_w_mod_Ta_high}). Similarly, at $Ri_0 = 1$ and $Ro = 0.3$, the baroclinic wave transitions from mode 2 to mode 3 with a decrease in the Rossby number, as illustrated in Figure \ref{fig:Vect_plot_w_high_Ta_mod} and Figure \ref{fig:Vect_plot_w_high_Ta_high}.

\textcolor{blue}{Two features of the $v$-contours are worth noting for the
plume--wave interaction. First, the
$|v|$-maxima are localised near the outer wall ($x \approx 0.18$~m), where the
horizontal temperature gradient is strongest, consistent with the thermal-wind
relation $\partial v/\partial z \propto \partial T/\partial x$; this
concentration is what makes the $v$-signature a sharper mode-identifier than
the $T$-contour, whose lobes fill the gap more diffusely. Second, the
zero-crossings of $v$ in $y$ are phase-shifted from the $T$-lobe centres by
approximately a quarter wavelength, so that the $|v|$-peaks --- and hence the
kinematic convergence zones of the wave --- coincide with the sharp
temperature fronts rather than with the hot or cold lobe centres. As will be
discussed in Sec.~\ref{sec:Rec_plume}, the plume clusters visible in the
$y$--$z$ snapshots of Figure~\ref{fig:YZ_vector_plot} are found precisely at these
$|v|$-peak locations, providing direct evidence that the plumes lock
kinematically onto the wave rather than driving it.}


The strength of the local temperature gradients across the thin jet-streams
is influenced by variations in $Ri_0$ and $Ro$. As $Ri_0$ decreases (i.e., as
the plume inlet velocity increases), the local horizontal temperature
gradient across the jet-stream increases: the higher inlet buoyancy flux
delivers more heat to the outer wall while the inner cold wall is unchanged,
so the outer--inner contrast $\Delta T(x)$ sharpens. The elevated fluid
temperatures at lower $Ri_0$ values are visible in Figure~\ref{fig:Vectr_plot}. The horizontal temperature gradient also sharpens as $Ro$ decreases: at
$Ri_0=4$ and $Ro=0.07$ (Figure~\ref{fig:Vect_plot_w_mod_Ta_high}) the gradients are stronger than for the
same $Ri_0$ at $Ro=0.10$ (Figure~\ref{fig:Vect_plot_w_mod_Ta_mod}) and $Ro=0.30$ (Figure~\ref{fig:Vect_plot_w_mod_Ta_low}); the same
trend is seen for $Ri_0=99$ and $Ri_0=1$. This strengthening of
$\partial T/\partial x$ with decreasing $Ro$ is required by thermal-wind
balance, $2\Omega\,\partial v/\partial z \sim (g/T_0)\,\partial T/\partial r$:
with the jet amplitude $|v|$ pinned by the vertically-integrated thermal-wind
constraint, $|\partial T/\partial r|\propto\Omega\propto Ro^{-1}$,
so faster rotation demands a proportionally sharper local horizontal gradient
to sustain the same vertical shear. Consistently, $|v|$ itself remains
comparable across the three $Ro$ values even as the local horizontal gradient
sharpens.

\subsection{\label{sec:CEOF}Analysis of the most energetic mode of the baroclinic wave}

 Complex Empirical Orthogonal Function (CEOF) analysis is employed to identify the most energetic mode of the large-scale propagating structures (i.e., baroclinic wave in our case) present in the flow (\cite{hannachi2007empirical}). In the CEOF analysis, any scalar function $f(\mathbf{x},t)$ is initially transformed into complex form $(F(\mathbf{x},t))$ by employing the Hilbert transform to obtain the imaginary part. The complex form is given as: 
\begin{eqnarray}
    F(\mathbf{x},t)=f(\mathbf{x},t)+i H[f](\mathbf{x},t)
\end{eqnarray}
where $H[f](\mathbf{x},t)$ is the Hilbert transform of the scalar function $f(\mathbf{x},t)$. Upon performing the CEOF analysis, the complex function $F(\mathbf{x},t)$ can be represented as follows:
\begin{eqnarray}
    F(\mathbf{x},t)&=&\sum_{K=1}^N {a}_K(t) {\psi}_K(\mathbf{x})
\end{eqnarray}
where $a_K(t)$ is the complex Principal components (PC), $\psi_K(\mathbf{x})$ is the complex eigenmodes, and $K$ is the mode number ($K=1$ for most energetic mode and $K=m$ for the least energetic mode). 

\begin{figure}[t]
\centering
\subfigure[]{%
\includegraphics[scale = 0.35]{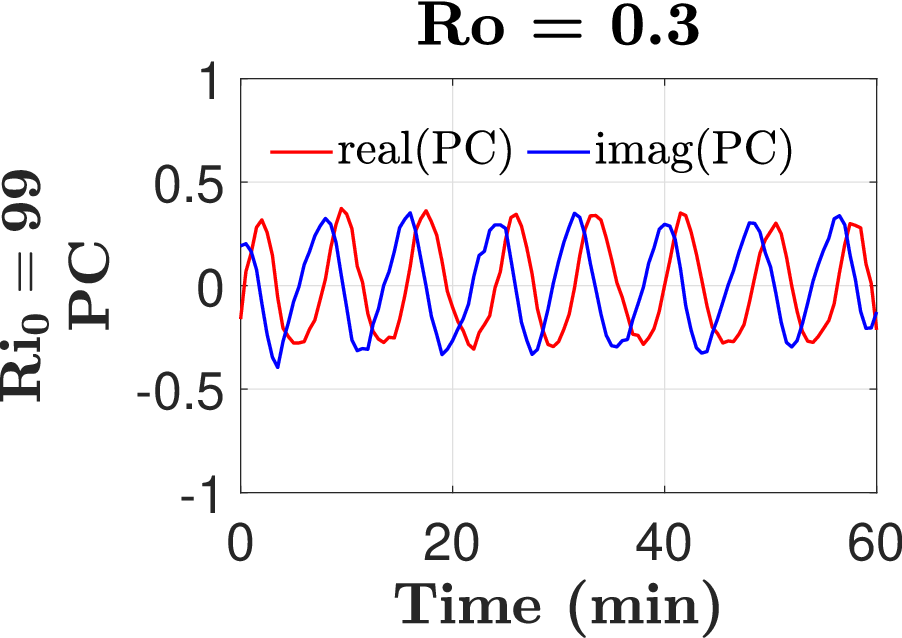}
\label{fig:PC_w_1.2_Ta_low}}
\subfigure[]{%
\includegraphics[scale = 0.35]{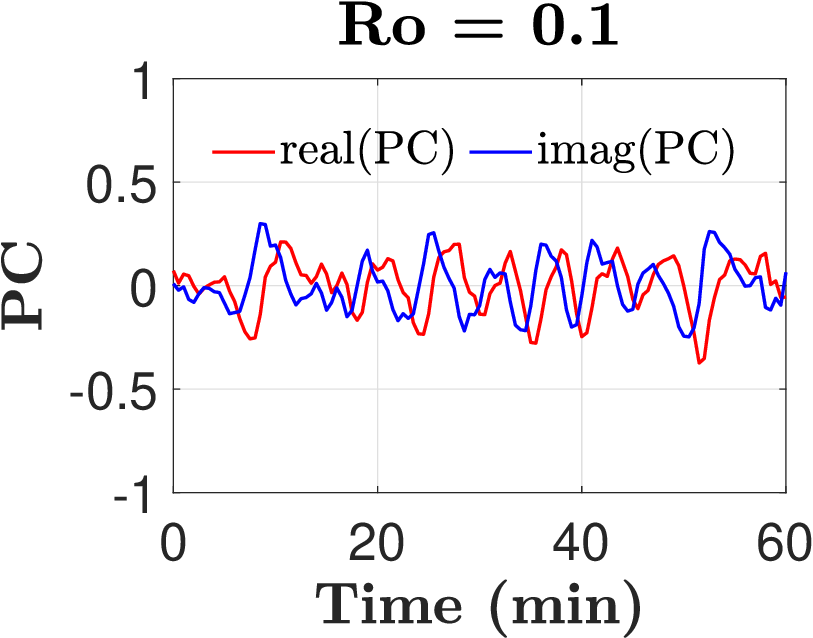}
\label{fig:PC_w_1.2_Ta_mod}}
\subfigure[]{%
\includegraphics[scale = 0.35]{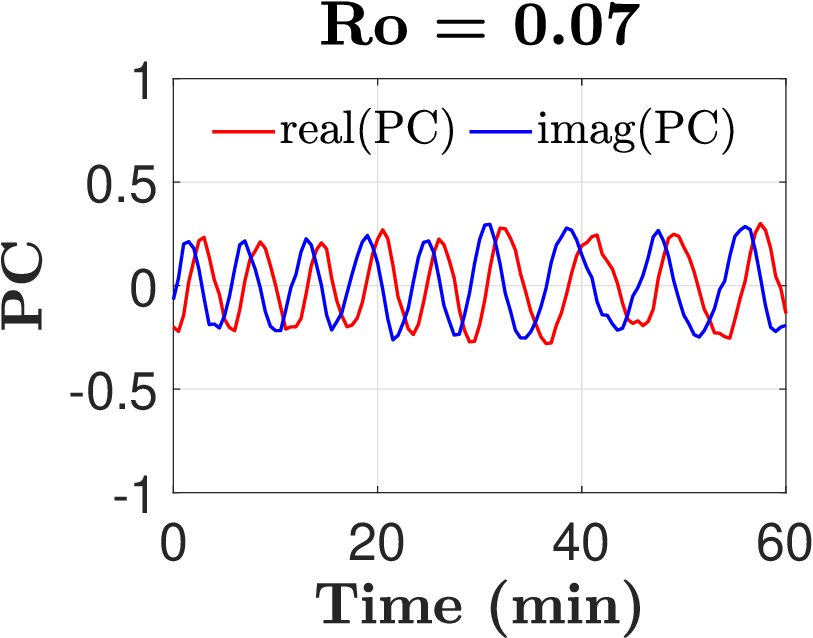}
\label{fig:PC_w_1.2_Ta_high}}
\\
\subfigure[]{%
\includegraphics[scale = 0.35]{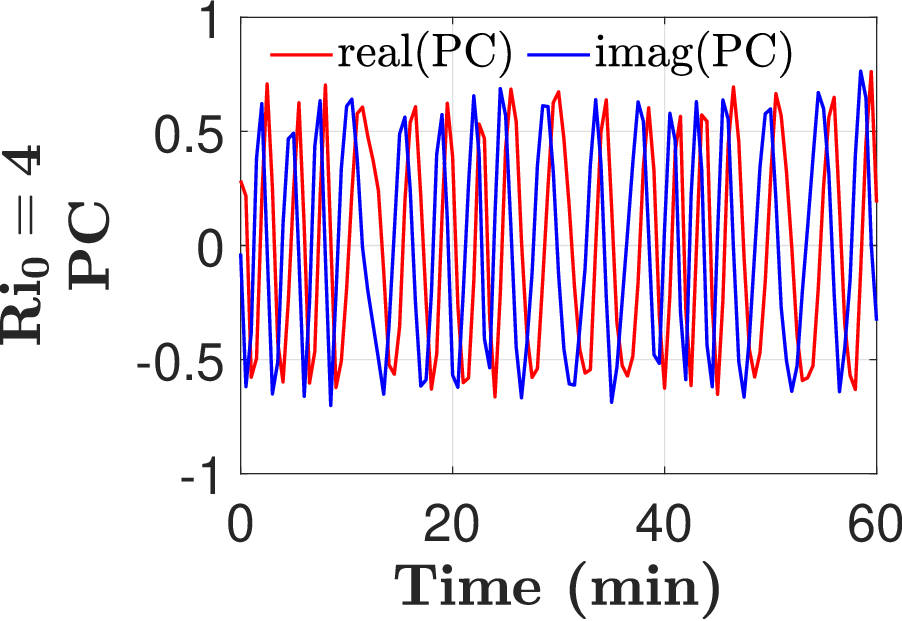}
\label{fig:PC_w_6_Ta_low}}
\subfigure[]{%
\includegraphics[scale = 0.35]{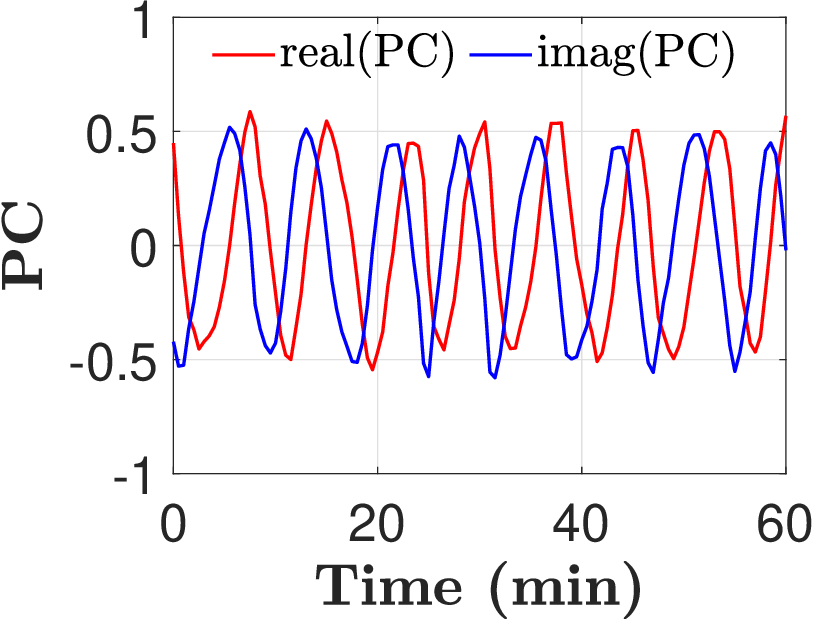}
\label{fig:PC_w_6_Ta_mod}}
\subfigure[]{%
\includegraphics[scale = 0.35]{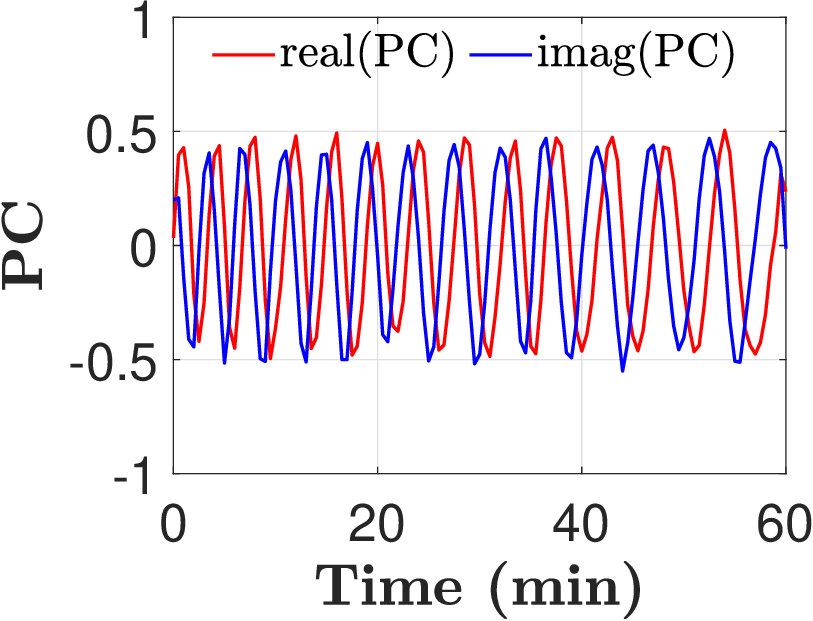}
\label{fig:PC_w_6_Ta_high}}
\\
\subfigure[]{%
\includegraphics[scale = 0.35]{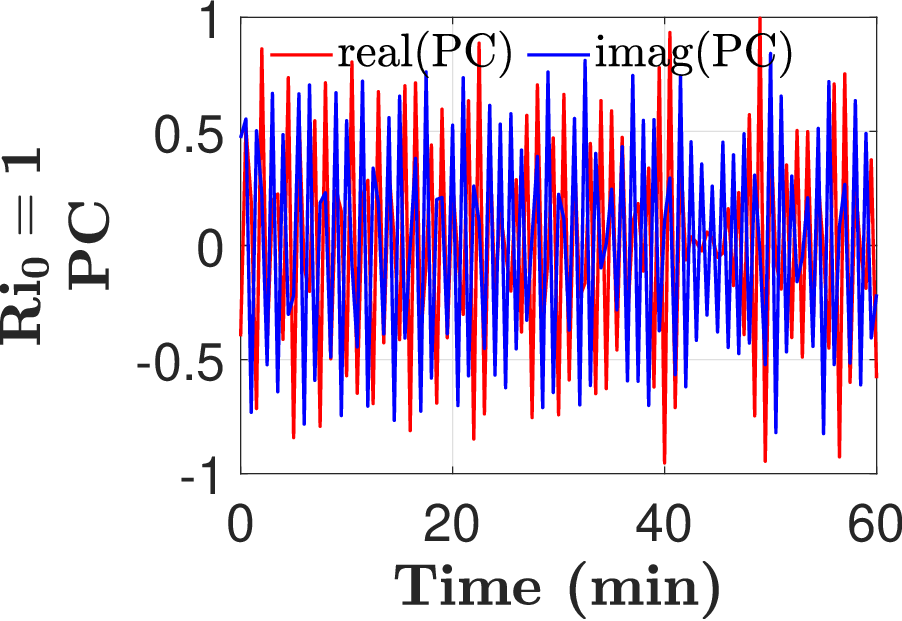}
\label{fig:PC_w_12_Ta_low}}
\subfigure[]{%
\includegraphics[scale = 0.35]{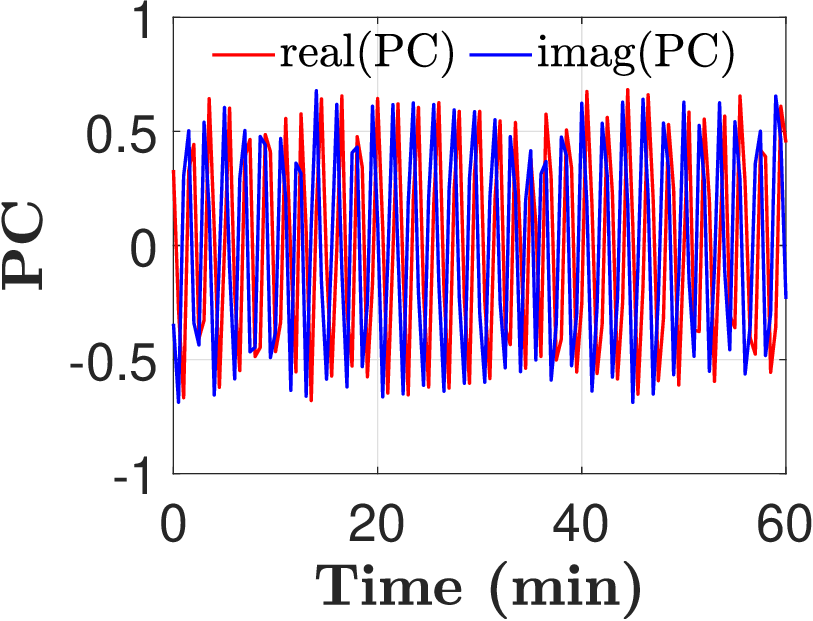}
\label{fig:PC_w_12_Ta_mod}}
\subfigure[]{%
\includegraphics[scale = 0.35]{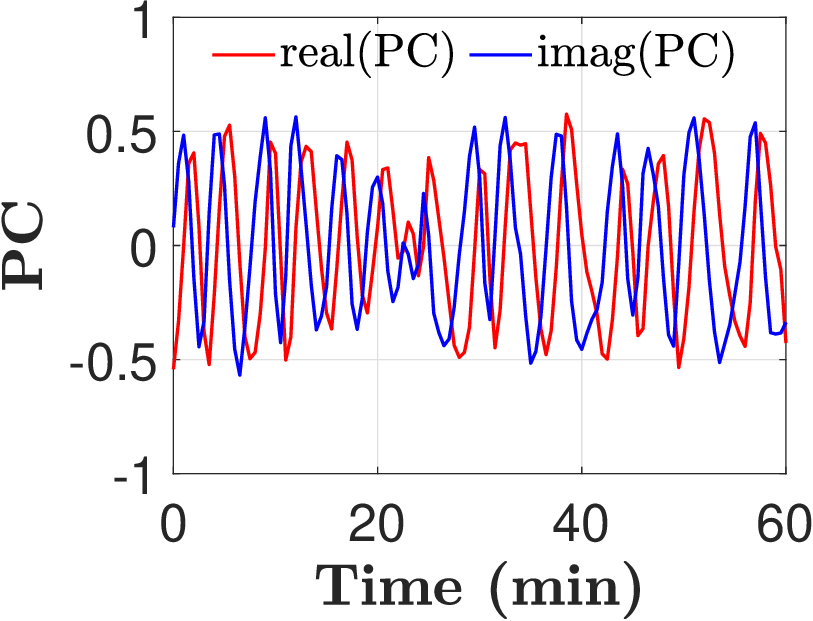}
\label{fig:PC_w_12_Ta_high}}

\caption{Time variation of the Principal component (PC) of the $1^{st}$ CEOF mode of the $x-$velocity for a range of $Ri_0$ and $Ro$. Here, $Ri_0=$ 99 and $Ro =$ 0.3, 0.1, 0.07 for [(a), (b), (c)], $Ri_0=$ 4 and $Ro =$ 0.3, 0.1, 0.07 for [(d), (e), (f)], and $Ri_0=$ 1 and $Ro =$ 0.3, 0.1, 0.07 for [(g), (h), (i)] respectively. The red and blue lines represent the real and imaginary parts of the PC of the $1^{st}$ CEOF mode respectively. The phase shift between the real and imaginary part infer that the baroclinic wave is propagating.}

\label{fig:PC_plot}
\end{figure}

The PCs $(a_K(t))$ represent the time-varying amplitude of the eigenmodes. As the CEOF analysis separates the real and imaginary part of the baroclinic wave with a phase difference of $\pi/2$, we can determine the phase speed

\begin{equation}
c = \frac{\pi}{2\tau_{lag} k}   
\end{equation}

of the baroclinic wave by computing the time lag ($\tau_{lag}$) between the real and imaginary part of the PC. Here, wave number is denoted by $k = 2\pi m/l_y$ rad/m or $k = m$, thus giving the phase speed

\begin{equation}
c = \frac{l_y}{4\tau_{lag} m} \quad \text{m/s} \quad \text{or} \quad c = \frac{\pi}{2\tau_{lag} m} \quad \text{rad/s}     
\end{equation}

A positive value of $c$ indicates that the wave is propagating in the anti-clockwise direction. A complete description of the COEF analysis is given in the work of Swarnakar \eal \cite{swarnakar2023numerical}.

The CEOF analysis is performed on the time series data of the $u$ (meridional) velocity on an $x$--$y$ plane at $z = 0.15$ m. The data used for the CEOF analysis comprises the  records from the last 1 hour of the simulation. Figure~\ref{fig:PC_plot} shows the real and imaginary PCs of the $1^{\mathrm{st}}$ CEOF mode of the $u$ velocity for cases with 
$Ri_0 = 99,\,4,\,1$ and $Ro = 0.3,\,0.1,\,0.07$. For $Ri_0 = 99$ and $Ro = 0.3$, the PC 
plot (Figure~\ref{fig:PC_w_1.2_Ta_low}) shows a regular pattern, which indicates that the 
wave is in a steady wave regime. However, when $Ro$ is decreased to 0.1 
(Figure~\ref{fig:PC_w_1.2_Ta_mod}), the PC pattern becomes irregular, indicating a change 
in the wave structure and its transition to a vacillating wave regime. This behaviour is consistent with a Hopf-bifurcated state: the irregular PC time series shows an amplitude modulation (envelope $\sim$1500\,s) superposed on the underlying wave oscillation, whose period $\sim$740\,s is consistent with the CEOF time-lag $\tau_{\mathrm{lag}}\approx185$\,s for this case ($c = 3.6\times10^{-4}$\,m\,s$^{-1}$, $m=3$); the coexistence of these two timescales is characteristic of a vacillating, torus (secondary Hopf) state\cite{strogatz1994nonlinear}; amplitude vacillation of this kind is well documented in the differentially heated rotating annulus \cite{fruh1997wave}.

In the case of $Ri_0 = 4$ and $Ro = 0.3$, the frequency and magnitude of the PC
time series (Figure~\ref{fig:PC_w_6_Ta_low}) are notably higher compared to the
$Ri_0 = 99$ case (Figure~\ref{fig:PC_w_1.2_Ta_low}). This arises mainly due to a
combination of higher thermal fluxes and a lower wave number (i.e.\ $m = 2$).
Moreover, for $Ri_0 = 4$, when $Ro$ is decreased to 0.1 from 0.3, the frequency
of the PC (Figure~\ref{fig:PC_w_6_Ta_mod}) decreases. This change is attributed
to the transition in the baroclinic wave mode from $m = 2$ to $m = 3$. However,
when $Ro$ is further decreased to 0.07, the frequency of the PC increases
(Figure~\ref{fig:PC_w_6_Ta_high}), despite the wave mode number remaining the
same (i.e.\ $m = 3$). The phase speed $(c)$ of the baroclinic wave for
$Ri_0 = 4$ varies non-monotonically with $Ro$: $c = 2.6 \times 10^{-3}$\,m/s at
$Ro = 0.3$, decreasing to $3.6 \times 10^{-4}$\,m/s at $Ro = 0.1$, and
recovering to $5.3 \times 10^{-4}$\,m/s at $Ro = 0.07$. The pronounced drop from $Ro = 0.3$ to $Ro = 0.1$ coincides with
the $m = 2\!\to\!m = 3$ mode transition; within the $m = 3$ mode
($Ro = 0.1$ and $0.07$) the phase speed shows a smaller,
non-monotonic variation that the mode change does not account for.
For $Ri_0 = 1$ and $Ro = 0.3$, the frequency of the PC is significantly higher
(Figure~\ref{fig:PC_w_12_Ta_low}), and it decreases monotonically as $Ro$ is
reduced to 0.1 (Figure~\ref{fig:PC_w_12_Ta_mod}) and further to 0.07
(Figure~\ref{fig:PC_w_12_Ta_high}). For $Ri_0 = 1$, the wave regime transitions
from $m = 2$ at $Ro = 0.3$ to $m = 3$ at lower $Ro$ values. The phase speed for
$Ri_0 = 1$ at $Ro = 0.3,\,0.1$, and 0.07 are $5.2 \times 10^{-3}$\,m/s,
$1.5 \times 10^{-3}$\,m/s, and $7.1 \times 10^{-4}$\,m/s, respectively,
reflecting both the mode transition and the increasing Coriolis parameter at
lower $Ro$. \textcolor{blue}{These speed values may be compared with linear baroclinic-instability theory: the most-unstable Eady mode is steered at mid-depth, so its phase speed equals the depth-mean zonal flow, which vanishes identically here under the zero-net-flux closure $\int_0^L\avg{v}\,dz=0$ (Eq.~\eqref{eq:v0_closure}). The wave is therefore expected to propagate slowly relative to the thermal-wind scale $U_y$, consistent with the measured $c\sim(1$--$9)\times10^{-3}\,\mathrm{m\,s^{-1}}$, i.e.\ $1$--$9\%$ of $U_y\approx2\times10^{-2}\,\mathrm{m\,s^{-1}}$.}

To summarize, we observed a transition in the baroclinic wave regime from mode $m = 2$ to mode $m = 3$ with the decrease in $Ro$ number for cases with $Ri_0 = 4$ and 1. Additionally, for $Ri_0 = 99$, the baroclinic wave changed from a steady regime to a vacillating regime at lower $Ro$ values. The strength of the zonal jet increases with a decrease in $Ri_0$ due to a higher local horizontal temperature gradient. At a fixed $Ri_0$, the strength of the zonal jet decreases with a decrease in the $Ro$ number. The frequency of the PC time series increases with a decrease in $Ri_0$ at a fixed $Ro$ number. This change in the frequency of the PC is possibly due to the increase in the thermal flux in the annulus with the increase in plume inlet velocity (or decrease in $Ri_0$). Additionally, the phase velocity of the baroclinic wave varies with $Ri_0$ and $Ro$ numbers. Across all cases, the leading CEOF mode captures 60-85\% of the variance, with mode~2 capturing 8-20\% and mode~3
capturing 3-10\%.  The dominance of mode 1 confirms that a single
wavenumber dominates each case.

\subsection{\label{sec:Rec_plume}Dynamics of forced plume near the outer wall}

\begin{figure}[!htbp]
  \centering
  \includegraphics[width=1.02\linewidth]{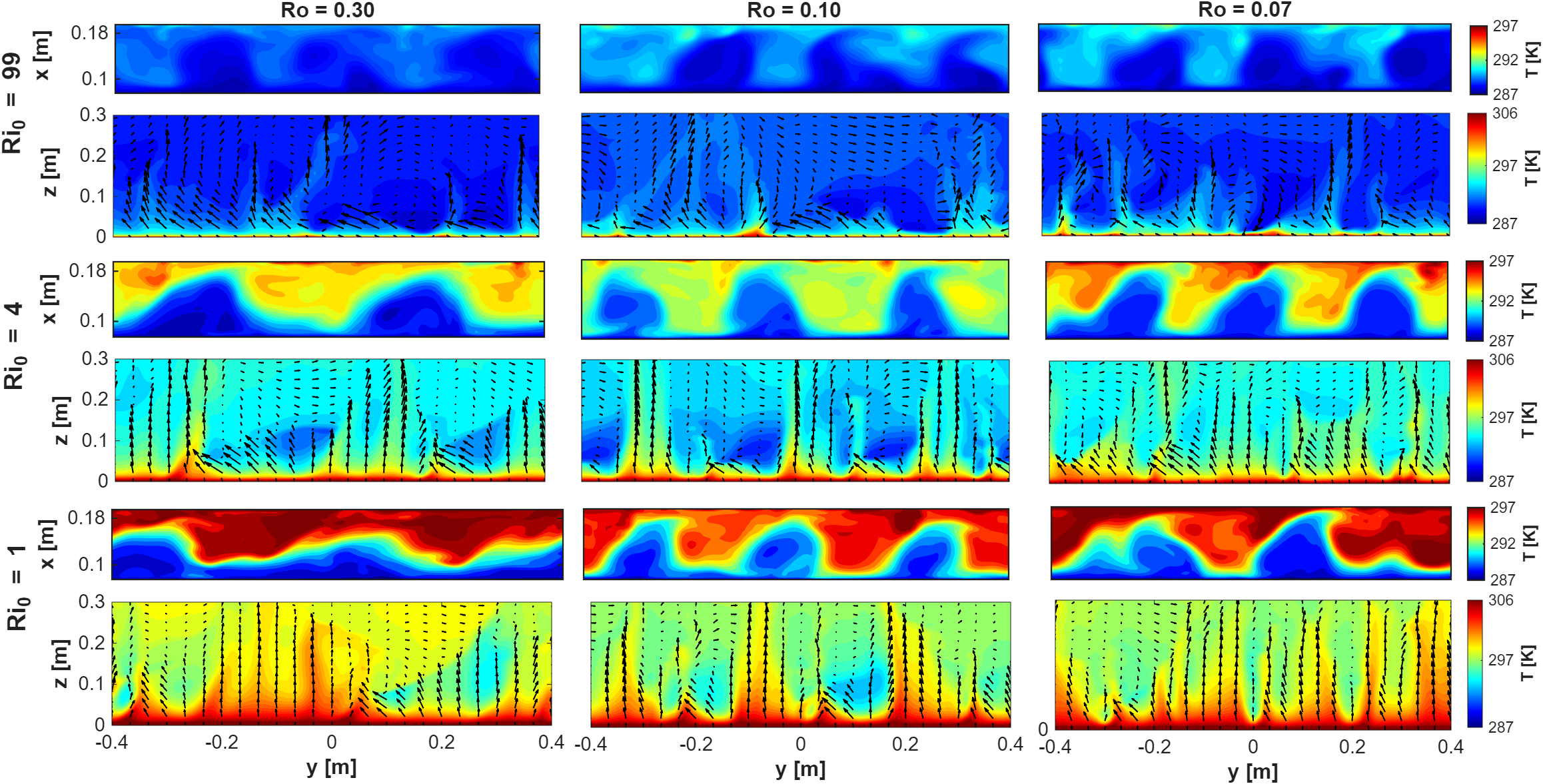}
  \caption{Instantaneous  temperature contours in the $x$-$y$ plane at $z$=0.15 m (1st, 3rd, 5th rows) and instantaneous velocity vectors superimposed on the temperature contours in the $y$--$z$ plane at $x=0.193$~m (2nd, 4th, 6th rows). First two rows correspond to $Ri_0$=99, third and fourth rows correspond to $Ri_0$=4, and fifth and sixth rows correspond to $Ri_0$=1. The $x$-$y$ plane primarily illustrates the baroclinic wave structure, while the $y$-$z$ plane reveals the associated plume clusters and vertical flow structures. Temperature is shown in K.
}
  \label{fig:YZ_vector_plot}
\end{figure}

In this section, we examine the influence of $Ri_0$ and $Ro$ on the forced plume at the outer edge of the annulus. Figure~\ref{fig:YZ_vector_plot} shows instantaneous temperature contours in the $x$--$y$ plane at $z=0.15$~m and instantaneous velocity vectors superimposed on the temperature contours in the $y$--$z$ plane at $x=0.193$~m, corresponding to the region over the heating zone. For each $(Ri_0,Ro)$ pair, both sections are extracted at the same instant and plotted on a common $y$-axis.

The bulk-region scaling of Sec.~\ref{sec:Rec_mean_field} breaks down in the immediate vicinity of the forced plume, where the relevant velocity is the inlet velocity $w$ and the relevant length is the width of the heated source strip $b$ rather than $U_{y}$ and $l_{x}$. To characterise the plume source more precisely, we use the volume flux ($Q_{0}$), kinematic momentum 
flux ($ M_{0}$) and kinematic buoyancy flux ($F_{0}$) per unit transverse length at the inlet~\citep{balasubramanian2018role},
\begin{equation}
    Q_{0} \;=\; w\,b,
    \qquad
    M_{0} \;=\; w^{2}\,b,
    \qquad
    F_{0} \;=\; w\,g'\,b \;=\; w\,\alpha g\,\Delta T_{p}\,b,
    \label{eq:M0F0}
\end{equation}

where $g' = \alpha g\,\Delta T_{p}$ is the reduced gravity. The Morton length scale, which separates the inertia-dominated (jet-like) and buoyancy-dominated (plume-like) 
regions of a forced plume~\citep{morton1956turbulent}, is
\begin{equation}
    L_{M} \;=\; \frac{M_{0}^{3/4}}{F_{0}^{1/2}}
    \;=\; \frac{w^{3/2}\,b^{3/4}}{(w\,g'\,b)^{1/2}}
    \;=\; b\,Ri_0^{-1/2}.
    \label{eq:LM}
\end{equation}
\citet{papanicolaou1988investigations} showed that the dimensionless distance 
$z/L_{M}$ from the source governs the local flow character: the flow is jet-like 
for $z/L_{M}<1$, exhibits forced-plume behaviour for $1<z/L_{M}<5$, and 
approaches a pure plume for $z/L_{M}>5$. To classify the source itself, 
\citet{hunt2005lazy} introduced the flux-balance parameter
\begin{equation}
    \Gamma_{0}
    \;=\; \frac{5\,Q_{0}^{2}\,F_{0}}{4\,\Gamma\,M_{0}^{5/2}},
    \label{eq:Gamma0}
\end{equation}
where $\Gamma$ is the entrainment coefficient appropriate to a forced 
plume~\citep{morton1956turbulent}. The source is a \emph{lazy plume} when 
$\Gamma_{0}>1$ (excess buoyancy relative to momentum), a \emph{pure plume} when 
$\Gamma_{0}=1$, and a \emph{forced plume} when $\Gamma_{0}<1$ (momentum in 
excess of buoyancy). In a physical sense, a lazy plume arises from a source with 
a deficit of momentum flux compared to a forced plume with the same source 
buoyancy and volume fluxes. With $b = 5$\,mm and $Ri_0 \in \{1,\,4,\,99\}$, Eq.~\eqref{eq:LM} gives 
$L_{M} \in \{5.0,\,2.5,\,0.5\}$\,mm. In every case $L_{M} \ll l_{z} = 0.30$\,m, 
confirming that the plume is buoyancy-dominated over essentially the entire depth of the baroclinic zone. The role of $w$ is therefore not to sustain a momentum jet but to set the buoyancy flux $F_{0}$ at the source.

When the ambient fluid is stably stratified, a further non-dimensional parameter 
is required. \citet{wong1988submerged} and \citet{bloomfield1998turbulent} 
introduced
\begin{equation}
    \sigma
    \;=\; \frac{M_{0}^{2}\,N^{2}}{F_{0}^{2}}
    \label{eq:sigma}
\end{equation}
where $N \;=\; \left(-\frac{g}{\rho_{0}}\frac{\mathrm{d}\rho}{\mathrm{d}z}
    \right)^{\!1/2}$ is the ambient buoyancy frequency. The limit $\sigma\gg 1$ corresponds 
to strong stratification and/or high inlet velocity and/or a small density 
deficit, while $\sigma\ll 1$ is the weakly stratified regime. It has been widely 
accepted that $\sigma^{1/2}$ controls the maximum rise height of the plume in a 
linearly stratified environment~\citep{bloomfield1998turbulent}.
To place the forced-plume buoyancy supply in perspective, we compare it with the equivalent natural plume. For a freely convecting strip of the same width $b$ and temperature contrast $\Delta T_{p}$, dimensional analysis~\citep{morton1956turbulent,turner1986turbulent} 
gives a characteristic source velocity
\begin{equation}
    w_{\mathrm{nat}} \;\sim\; (g'\,b)^{1/2},
    \label{eq:wnat}
\end{equation}
so that the natural-plume buoyancy flux per unit length is
\begin{equation}
    F_{0,\mathrm{nat}} \;\sim\; w_{\mathrm{nat}}\,g'\,b \;=\; (g'\,b)^{3/2}.
    \label{eq:F0nat}
\end{equation}
The ratio of the forced to the natural-plume buoyancy flux is therefore
\begin{equation}
    \frac{F_{0}}{F_{0,\mathrm{nat}}}
    \;=\; \frac{w\,g'\,b}{(g'\,b)^{3/2}}
    \;=\; \frac{w}{(g'\,b)^{1/2}}
    \;=\; Ri_0^{-1/2}.
    \label{eq:Fratio}
\end{equation}
This ratio has a clean physical interpretation: at $Ri_0 = 1$ the imposed inlet velocity $w$ coincides with the natural free-fall velocity $w_{\mathrm{nat}}$, and the forced and 
natural plumes deliver identical buoyancy flux at the source. For $Ri_0 < 1$ the forced plume is momentum-driven and overrides the local stratification; for $Ri_0 > 1$ the imposed $w$ is sub-buoyant and the source buoyancy flux falls progressively below the 
natural reference. In the present study,
\begin{equation}
    \frac{F_{0}}{F_{0,\mathrm{nat}}}
    \;=\;
    \begin{cases}
        \;0.10 & \text{at } Ri_0 = 99,\\[3pt]
        \;0.50 & \text{at } Ri_0 = 4,\\[3pt]
        \;1.00 & \text{at } Ri_0 = 1,
    \end{cases}
    \label{eq:Fratio_num}
\end{equation}
with corresponding absolute buoyancy fluxes 
$F_{0} = w\,g'\,b \in \{1.7,\,8.5,\,17\}\times10^{-7}$\,m$^{3}$\,s$^{-3}$. Thus the 
$Ri_0 = 1$ case sits precisely at the threshold where the forced plume matches the buoyancy supply of a freely convecting strip of the same $\Delta T_{p}$, while the higher-$Ri_0$ cases correspond to forcing levels progressively weaker than that natural 
reference.

This scaling underpins the plume behaviour seen in Figure~\ref{fig:YZ_vector_plot}. 
For $Ri_0 \gg 1$ the inlet inertia is weak: the buoyancy flux $F_0$ is small, the plume is rapidly arrested by the stable background stratification within a short rise height, and no columnar structure forms. The plume is instead swept laterally by the zonal flow. For $Ri_0 \lesssim \mathcal{O}(1)$ the inlet momentum is large enough to carry the plume across the full baroclinic zone, producing the sustained columnar plumes. The intermediate case $Ri_0 = 4$ yields columnar plumes of intermediate vertical extent.

For a fixed $Ro$, as $Ri_0$ decreases, the plumes take a prominent columnar shape extending to the top of the domain, a feature observed across all $Ro$ values. This indicates that as the plume receives more momentum (equivalently, more buoyancy flux 
$F_0$), it persists for longer and retains its columnar structure while effectively transporting heat vertically. Consequently, the annulus temperature is higher at low 
$Ri_0$ values, as clearly shown in Figure~\ref{fig:YZ_vector_plot}. In particular, for $Ri_0 = 99$ and $Ro = 0.3$, the temperature contour 
shows that the forced plume does not exhibit a strong updraft and is diffused along the $y$-direction. Columnar structures are absent, and the plume is advected by the zonal flow present near the outer wall. At lower $Ro$ values of $0.1$ and 
$0.07$, the influence of the zonal flow on the plume diminishes because the weakening of the horizontal temperature gradient 
$(\partial T/\partial x)$ reduces the zonal velocity near the outer wall. Accordingly, for $Ri_0 = 99$, the columnar plume structure becomes more noticeable at $Ro = 0.1$ and 
$0.07$ compared with $Ro = 0.3$.

For $Ri_0 = 4$ and $1$ at $Ro = 0.3$, the higher inlet momentum results in the formation of coherent columnar structures that extend to the top of the annulus and transport heat and mass effectively. These 
coherent plume structures persist even at higher rotation rates (lower $Ro$) owing to 
the larger momentum flux injected at the inlet, as shown in 
Figure~\ref{fig:YZ_vector_plot}. Additionally, the influence of the zonal flow on the plume structure diminishes at lower $Ri_0$ because the stronger buoyant plume enhances thermal mixing near the outer wall, thereby reducing the local radial temperature gradient that drives the thermal-wind circulation. Moreover, for $Ri_0 = 4$ and $1$, the plume structures organise into clusters 
corresponding to the baroclinic wave modes. A similar correspondence between the number 
of plume clusters along the zonal direction and the baroclinic wave mode number was reported 
by~\citet{swarnakar2023numerical}. For $Ri_0 = 99$, coherent columnar plumes do not develop, thereby preventing the formation of plume clusters.


The combined presentation of Figure~\ref{fig:YZ_vector_plot} makes the
correspondence clearer. For each $(Ri_0,\Ro)$ pair the upper panel is a
horizontal $x$--$y$ slice through the baroclinic zone and the lower panel is the
$y$--$z$ section through the heating strip, both extracted at the same instant and
plotted on a common $y$-axis, so
that the wave and the plume field can be read against one another. At $Ri_0=4$ and
$1$ the columnar plumes in the lower panels are not distributed uniformly along $y$
but gather into groups whose number and spacing match the alternating warm and cold
tongues of the wave directly above, with the group centres under the warm phase; the
count in both panels changes together as the mode number increases with rotation.
The $Ri_0=99$ row isolates which structure is dependent: the wave is present and
countable there, but no columnar plumes form, and the clustering vanishes while the
wave persists. The same correspondence was established objectively for the rotating cylindrical annulus configuration with Ekman layers by \citet[Fig.~9]{swarnakar2023numerical}, whose leading complex
empirical orthogonal functions of temperature and of vertical velocity shared the
same azimuthal wavenumber.


The matching of plume-cluster number to the baroclinic-wave mode is a
kinematic consequence of the wave's spatial structure rather than a
dynamical resonance. The columnar plumes act as passive markers,
continuously injected at the heating strip and rising through a
baroclinic wave whose horizontal-convergence field has $m$-fold symmetry
along the zonal direction by construction; the plume material
accumulates in the wave's $m$ convergence zones per wavelength $l_{y}$. The plume sets and maintains the bi-directional mean state, but does not directly excite the wave: the baroclinic instability of that mean state selects its own wavelength through $\lambda_{\max} = 2\pi NH/f$, and the plume's narrow footprint ($b/l_y \approx 6\times10^{-3}$) projects only weakly onto the domain-filling mode $k_m = 2\pi m/l_y$, with no resonant or parametric coupling evident in the momentum and energy budgets. The wave therefore requires no amplification from the plume forcing; once established, its zonal-varying divergence organises the plume distribution, which is enslaved to a wave it does not drive.  Whether the imprint appears at all is set primarily by the existence of a coherent columnar plume to act as a marker: at $Ri_{0}=4$ and $1$ such columns are present and the clustering is clear, whereas at $Ri_{0}=99$ no columnar plume forms and the buoyant fluid is arrested and swept laterally, therefore, there is no coherent tracer to organise and the clusters are absent. Given a columnar plume, the \emph{sharpness} of the imprint is governed by the ratio of the plume transit time across the baroclinic zone, $\tau_{p}=l_{z}/w$, to the wave period $T_{w}=2\pi/(c\,k_{m})$, the imprint being sharp for $\tau_{p}\lesssim T_{w}$ and washed out for $\tau_{p}\gg T_{w}$. For the present runs $\tau_{p}/T_{w}\approx 0.07\text{--}0.5$ in every case, so the plume crosses the baroclinic zone in a fraction of a wave period and the wave pattern is effectively frozen during transit; the columnar plumes at $Ri_{0}=4$ and $1$ are accordingly imprinted sharply. For $Ri_0=1$ the plume is sustained-columnar and, as per the regime map (Figure~\ref{fig:regime_map}), essentially insensitive to rotation across all three $\Ro$; the morphological change with $\Ro$ is therefore weak and need not be monotone. 
The quantitative kinematic theory of
this transit-time locking, including the explicit dependence of the
clustering contrast on the ratio $\tau_{p}/T_{w}$ and the closed-form
$y$-distribution of plume tracer, will be taken in future works.


In summary, at $Ri_0 = 4$ and $1$, coherent columnar plume structures develop and 
drive higher temperatures within the rectangular annulus. At $Ri_0 = 99$ no coherent 
columnar structures are present, and the plume dynamics are dominated by lateral 
advection by the zonal flow. The influence of reducing $Ro$ on plume morphology is most 
pronounced at high $Ri_0$; at lower $Ri_0$ the stronger inlet momentum flux renders the 
plume structure largely insensitive to the rotation rate.

\begin{figure}[t]
\centering
\subfigure[]{%
\includegraphics[scale = 0.4]{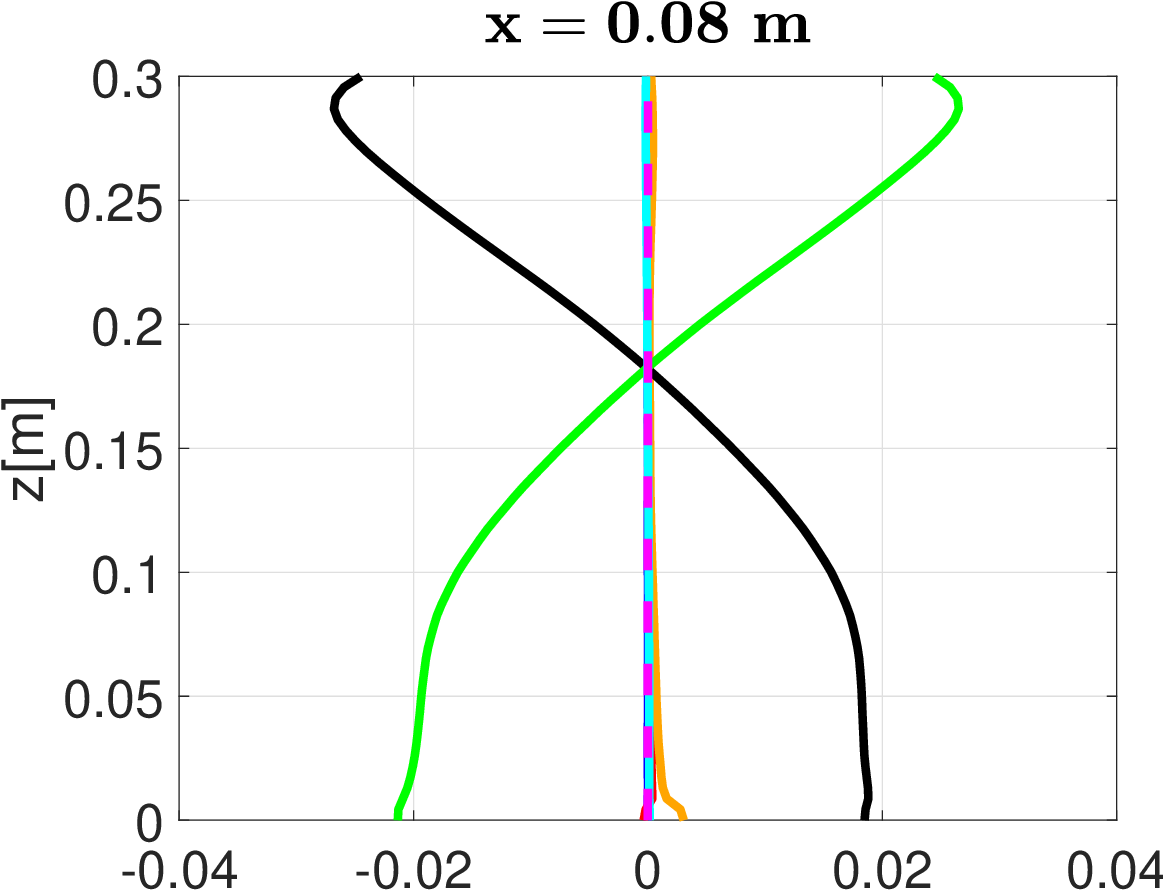}
\label{fig:x_mom_bdgt_a}}
\subfigure[]{%
\includegraphics[scale = 0.4]{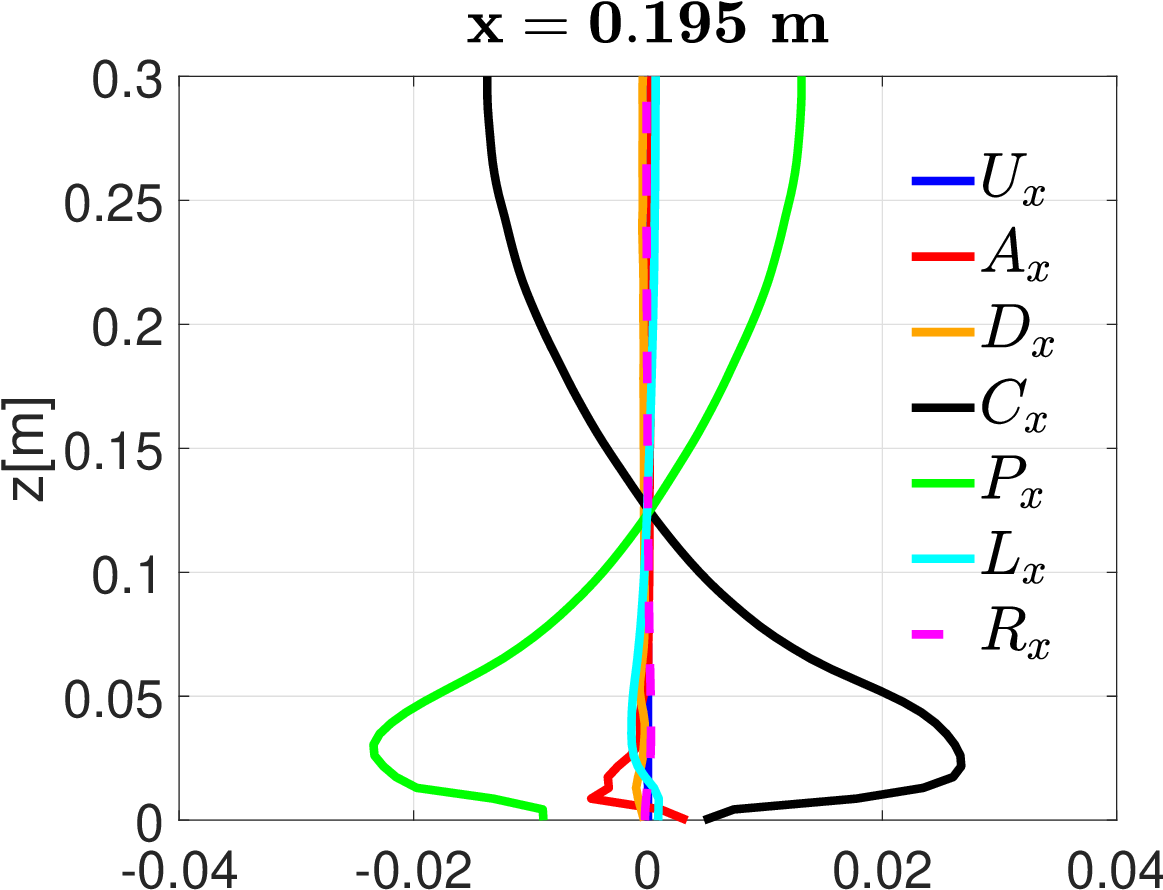}
\label{fig:x_mom_bdgt_b}}

\caption{Vertical variation of the $x-$momentum budget terms at $x=$ 0.08 m, near the cold wall (a), and $x=$ 0.195 m, over the forced heated plume (b), for $Ri_0 = 4$ and $Ro = 0.1$. Blue, red, orange, black, green, cyan, and magenta lines are for unsteady term $(U_x)$, advection term $(A_x)$, the divergence of Reynolds stress term $(D_x)$, Coriolis term $(C_x)$, pressure gradient term $(P_x)$, laplacian term $(L_x)$, and residual term $(R_x)$ respectively. Unit is $m^{2}s^{-1}$.} 
\label{fig:x_mom_bdgt}
\end{figure}

\subsection{\label{sec:Rec_budget}Momentum budget and entrainment rate of plume}

In this section, we examine the budget for $x$ and $z-$momentum equations to understand the contribution of each term in the budget towards the average flow dynamics of the plume. In the discussion below, the operator $\avg{\cdot}$ represents averaging over time and $y-$direction. 

\subsubsection{Analysis of x-momentum budget}

The geostrophic balance anticipated from the scaling analysis of
Sec.~\ref{sec:Rec_mean_field} - namely
$v^{*} = \partial p^{*}/\partial x^{*}$ at leading order in $\Ro$ and $\Ek$ -- can now be verified directly from the $x$-momentum budget. The terms of the $x-$momentum equation are summarized below in Eqs. \eqref{eqn:Unsteady_x} - \eqref{eqn:laplacian_x}. 

\begin{eqnarray}
    \label{eqn:Unsteady_x}
     \text{Unsteady term}~(U_x) &=& \frac{\partial \avg u}{\partial t} \\
    \label{eqn:advection_x}
    \nonumber \text{Advection term}~(A_x) &=& \avg{\pd{u u}{x}} + \avg{\pd{wu}{z}} - \pd{\avg{u'u'}}{x} \\  &&- \pd{\avg{u'w'}}{z} \\
    \label{eqn:divReyStress_x}
    \text{Divergence of Reynolds  stress}~(D_x) &=& \pd{\avg{u'u'}}{x} + \pd{\avg{u'w'}}{z} \\
    \label{eqn:coriolis}
    \text{Coriolis term}~(C_x) &=& - f \avg{v} \\
    \label{eqn:pgrad_x}
    \text{Pressure gradient term} (P_x) &=& \pd{\avg{p}}{x} \\
    \label{eqn:laplacian_x}
    \text{Viscous Laplacian term}~(L_x) &=& -\nu \left( \frac{\partial^2 \avg{u}}{\partial x^2} + \frac{\partial^2 \avg{u}}{\partial z^2} \right)
\end{eqnarray}

Based on the above definitions, the residual of $x-$ momentum equation budget will be:
\begin{eqnarray}
    R_x =  U_x + A_x + C_x + D_x - P_x - L_x.
\end{eqnarray}

\begin{figure}
\centering
\subfigure[]{%
\includegraphics[scale = 0.33]{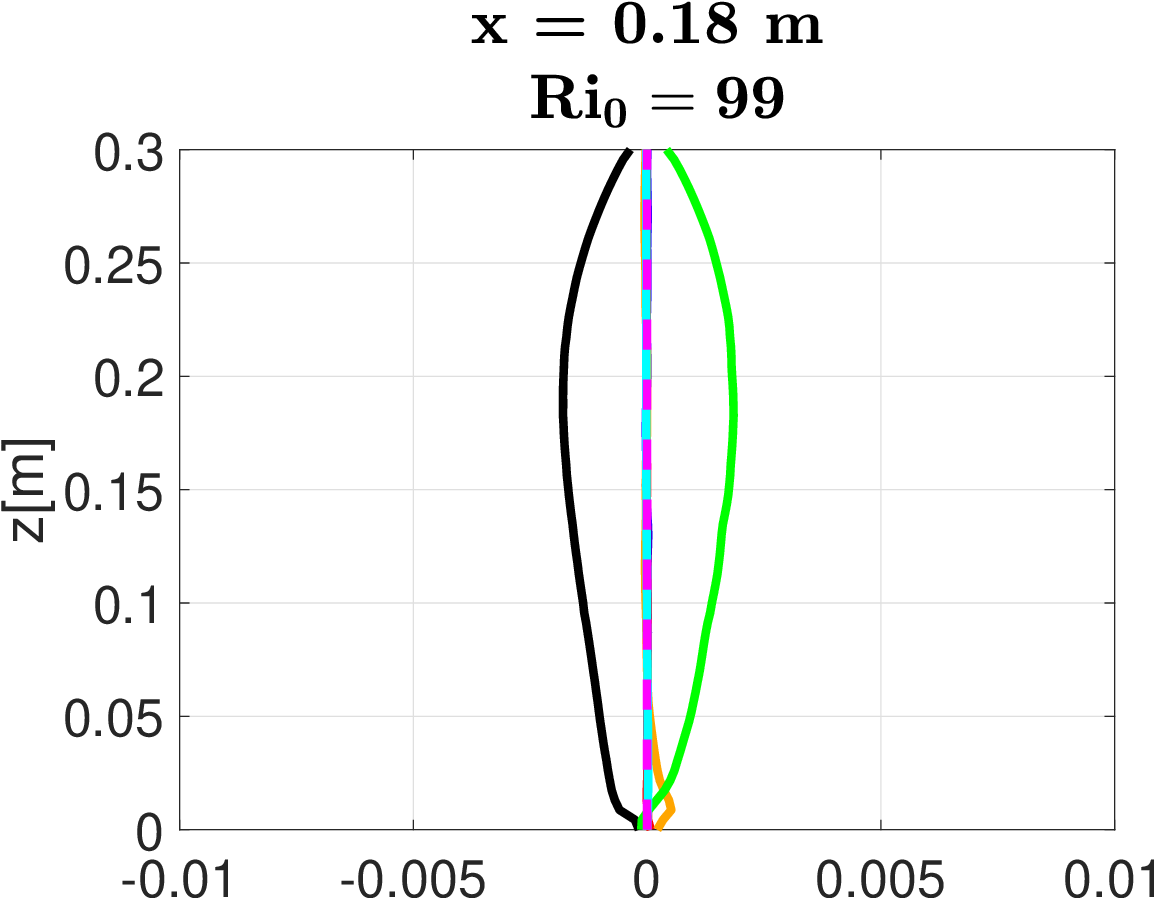}
\label{fig:z_mom_bdgt_aa}}
\subfigure[]{%
\includegraphics[scale = 0.33]{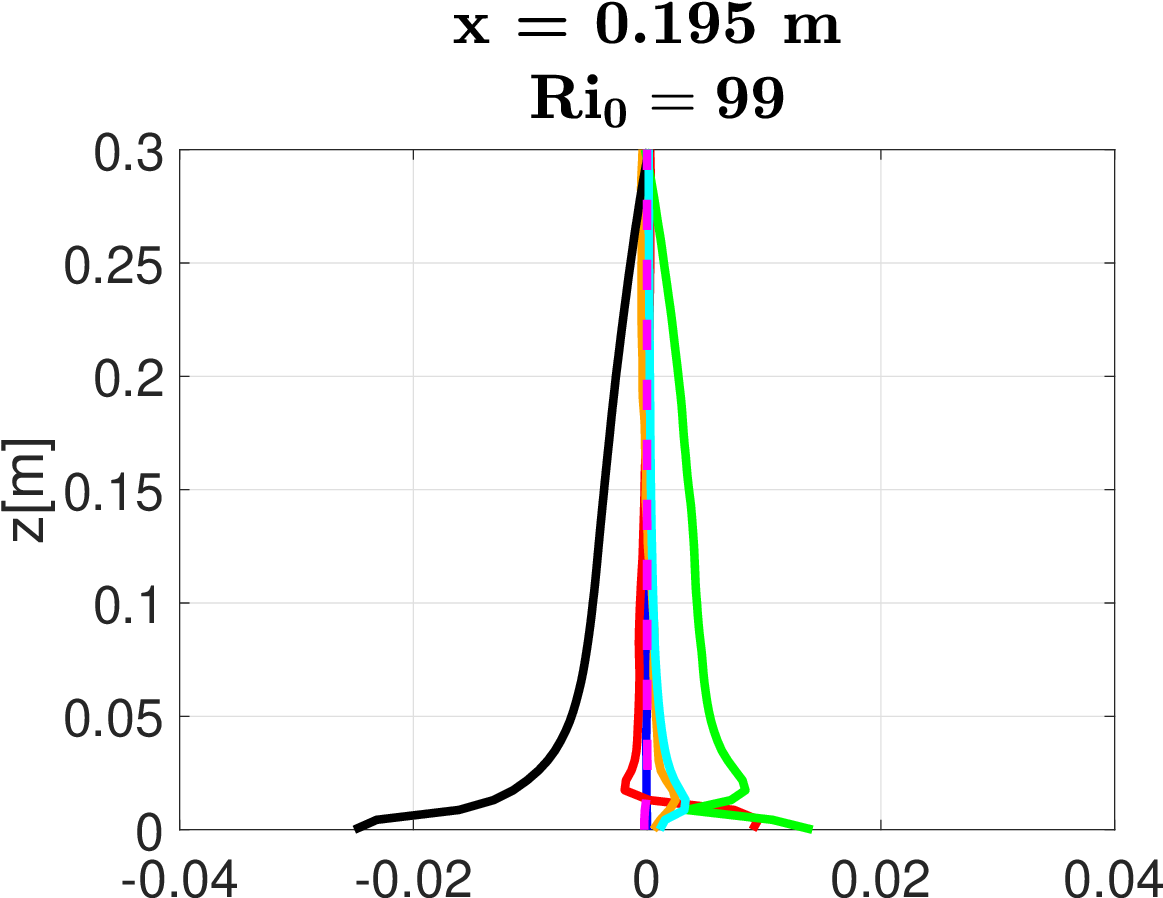}
\label{fig:z_mom_bdgt_a}}
\\
\subfigure[]{%
\includegraphics[scale = 0.33]{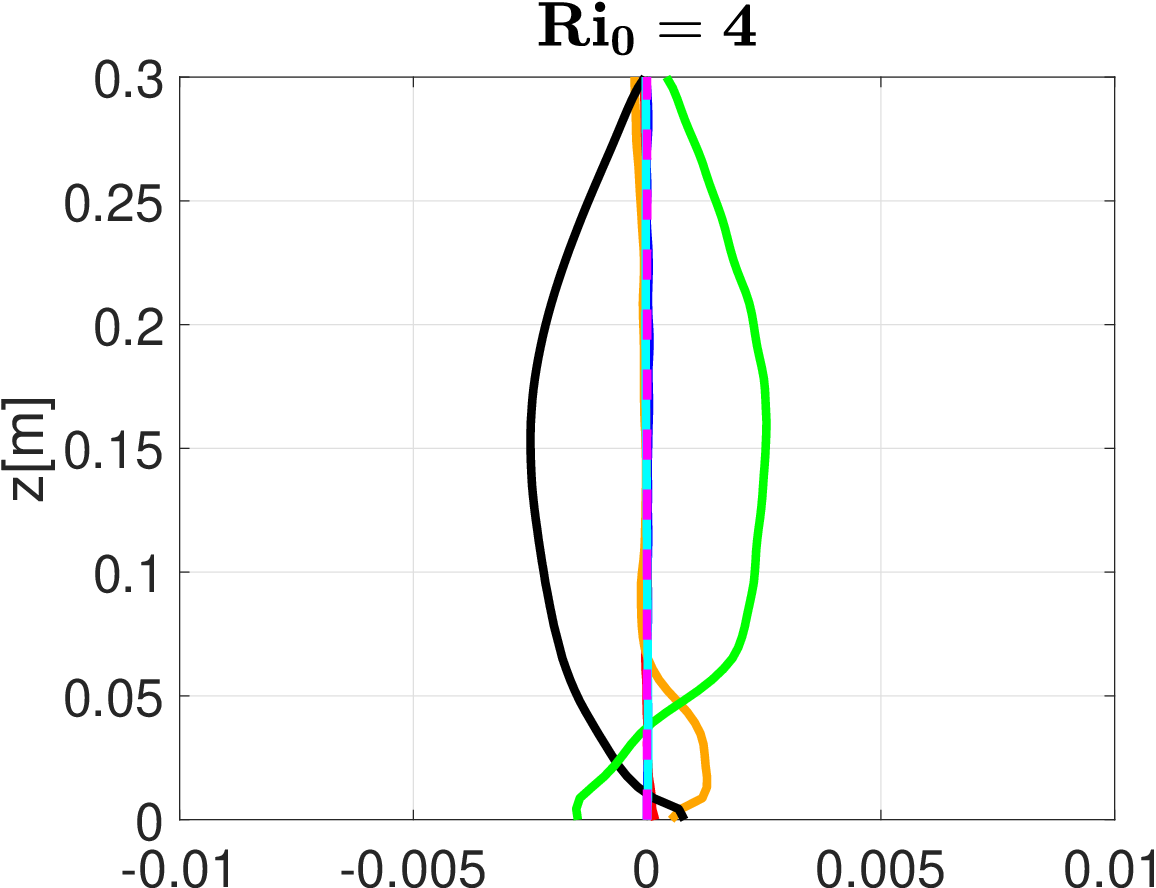}
\label{fig:z_mom_bdgt_bb}}
\subfigure[]{%
\includegraphics[scale = 0.33]{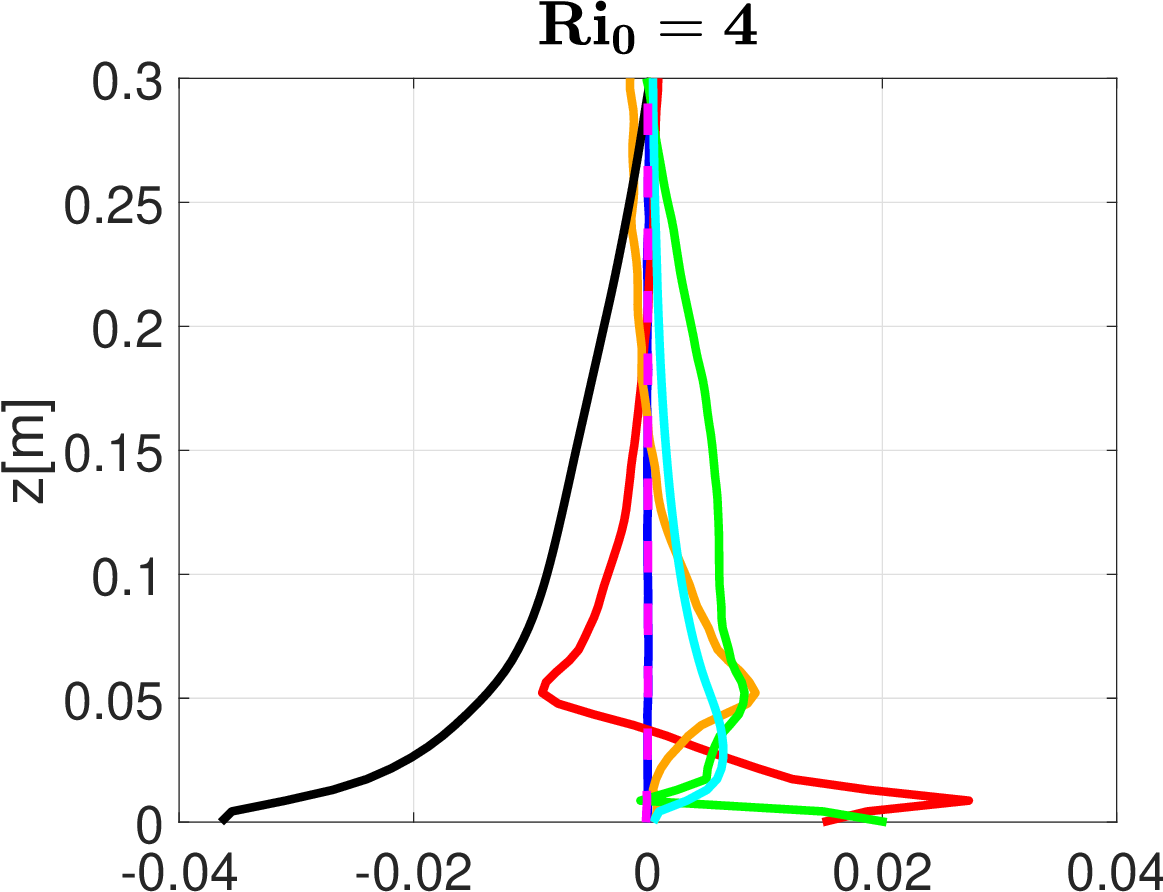}
\label{fig:z_mom_bdgt_b}}
\\
\subfigure[]{%
\includegraphics[scale = 0.323]{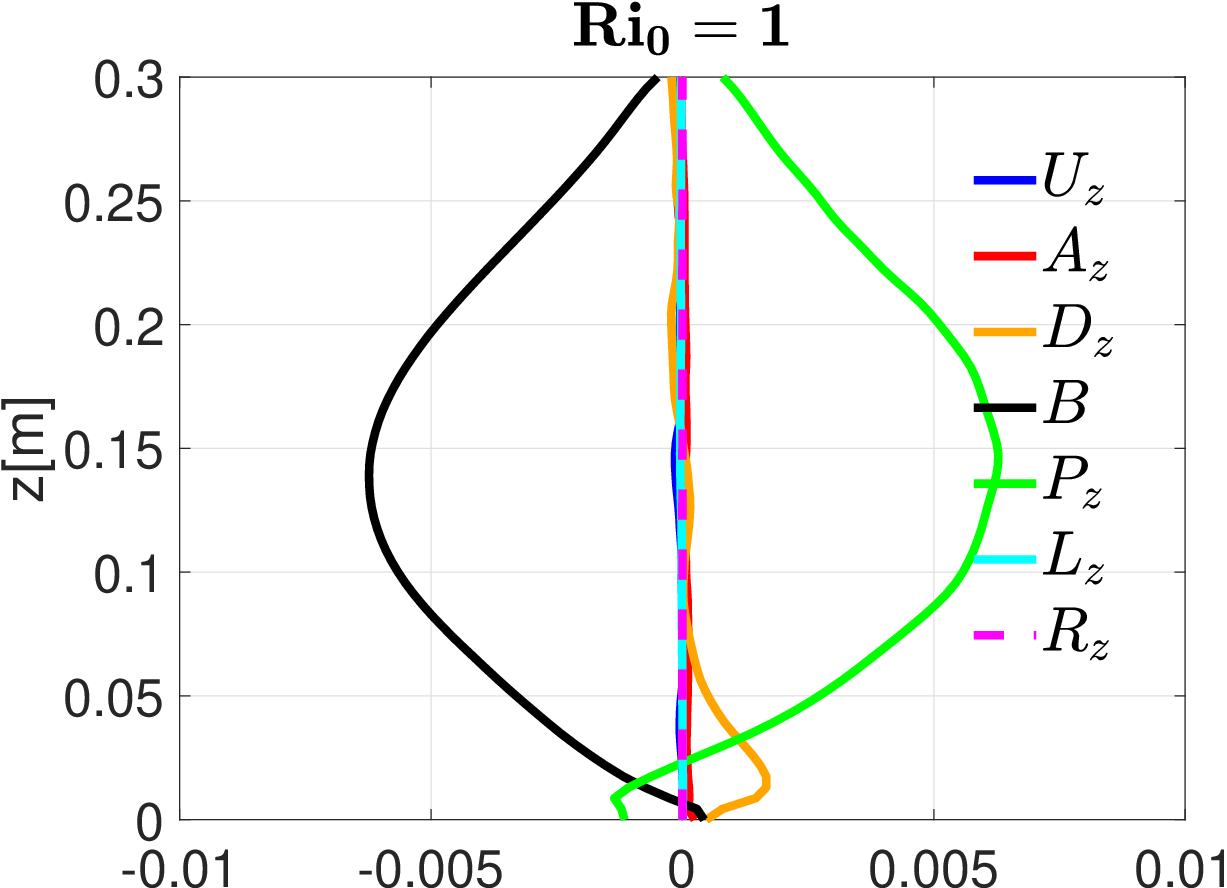}
\label{fig:z_mom_bdgt_cc}}
\subfigure[]{%
\includegraphics[scale = 0.323]{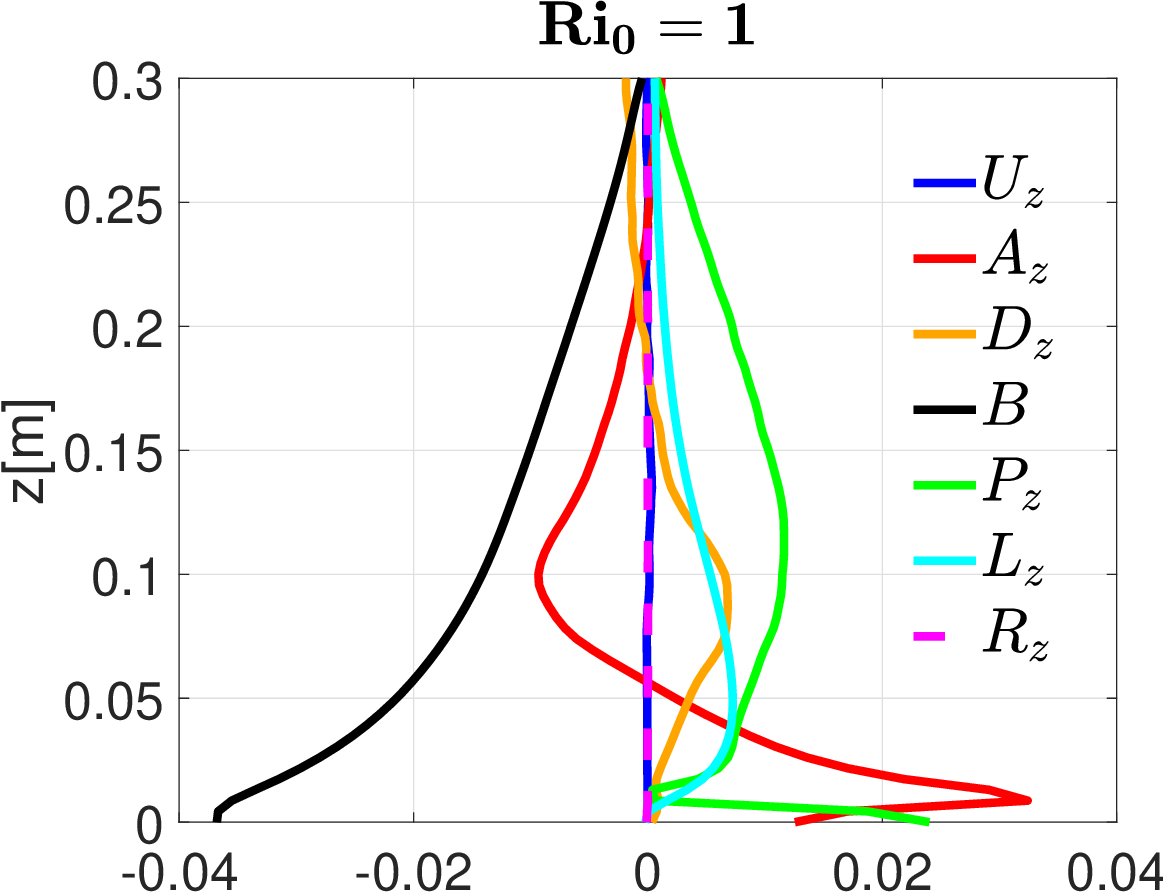}
\label{fig:z_mom_bdgt_c}}

\caption{Vertical variation of the $z-$momentum budget terms at $x=$ 0.18 m (away from the forced plume) and $x=$ 0.195 m (over the forced plume) for $Ro = 0.1$ and $Ri_0 = 99$ [(a), (b)], $Ri_0 = 4$ [(c), (d)], and $Ri_0 = 1$ [(e), (f)]. Blue, red, orange, black, green, cyan, and magenta lines are for unsteady term $(U_z)$, advection term $(A_z)$, divergence of Reynolds stress term $(D_z)$, buoyancy term $(B)$, pressure gradient term $(P_z)$, laplacian term $(L_z)$, and residual term $(R_z)$ respectively. Unit is $m^{2}s^{-1}$.} 
\label{fig:z_mom_bdgt}
\end{figure}

The residual should vanish at statistically stationary state.
Eq.~\eqref{eq:nondim_xmom} predicts an $\mathcal{O}(1)$ balance between the Coriolis
and pressure-gradient terms, with inertial corrections no larger than
$\Ro\lesssim0.3$ and viscous corrections of order $\Ek=\mathcal{O}(10^{-5})$.
Figure~\ref{fig:x_mom_bdgt} shows the vertical variation of the budget
terms at $x=0.08$~m (Figure~\ref{fig:x_mom_bdgt_a}) and $x=0.195$~m
(Figure~\ref{fig:x_mom_bdgt_b}) for the representative case $Ri_0=4$,
$\Ro=0.1$, and confirms these predictions: $|C_x|\simeq|P_x|\approx
2\times10^{-2}\,\mathrm{m^2\,s^{-2}}$ dominates the balance across the
bulk, the advection term $A_x$ is confined to $z<0.05$~m with a local
peak of $\approx3\times10^{-3}\,\mathrm{m^2\,s^{-2}}$ (about $15\%$ of
$|P_x|$ locally, $<2\%$ in the bulk) and is visible primarily at
$x=0.195$~m, the viscous Laplacian is $\mathcal{O}(\nu\avg{u}/l_x^2)\sim
10^{-6}\,\mathrm{m^2\,s^{-2}}$ (an $\mathcal{O}(\Ek)\sim10^{-5}$ correction), and
the residual $R_x$ remains below $1\%$ of $|P_x|$. These ratios match
the scaling estimates of Sec.~\ref{sec:Rec_mean_field}.

The $C_x\!\simeq\!-P_x$ balance is the horizontal geostrophic constraint:
the mean zonal flow (Figure~\ref{fig:Vectr_plot}) is aligned with the
constant-pressure lines, and combining it with hydrostatic balance in the
vertical yields the thermal-wind relation
$\partial v/\partial z=(g\alpha/f)\,\partial T/\partial x$, which sets the
strengthening of the mean zonal jet with the local horizontal temperature
gradient. The vertical shear implied by this relation produces positive
$\avg{v}$ in the upper half of the domain and negative $\avg{v}$ in the
lower half (Figure~\ref{fig:VMean_w_6_Ta_mod}), so $C_x=-f\avg{v}$
changes sign with height, and $P_x=-C_x$ mirrors it at both $x$
locations. The same behaviour is observed at the remaining $(Ri_0,\Ro)$
combinations.

\subsubsection{Analysis of z-momentum budget}

The vertical counterpart of this balance is hydrostatic.  In
Eq.~\eqref{eq:nondim_zmom} the buoyancy coefficient $\Ra\,\Ek^{2}/(\Pran\,\Ro) = l_{x}/l_{z} \approx 0.4$ is $\mathcal{O}(1)$, of
the same order as the vertical pressure-gradient coefficient, while the inertial and viscous coefficients are $\mathcal{O}(\Ro)$ and $\mathcal{O}(\Ek)$
respectively.  The leading-order vertical balance is therefore $\partial p^{*}/\partial z^{*} \simeq (l_{x}/l_{z})\,\theta^{*}$.
Figure~\ref{fig:z_mom_bdgt} confirms that, in the bulk and along the plume above the inlet, the buoyancy and vertical pressure-gradient terms are indeed the dominant pair; the advection
term becomes comparable only in the near-inlet region, consistent with the $\mathcal{O}(\Ro)$ estimate and with the breakdown of the bulk scaling near the source discussed in Sec.~\ref{sec:Rec_plume}.

In this section, we analysis the $z$-momentum equation, for which the budget terms are:
\begin{eqnarray}
    \label{eqn:Unsteady_z}
     \text{Unsteady term}~(U_z) &=& \frac{\partial \avg w}{\partial t} \\
    \label{eqn:advection_z}
    \nonumber \text{Advection term}~(A_z) &=& \avg{\pd{u w}{x}} + \avg{\pd{ww}{z}} - \pd{\avg{u'w'}}{x}  \\ && - \pd{\avg{w'w'}}{z} \\
    \label{eqn:divReyStress_z}
    \text{Divergence of Reynolds  stress}~(D_z) &=& \pd{\avg{u'w'}}{x} + \pd{\avg{w'w'}}{z} \\
    \label{eqn:Buoyancy}
    \text{Buoyancy term}~(B) &=& g \left (\avg{\rho_k} - \avg{\overline{\rho_k}} \right) \\
    \label{eqn:pgrad_z}
    \text{Pressure gradient term}~(P_z) &=& \pd{\avg{p}}{z} \\
    \label{eqn:laplacian_z}
    \text{Viscous Laplacian term}~(L_z) &=& -\nu \left( \frac{\partial^2 \avg{w}}{\partial x^2} + \frac{\partial^2 \avg{w}}{\partial z^2} \right)
\end{eqnarray}

In Eq. \ref{eqn:Buoyancy}, the term $\overline{\rho_k}$ is the average of $\rho_k$ at every $z$ location. The buoyancy term, $g(\avg{\rho_k} - \avg{\overline{\rho_k}})$ is equivalent to $-g \alpha (\avg{T} - \avg{\overline{T}})$. The pressure gradient term $(P_z)$ consists of vertical gradient of mean non-hydrostatic pressure $\left (\pd{\avg{p}}{z}\right)$. The residual of $z-$momentum equation budget terms is given as,
\begin{eqnarray}
    R_z =  U_z + A_z + D_z - B - P_z - L_z.
\end{eqnarray}

Figure \ref{fig:z_mom_bdgt} shows the vertical variation of $z-$momentum budget terms for $Ro =$ 0.1 and $Ri_0 =$ 99, 4, and 1 close to the outer wall of the annulus ($x=0.18$ m) and at the outer wall of the annulus ($x=0.195$ m). For other combinations of $Ro$ and $Ri_o$, the balance of the $z-$momentum budget terms is similar to that shown in  Figure \ref{fig:z_mom_bdgt} and hence not shown here for brevity. Moreover, the vertical momentum balance in the bulk region has not been plotted due to the small magnitudes of $\avg{w}$ (i.e. local hydrostatic balance) observed there. The unsteady term $(U_z)$ is close to zero for all the cases as shown in Figure \ref{fig:z_mom_bdgt}, which suggests that the simulations have converged. The residual term $R_z$ is again negligible compared to the dominant terms in the budget. 
At $x=0.18$ m, just outside the forced plume, it appears that the vertical momentum balance is primarily between buoyancy and pressure gradient for all values of $Ri_0$ (Figures \ref{fig:z_mom_bdgt_aa}, \ref{fig:z_mom_bdgt_bb}, and \ref{fig:z_mom_bdgt_cc}). We observe that the vertical pressure gradient at the outer wall ($x=0.195$ m) is similar to that of $x=0.18$ m, which is not surprising, given the slenderness of the plume. At $Ri_0 = 99$ and $Ro = 0.1$, for $z$ in the range of 0.02 m to 0.3 m, the buoyancy term balances the pressure gradient term, and the remaining terms are close to zero (Figure \ref{fig:z_mom_bdgt_a}). Below $z=$ 0.02 m, mainly the advection term and pressure gradient term balance the buoyancy term, and the importance of the viscous, laplacian, and divergence of Reynolds stress terms is relatively less. The magnitude of the advection term (red line in Figure \ref{fig:z_mom_bdgt_c}) is highest for $Ri_0 = 1$, which is consistent with the higher plume inlet velocity at this $Ri_0$. The pressure gradient term $(P_z)$ in the axial direction is much higher at lower $Ri_0$ values. This leads to the existence of sustained plume structures up to higher $z$ values at lower $Ri_0$ values (as shown in Figure \ref{fig:YZ_vector_plot}). 

Analyzing the buoyancy term, we first observe that $B$ is weaker at $Ri_0 = 99$, compared to $Ri_0=4$ and $Ri_0=1$. This indicates that $\rho_k$ is closer to $\bar{\rho}_k$ for low inlet velocities. This is not so surprising, since the lower buoyancy flux at the inlet will lead to lower heating of the overall flow domain. The buoyancy at the inlet is similar for $Ri_0=4$ and $Ri_0=1$, which indicates that at higher inflow, the inlet buoyancy $B=g(\avg{\rho_k}-\avg{\bar{\rho}_k})|_{inlet}$ almost reaches the limit $-g\alpha \Delta T_x=-0.0405 \textrm{m}^2/\textrm{s}$ prescribed by the imposed horizontal temperature difference $\Delta T_x$. To understand the evolution of $B$ with height along the plume, it is important to note that the rate of decrease of integral buoyancy flux of the plume is proportional to the product of background thermal stratification and vertical velocity of the plume $\avg{w}\partial \bar{T}/\partial z$. Since thermal stratification is stable ($\partial \bar{T}/\partial z<0$) and $\avg{w}>0$ within the plume, therefore it follows that the integral buoyancy flux will decrease with height. In Figure \ref{fig:z_mom_bdgt}, it can be observed that the vertical advection term $A_z$ (representing vertical acceleration of plume) peaks just above the inlet to balance the large $B$ in that region. The advection then reduces with height due to the adverse mean vertical pressure gradient ($P_z>0$) and viscous stress. Above a threshold height, the plumes undergo negative deceleration, or detrainment. At larger heights, the budget is again dominated by a balance between the mean buoyancy and mean pressure gradient, and the other terms in the budget play a secondary role in the flow dynamics. The threshold detrainment height is much less for $Ri_0=99$, compared to $Ri_0=1$ and $Ri_0=4$ due to the lower inlet buoyancy and momentum flux of the plume at high values of $Ri_0$. Moreover, the surrounding temperature field is much colder for lower $Ri_0$. Therefore the ambient stable stratification near the inlet is much stronger for $Ri_0=99$ compared to the other cases, which leads to a quicker reduction of buoyancy flux with height, as well as lower vertical acceleration.  These trends are also evident from the isocontours of instantaneous temperature of the plume in Figure \ref{fig:YZ_vector_plot}. 

\begin{figure}
\centering
\subfigure[]{%
\includegraphics[scale = 0.5]{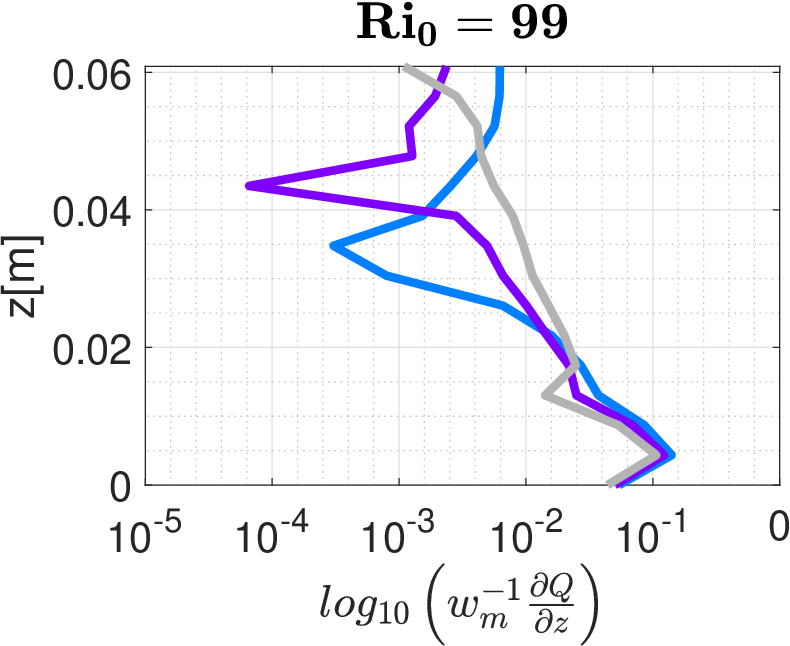}
\label{fig:entrnmnt_a}}
\subfigure[]{%
\includegraphics[scale = 0.5]{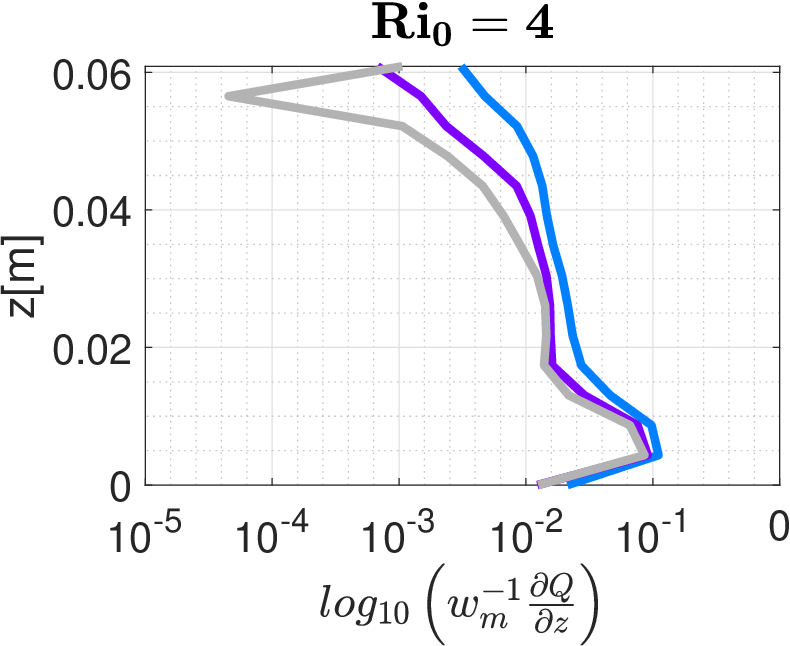}
\label{fig:entrnmnt_b}}
\subfigure[]{%
\includegraphics[scale = 0.5]{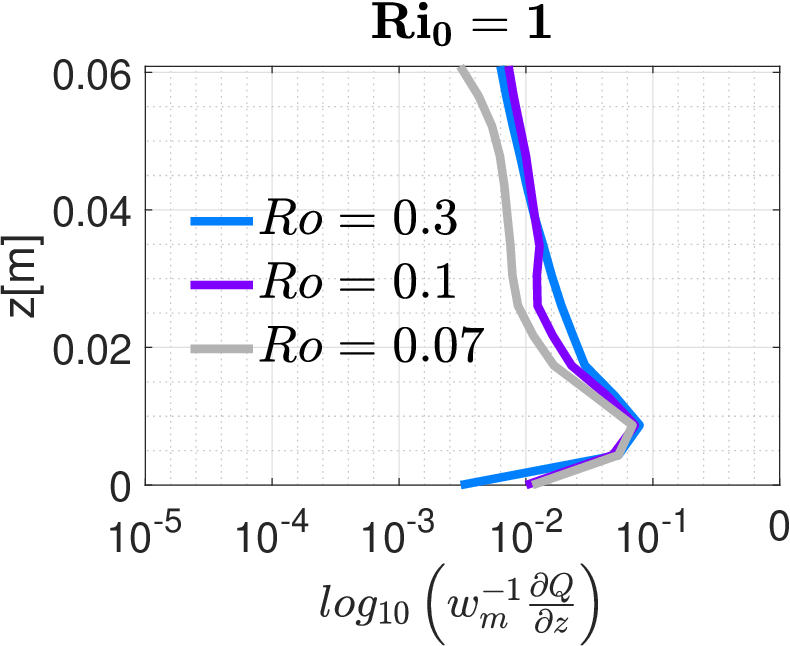}
\label{fig:entrnmnt_c}}
\caption{Log plot of the forced heated plume entrainment coefficient $(\Gamma)$, up to $z = 0.06$ m, for $Ri_0 = 99$ (a), $Ri_0 = 4$ (b), and $Ri_0 = 1$ (c) at various $Ro$ values. Here $\Gamma = \avg{w}_{m}^{-1}\frac{dQ}{dz}$ and $Q = \int_{x_l}^{x_u} \avg{w} \,dx$, where $\avg{w}$ is the time and $y-$direction averaged $z-$direction velocity, $x_u = 0.195$ m and $x_l = 0.185$ m are the upper and lower limits of integration along the $x-$direction, and $\avg{w}_{m}$ is the maximum time and $y-$direction averaged axial velocity at each $z$ levels within the integration limits. Negative $\avg{w}$ velocity values are ignored while integrating.}

\label{fig:entrnmnt_coeff}
\end{figure}

\subsection{Entrainment characteristics of forced plume}

We next examine the entrainment characteristics of the forced planar plume. According to the entrainment hypothesis \cite{morton1956turbulent} for an isolated planar plume, the maximum mean vertical velocity $\avg{w}_{m}(z)$ of the plume is related to the mean entrainment rate of the plume via the relationship $dQ/dz=\Gamma(z) \avg{w}_{m}$, where $\Gamma$ is the entrainment rate coefficient and $Q(z) = \int_{x_l}^{x_u} \avg{w}(x,z) \,dx$ is the volume flow rate. We use the integration limits i.e., $x_u = 0.195 ~\text{m}$ and $x_l=0.185~\text{m}$, and we consider only positive values of $\avg{w}$ while integrating, so that the integral is only over the plume updraft.

The entrainment coefficient is obtained by rearranging the entrainment relation as $\Gamma(z)=\avg{w}_{m}^{-1} dQ/dz$. Figure \ref{fig:entrnmnt_coeff} shows the vertical variation of $\Gamma(z)$ for $Ri_0 = 99, 4,$ and $1$ for different $Ro$. It is observed that $\Gamma(z)$ is maximum near the source of the forced plume for $Ri_0 = 99$ and decreases with height. For $Ri_0 = 99$ (Figure \ref{fig:entrnmnt_a}), $\Gamma(z)$ decreases more rapidly compared to lower values of $Ri_0$, which is consistent with the break up of plume structures in Figure \ref{fig:YZ_vector_plot}. At this inlet velocity, $\Gamma(z)$ is quite sensitive to $Ro$. Specifically, at larger rotation rates (smaller $Ro$), $\Gamma(z)$ exhibits a slower decline with increasing height. For $Ri_0 =$ 4 (Figure \ref{fig:entrnmnt_b}) and $Ri_0 = $ 1 (Figure \ref{fig:entrnmnt_c}), the entrainment rate coefficient $\Gamma(z)$ shows a less pronounced decrease with increasing height. At these higher inlet velocities, $\Gamma(z)$ also appears to be less sensitive to the frame rotation rate ($Ro$). Thus, it appears that at low $Ri_0$, frame rotation makes the plumes more coherent along $z$, allowing the plumes to detrain at a higher height. 

To summarize, the momentum budget along the $x$ direction displays local geostrophy throughout the flow domain, and the iscontours of mean zonal velocity field appears to be consistent with geostrophic balance. The momentum budget along $z$ within the plume suggests that, the buoyancy near the inlet drives the vertical momentum. At higher height, the stable background stratification along with the leads to detrainment of the plume. Frame rotation rate also appears to play an important role in determining the entrainment characteristics of the plumes, and may either hinder detrainment (at low inlet velocities) or aid detrainment (at high inlet velocities). 

The opposite effects of rotation on plume detrainment at low and high
$Ri_{0}$ admit a compact physical interpretation in terms of the local
plume Rossby number,
\begin{equation}
\mathrm{Ro}_{p}\;=\;\frac{w}{2\Omega\,b},
\label{eq:Rop}
\end{equation}
the ratio of the plume's vertical inertia to the Coriolis acceleration
acting on it. Two distinct rotational mechanisms operate, selected by
$\mathrm{Ro}_{p}$. For $\mathrm{Ro}_{p}\ll 1$ the plume is strongly
rotationally constrained: the horizontally convergent inflow required
for lateral entrainment of ambient fluid is opposed by the Coriolis
force, ambient exchange is suppressed, and a weak plume that would
otherwise be eroded is instead \emph{protected}, so that rotation
\emph{hinders} detrainment. For $\mathrm{Ro}_{p}=\mathcal{O}(1)$ the
dominant effect is instead the direct Coriolis force on the plume's
substantial vertical momentum: the rising jet is deflected off the
vertical, the buoyancy is no longer aligned with the trajectory, the
entrainment surface increases, and rotation \emph{aids} detrainment. 
The crossover sits at $\mathrm{Ro}_{p}\sim 1$. Evaluating Eq.~\eqref{eq:Rop} for the present nine cases places the
$Ri_{0}=99$ runs deep in the first regime ($\mathrm{Ro}_{p}\approx
0.066$--$0.135$, $\mathrm{Ro}_{p}\!\ll\!1$), the $Ri_{0}=4$ runs
moderately within it ($\mathrm{Ro}_{p}\approx 0.33$--$0.67$, all
below the crossover), and the $Ri_{0}=1$ runs straddling the crossover
($\mathrm{Ro}_{p}\approx 0.66$--$1.35$). The sensitivity of $\Gamma(z)$
to $\Ro$ visible in Figure~\ref{fig:entrnmnt_coeff} follows this
ordering: strongest at $Ri_{0}=99$, where the plume sits deep in the
suppression regime and every reduction in $\Ro$ tightens the
rotational protection of the rising column; intermediate at
$Ri_{0}=4$, where the hindrance mechanism is weaker;
and weakest at $Ri_{0}=1$, where the range of $\Ro$ carries $\mathrm{Ro}_p$
across the crossover so that suppression at low $\mathrm{Ro}_p$ and
Coriolis-aided detrainment at $\mathrm{Ro}_p\gtrsim 1$ partly cancel.
The integral plume theory that turns this scaling into a quantitative
entrainment law $\Gamma(z;\mathrm{Ro}_{p})$, including the closed-form
detrainment height $h_{\mathrm{tr}}(\mathrm{Ro}_{p})$ and the
calibration against the simulated $\Gamma(z)$ of
Figure~\ref{fig:entrnmnt_coeff}, will be taken up in future work.

\begin{figure}
\centering
\subfigure[]{%
\includegraphics[scale = 0.1]{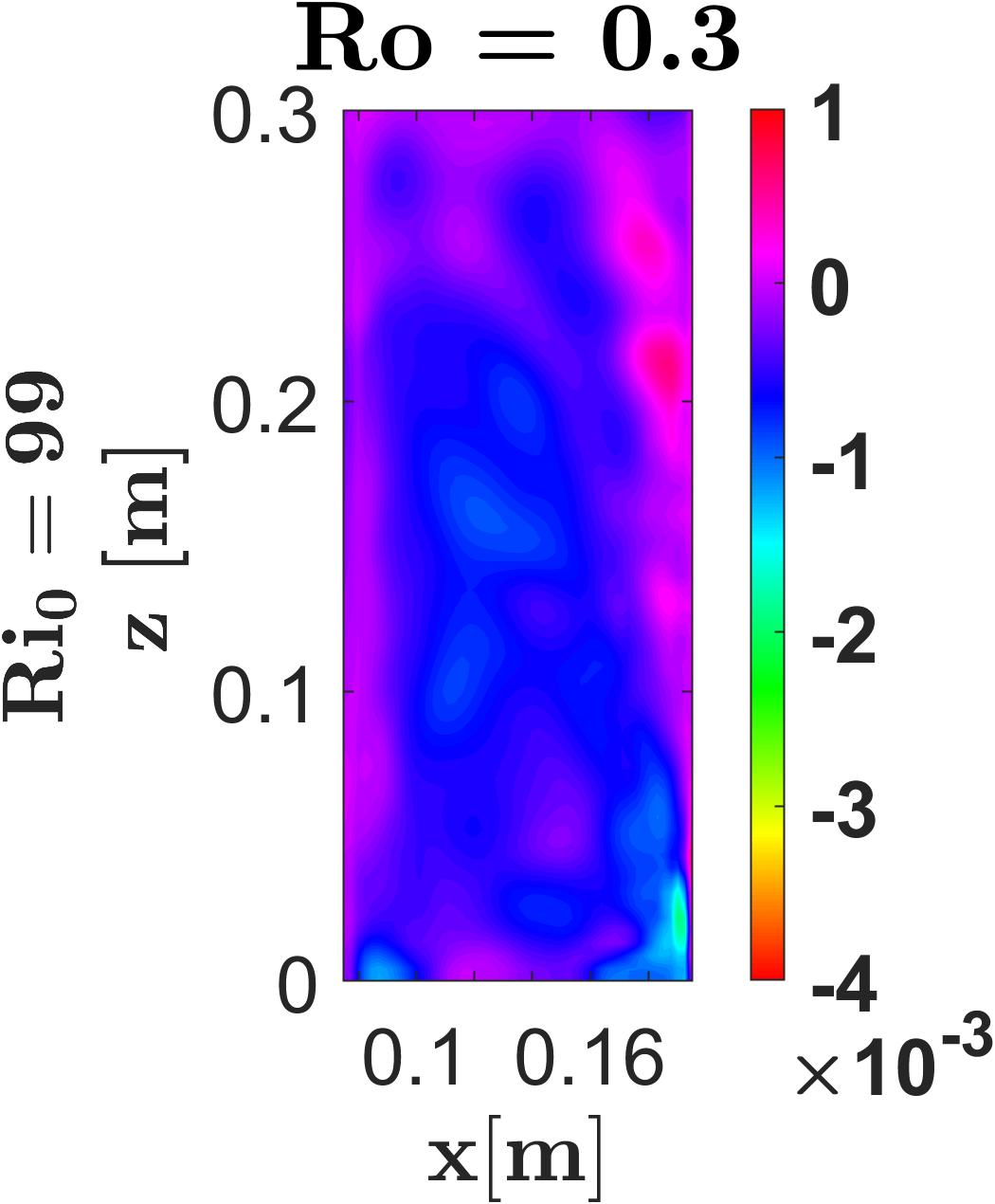}
\label{fig:x_enrgyflux_w_1.2_Ta_low}}
\hspace{1.2cm}
\subfigure[]{%
\includegraphics[scale = 0.1]{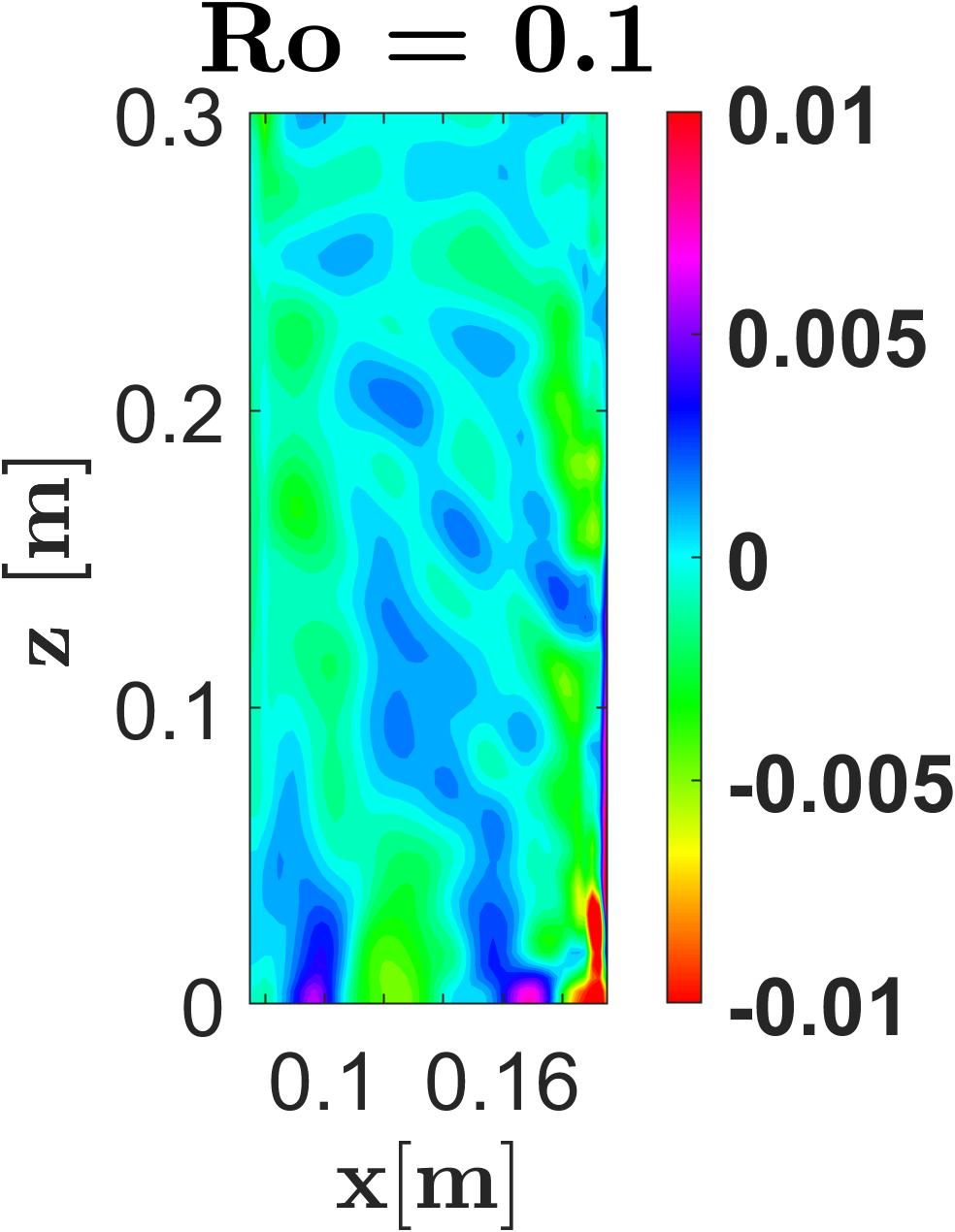}
\label{fig:x_enrgyflux_w_1.2_Ta_mod}}
\hspace{1.2cm}
\subfigure[]{%
\includegraphics[scale = 0.1]{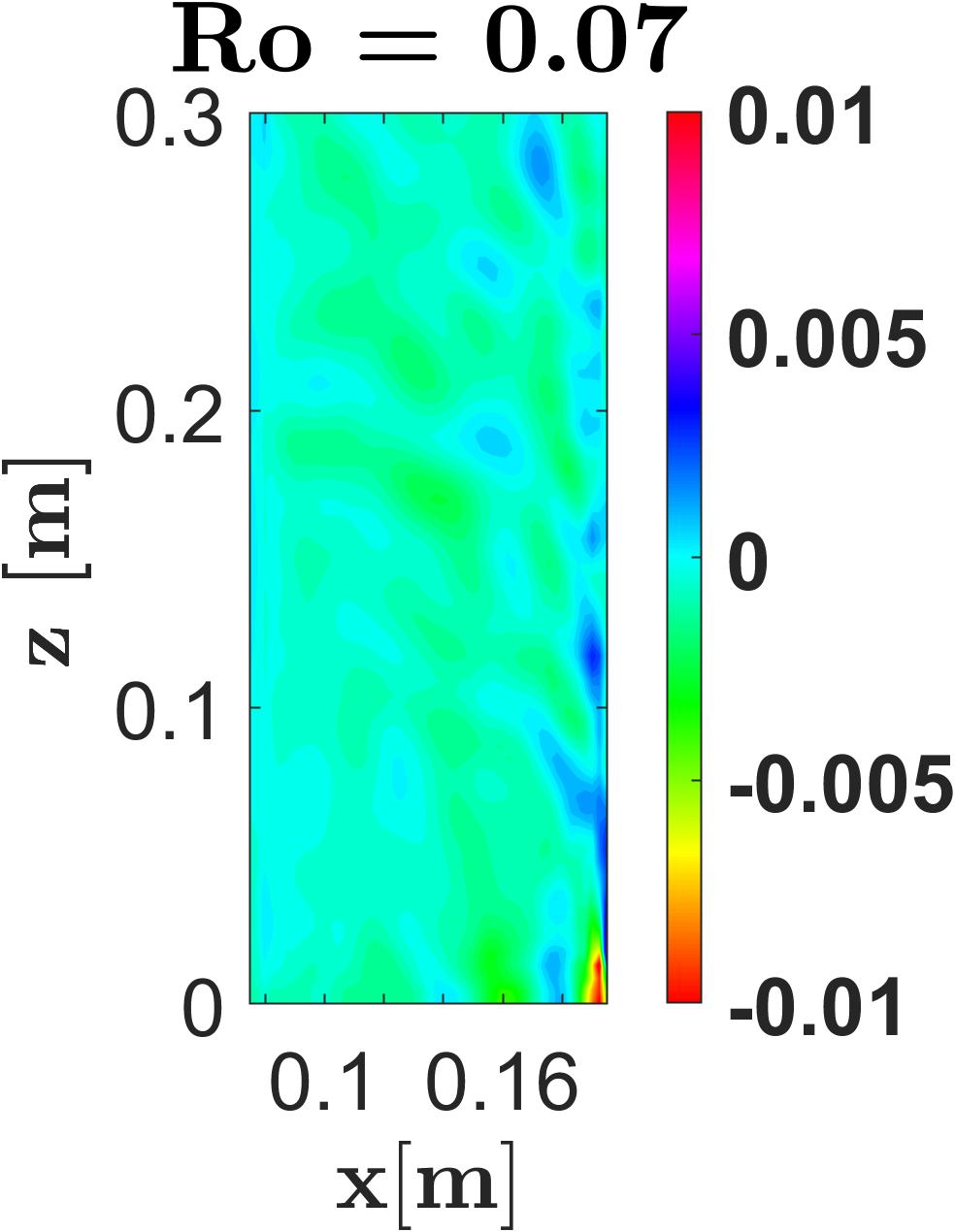}
\label{fig:x_enrgyflux_w_1.2_Ta_high}}
\\
\subfigure[]{%
\includegraphics[scale = 0.09]{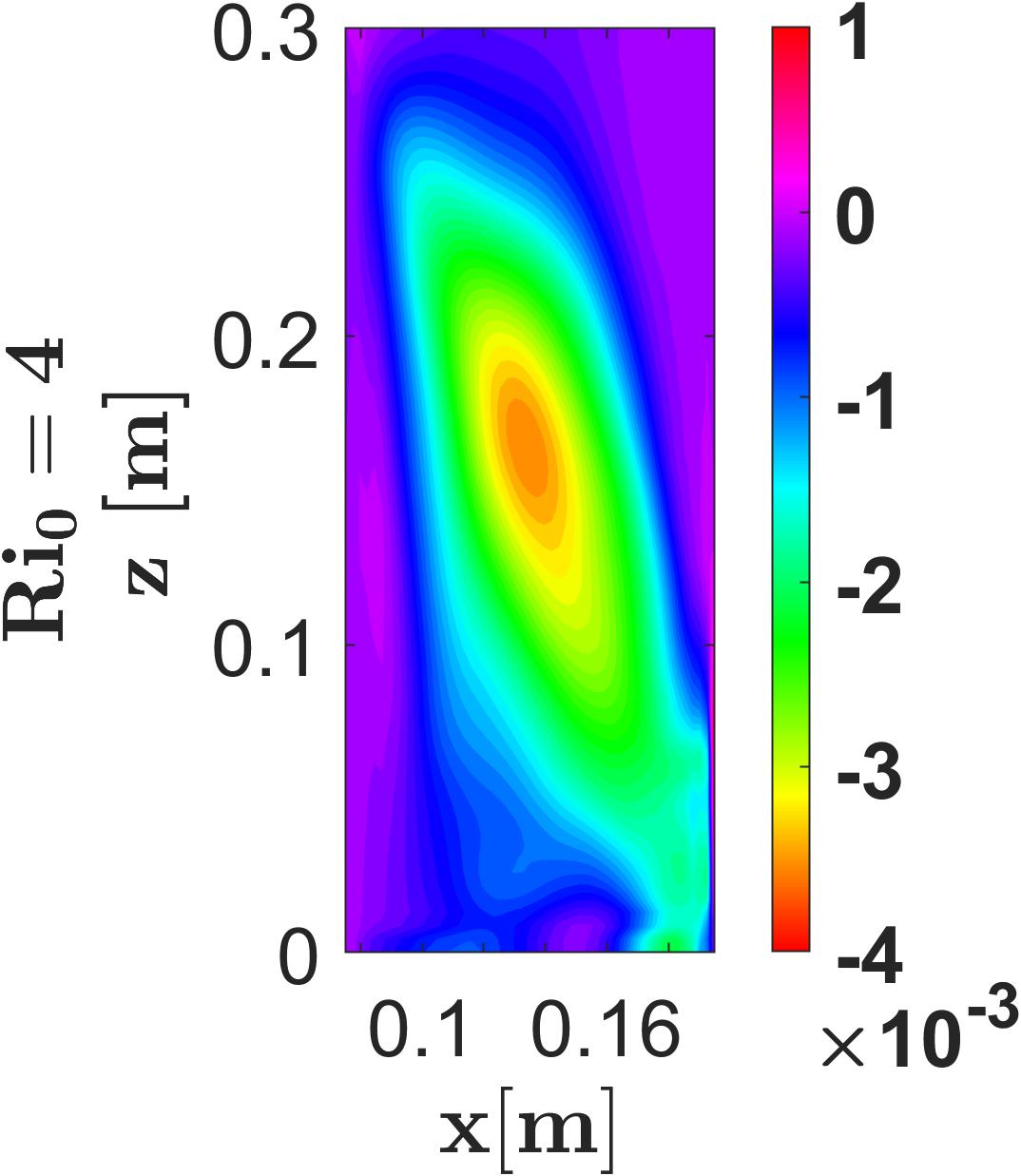}
\label{fig:x_enrgyflux_w_6_Ta_low}}
\hspace{1.7cm}
\subfigure[]{%
\includegraphics[scale = 0.09]{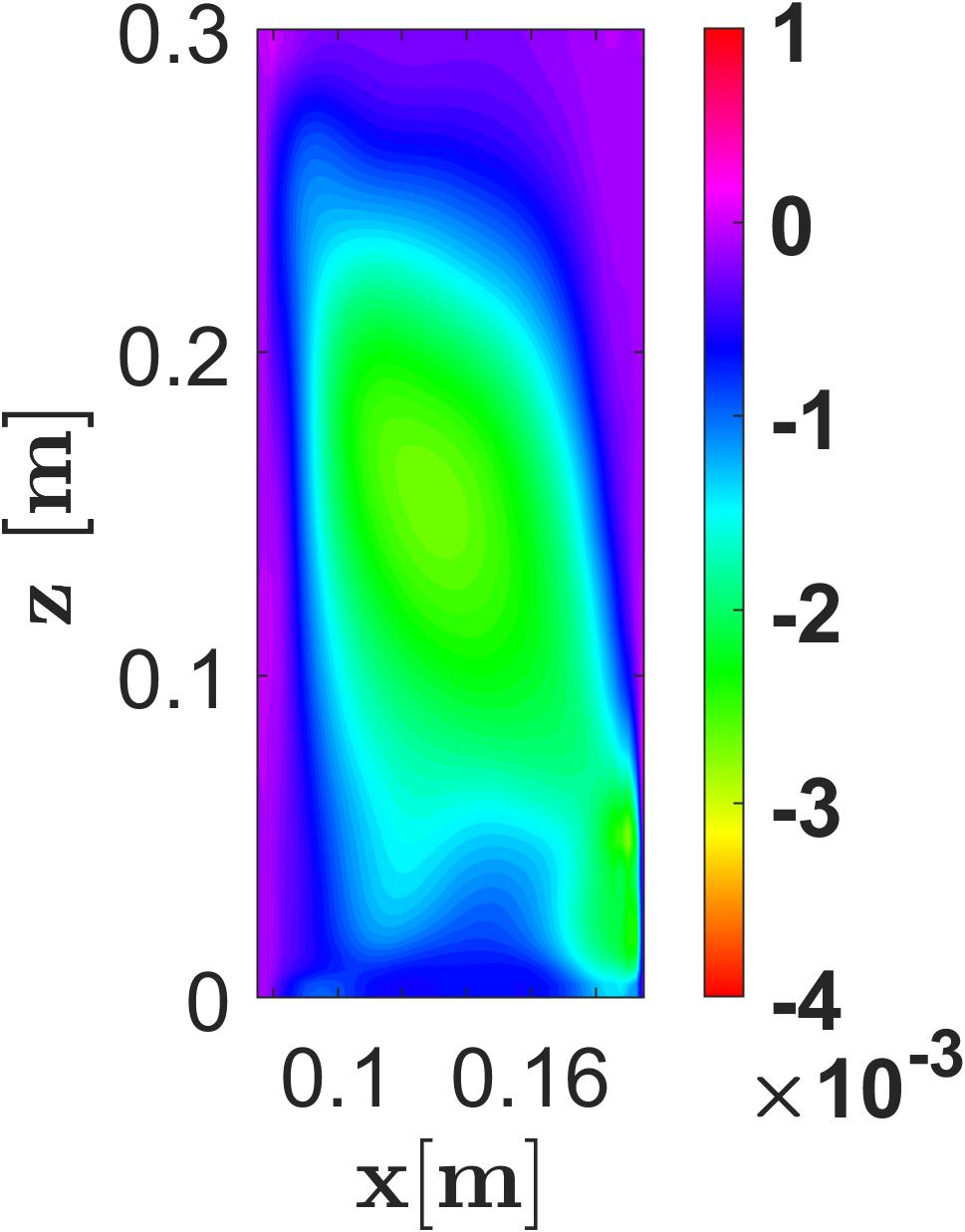}
\label{fig:x_enrgyflux_w_6_Ta_mod}}
\hspace{1.7cm}
\subfigure[]{%
\includegraphics[scale = 0.09]{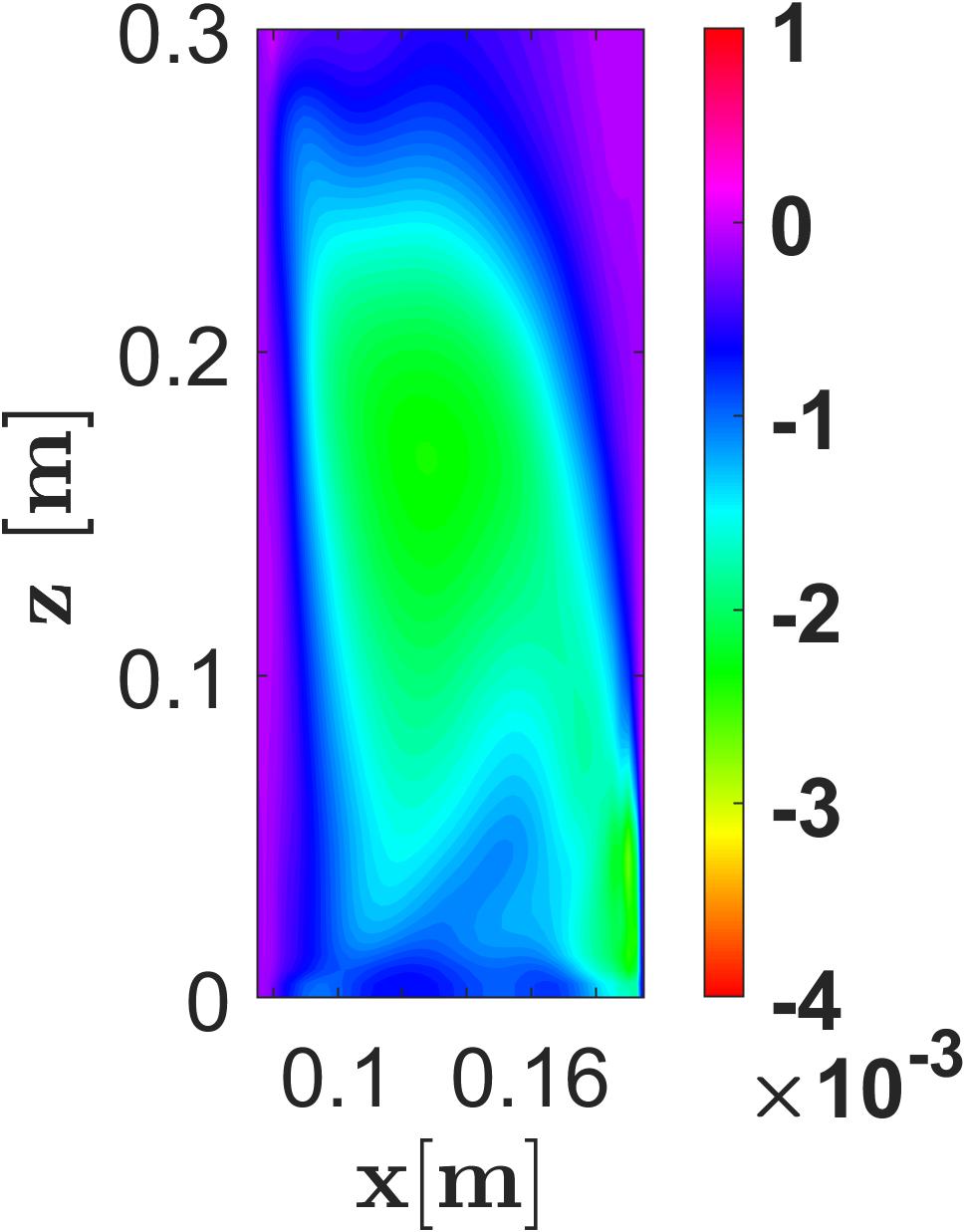}
\label{fig:x_enrgyflux_w_6_Ta_high}}
\\
\subfigure[]{%
\includegraphics[scale = 0.09]{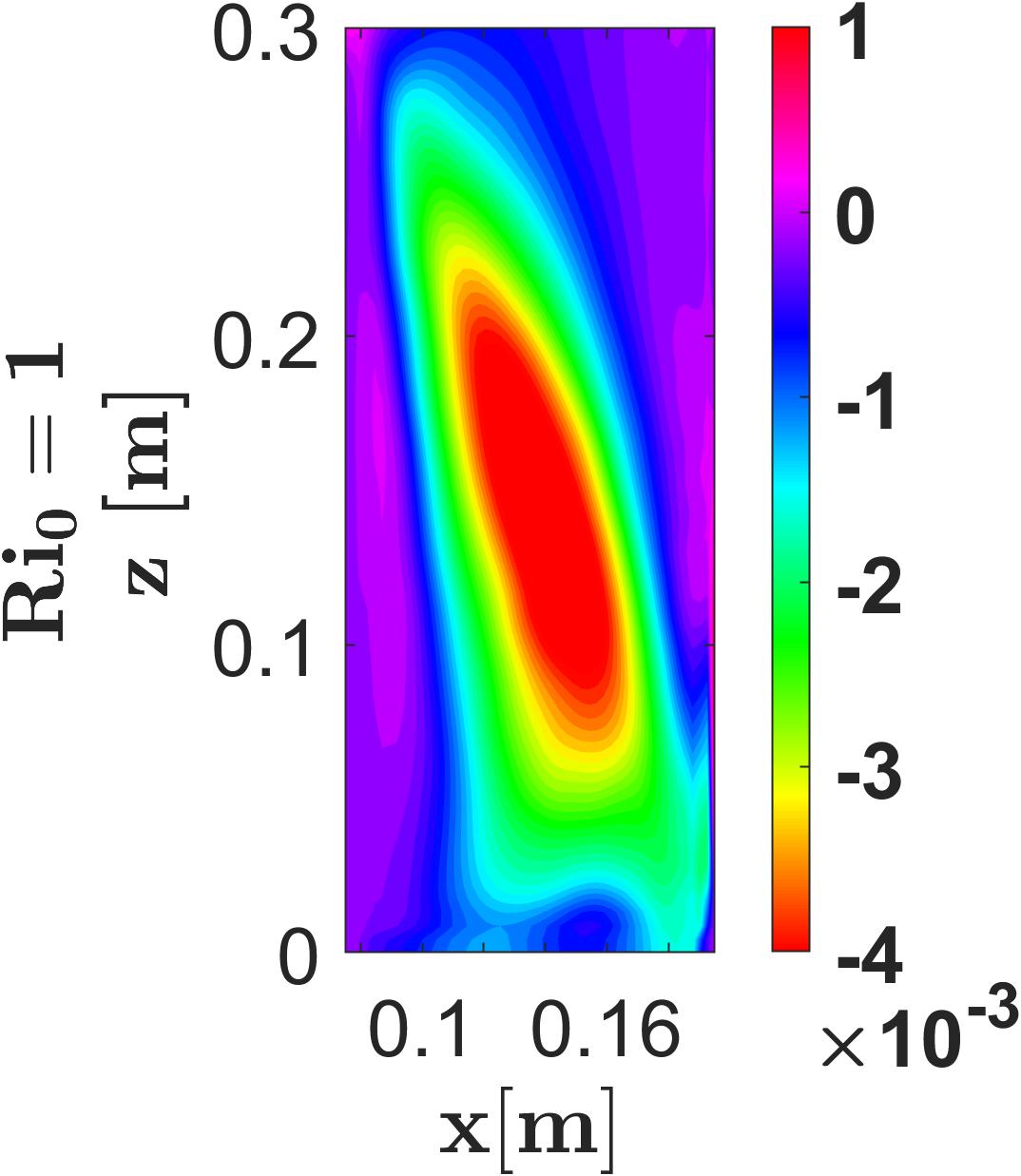}
\label{fig:x_enrgyflux_w_12_Ta_low}}
\hspace{1.5cm}
\subfigure[]{%
\includegraphics[scale = 0.09]{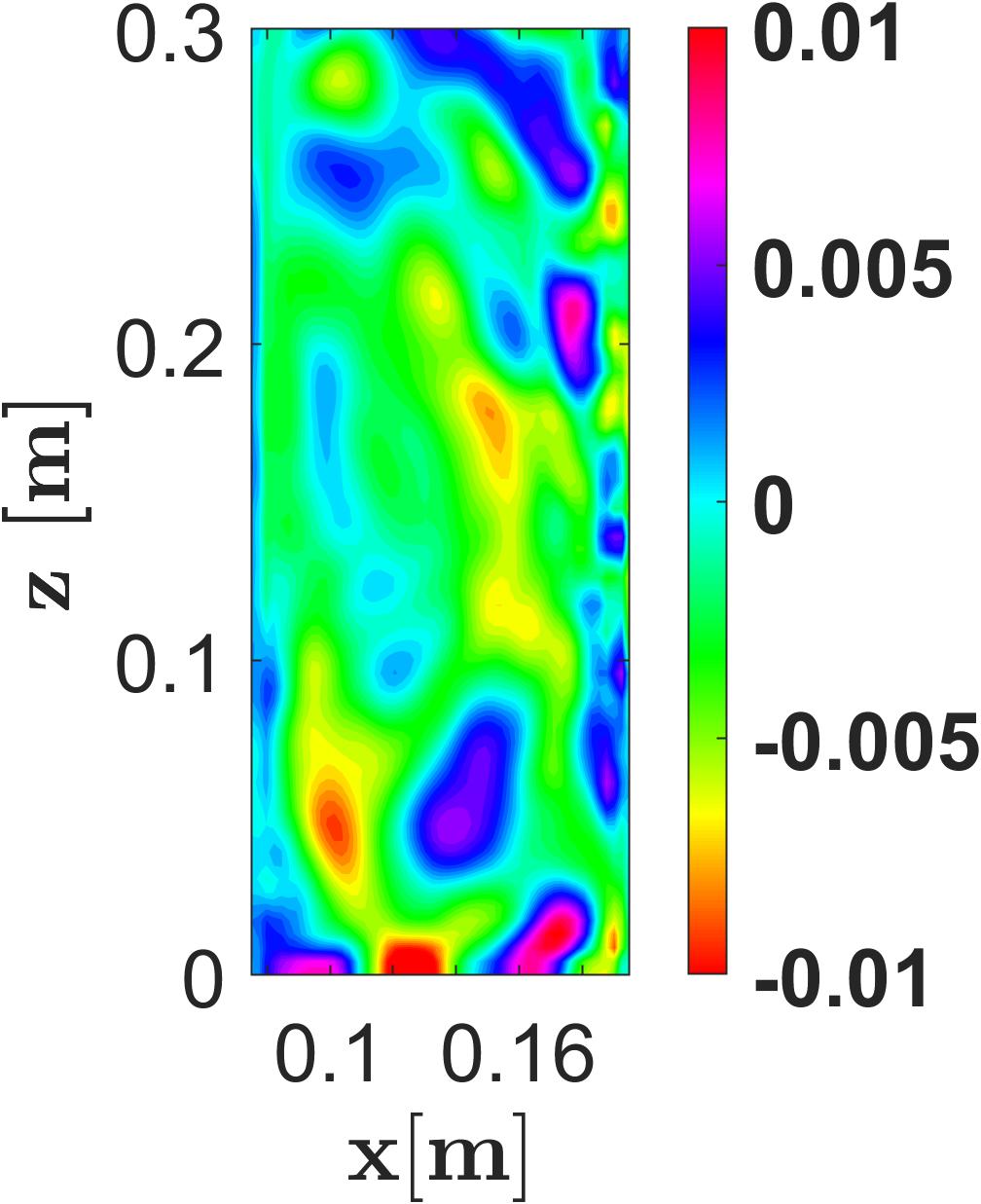}
\label{fig:x_enrgyflux_w_12_Ta_mod}}
\hspace{1.5cm}
\subfigure[]{%
\includegraphics[scale = 0.09]{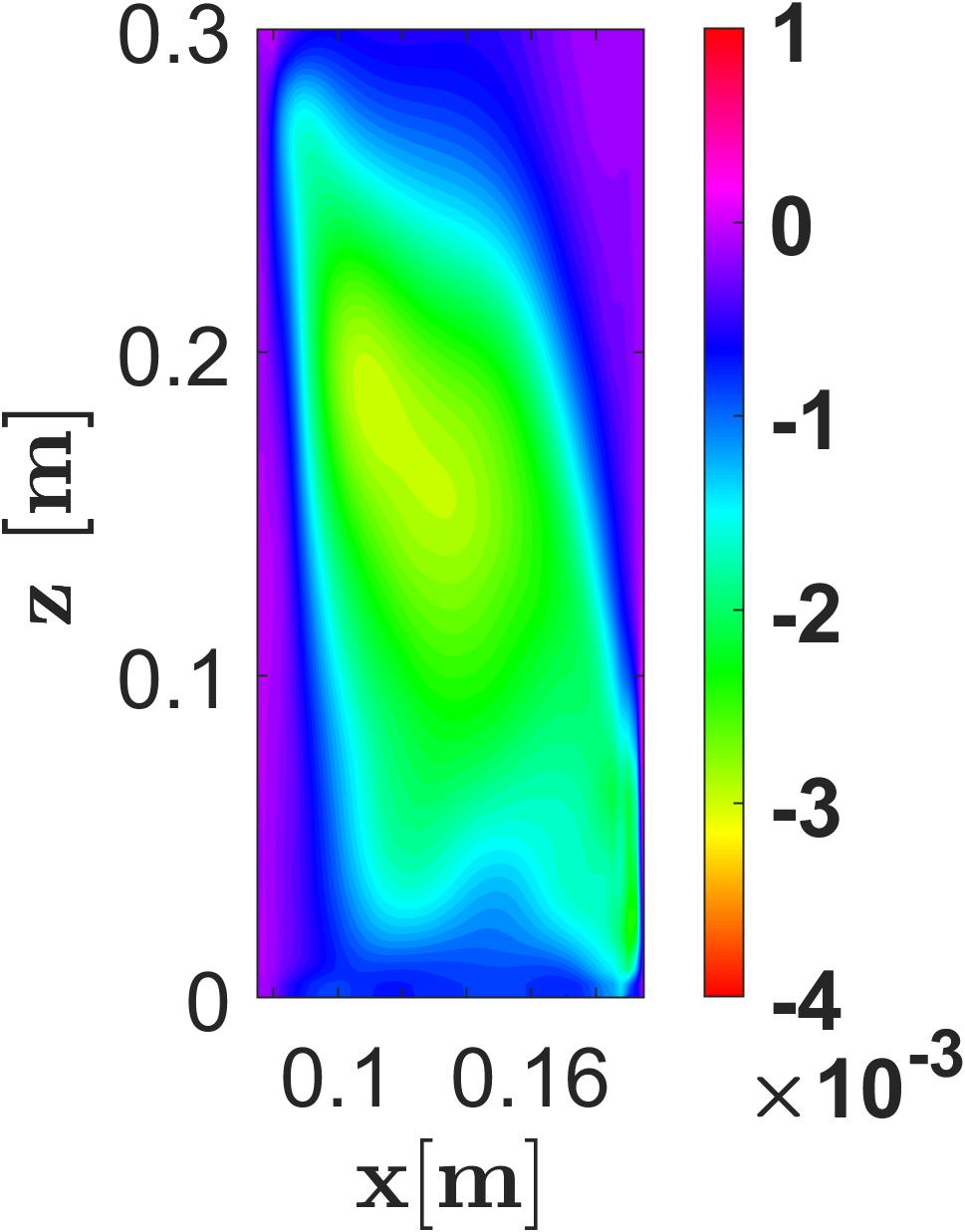}
\label{fig:x_enrgyflux_w_12_Ta_high}}

\caption{Contours of time and $y-$direction averaged turbulent flux of temperature $\avg{u'T'}$ in $x-$direction, for a range of $Ri_0$ and $Ro$. $\avg{u'T'}$ represents heat transport in $x-$direction. Here, $Ri_0=$ 99 and $Ro =$ 0.3, 0.1, 0.07 for [(a), (b), (c)], $Ri_0=$ 4 and $Ro =$ 0.3, 0.1, 0.07 for [(d), (e), (f)], and $Ri_0=$ 1 and $Ro =$ 0.3, 0.1, 0.07 for [(g), (h), (i)] respectively.}

\label{fig:x_enrgyflux_plot}
\end{figure}

\textcolor{blue}{It is worth distinguishing the present columnar plumes from the columnar structures of rotating Rayleigh--B\'enard convection. Here columnarity is \emph{momentum-driven} and is set by the source Richardson number through $L_M=b\,Ri_0^{-1/2}$ and $M_0\propto1/Ri_0$; it is therefore mechanistically distinct from the Taylor--Proudman columns that arise from the rotational constraint at small convective Rossby number \cite{zhong2009prandtl, stellmach2014approaching}. Our regime is moreover only moderately rotationally constrained ($\Ro\approx0.07$--$0.3$, $Ro_c\approx0.14$--$0.7$), and the Ekman pumping that organises and clusters vortex columns in rotating convection \cite{chong2020vortices, ding2021inverse} and underlies the large-scale-circulation clustering of columns\cite{ahlers2009heat}, is absent here by construction. Accordingly, the increase in plume coherence with decreasing $\Ro$ at high $Ri_0$ is \emph{not} Taylor-column formation but the Coriolis suppression of lateral entrainment ($\mathrm{Ro}_{p}\ll1$), which protects an otherwise weak plume from erosion; and the plume clustering we report is a kinematic enslavement of the columns to the $m$-fold convergence field of the independently generated baroclinic wave, not a rotational vortex self-organisation. The $Ri_0=99$ plume corresponds to forcing levels progressively weaker than naturally-convective limit ($F_0/F_{0,\mathrm{nat}}=0.10$), $Ri_0=4$ is intermediate ($F_0/F_{0,\mathrm{nat}}=0.50$), and $Ri_0=1$ sits at the natural-convection threshold ($F_0/F_{0,\mathrm{nat}}=1$). This rotational sensitivity should therefore not be read as standard rotating-convection physics.
}

\subsection{\label{sec:Rec_heat}Heat flux analysis}

In this section, we investigate the turbulent heat flux in the $x$-direction, $\avg{u'T'}$, which was introduced earlier as the covariance between the fluctuating horizontal velocity and temperature. Figure \ref{fig:x_enrgyflux_plot} shows the contours of $\avg{u'T'}$ in a $x-z$ plane for a combination of $Ri_0$ and $Ro$ values. The negative (positive) value of $\avg{u'T'}$ indicates that the warm (cold) fluid is moving toward the inner (outer) wall. We observe that for cases shown in Figure \ref{fig:x_enrgyflux_w_1.2_Ta_low}, Figures \ref{fig:x_enrgyflux_w_6_Ta_low}-\ref{fig:x_enrgyflux_w_6_Ta_high}, Figures \ref{fig:x_enrgyflux_w_12_Ta_low} and \ref{fig:x_enrgyflux_w_12_Ta_high}, the heat transport is mainly through the bulk region of the annulus. Swarnakar \eal\cite{swarnakar2023numerical} also observed the heat transport mostly through the bulk region of the annulus at various rotation rates. However, in the rectangular annulus we observe that some cases display significantly high heat transport and irregular contours of $\avg{u'T'}$ as shown in Figure \ref{fig:x_enrgyflux_w_1.2_Ta_mod}, Figure \ref{fig:x_enrgyflux_w_1.2_Ta_high}, and Figure \ref{fig:x_enrgyflux_w_12_Ta_mod}. 

We first discuss the cases where the contours of $\avg{u'T'}$ show heat transport in the bulk region, as mentioned above. In these cases, the magnitude of heat transport in the $x-$direction depends on $Ri_0$. The heat flux is maximum for $Ri_0 = 1$ and $Ro = 0.3$ case (Figure \ref{fig:x_enrgyflux_w_12_Ta_low}), and decreases with an increase in the $Ri_0$ number (Figure \ref{fig:x_enrgyflux_w_6_Ta_low} and \ref{fig:x_enrgyflux_w_1.2_Ta_low}) because of the decrease in the plume inlet velocity which causes a reduction in the thermal flux entering the annulus \cite{banerjee2021investigation}. Irrespective of the $Ri_0$ and $Ro$ values for these cases, the heat transport is mostly concentrated in the bulk region of the annulus. For $Ri_0 = 4$, we observed a reduction in the $\avg{u'T'}$ with a decrease in the $Ro$ number from $Ro = 0.3$ to $Ro = 0.1$. However, the reduction in heat transport is considerably smaller as $Ro$ decreases from 0.1 to 0.07. 


For the cases in Figures~\ref{fig:x_enrgyflux_w_1.2_Ta_mod},
\ref{fig:x_enrgyflux_w_1.2_Ta_high} and \ref{fig:x_enrgyflux_w_12_Ta_mod},
the $\avg{u'T'}$ field is qualitatively different from the organised
transport seen at stronger forcing. Two features stand out: large local
magnitudes confined to the bottom outer corner, and patches of alternating
sign, predominantly negative, scattered through the bulk. Both follow from
the same cause. At $Ri_0=99$ the source buoyancy flux is weakest
($F_0\propto Ri_0^{-1/2}$), and at $\Ro\in\{0.1,0.07\}$ rotation further
limits the fluctuation amplitude, so the eddy kinetic energy in the bulk is
small. The flux $\avg{u'T'}=\rho\,\sigma_u\sigma_T$ therefore depends on the
$u'$--$T'$ phase relation rather than on the fluctuation intensity alone:
where the two fields are close to being in quadrature, the product averages
towards zero and the residual is set by the sampling error of the finite
averaging window. The bulk contours are consequently a noise signature, not
a transport pattern, and their sign should not be read physically. Near the
bottom outer corner the fluctuations are intense but equally poorly
correlated, so the same weak $\rho$ multiplies much larger
$\sigma_u\sigma_T$ and leaves a large-magnitude residual there. The two sub-groups differ in why the correlation is weak. In
Figures~\ref{fig:x_enrgyflux_w_1.2_Ta_mod} and
\ref{fig:x_enrgyflux_w_1.2_Ta_high} the plume has lost its columnar form and
the waves are distorted (Figures~\ref{fig:YZ_vector_plot} and
\ref{fig:Vectr_plot}), so there is no persistent structure to hold $u'$ and
$T'$ in phase. For Figure \ref{fig:x_enrgyflux_w_12_Ta_mod}, the waves and the plume show coherent patterns, but it is not enough to provide considerable thermal flux in the bulk.



The overall weak dependence of heat transport on the rotation rate $(\Omega)$ was also observed by Banerjee \eal\cite{banerjee2021investigation}, especially for cases with $\Omega > 0.89$ rad/s, which corresponds to $Ro > 0.3$ cases in this work.  
In addition, for $Ri_0 = 1$, we could see a reduction in the heat transport between the cases with $Ro = 0.3$ (Figure \ref{fig:x_enrgyflux_w_12_Ta_low}) and $Ro = 0.07$ (Figure \ref{fig:x_enrgyflux_w_12_Ta_high}). It is also observed that cases having baroclinic waves with mode $m = 2$ have higher heat transport (Figures \ref{fig:x_enrgyflux_w_6_Ta_low} and \ref{fig:x_enrgyflux_w_12_Ta_low}) in the bulk region as compared to cases with $m = 3$ modes baroclinic wave. This indicates that better heat transport occurs when the baroclinic wave is in a steady low wave number regime and as the wave modes $(m)$ increases or the baroclinic wave becomes irregular the heat transport in the annulus reduces. Swarnakar \eal\cite{swarnakar2023numerical} also reported a reduction in heat transport for cases with unsteady and irregular baroclinic wave regimes.



\begin{figure}[!htbp]
  \centering
  \includegraphics[width=0.85\linewidth]{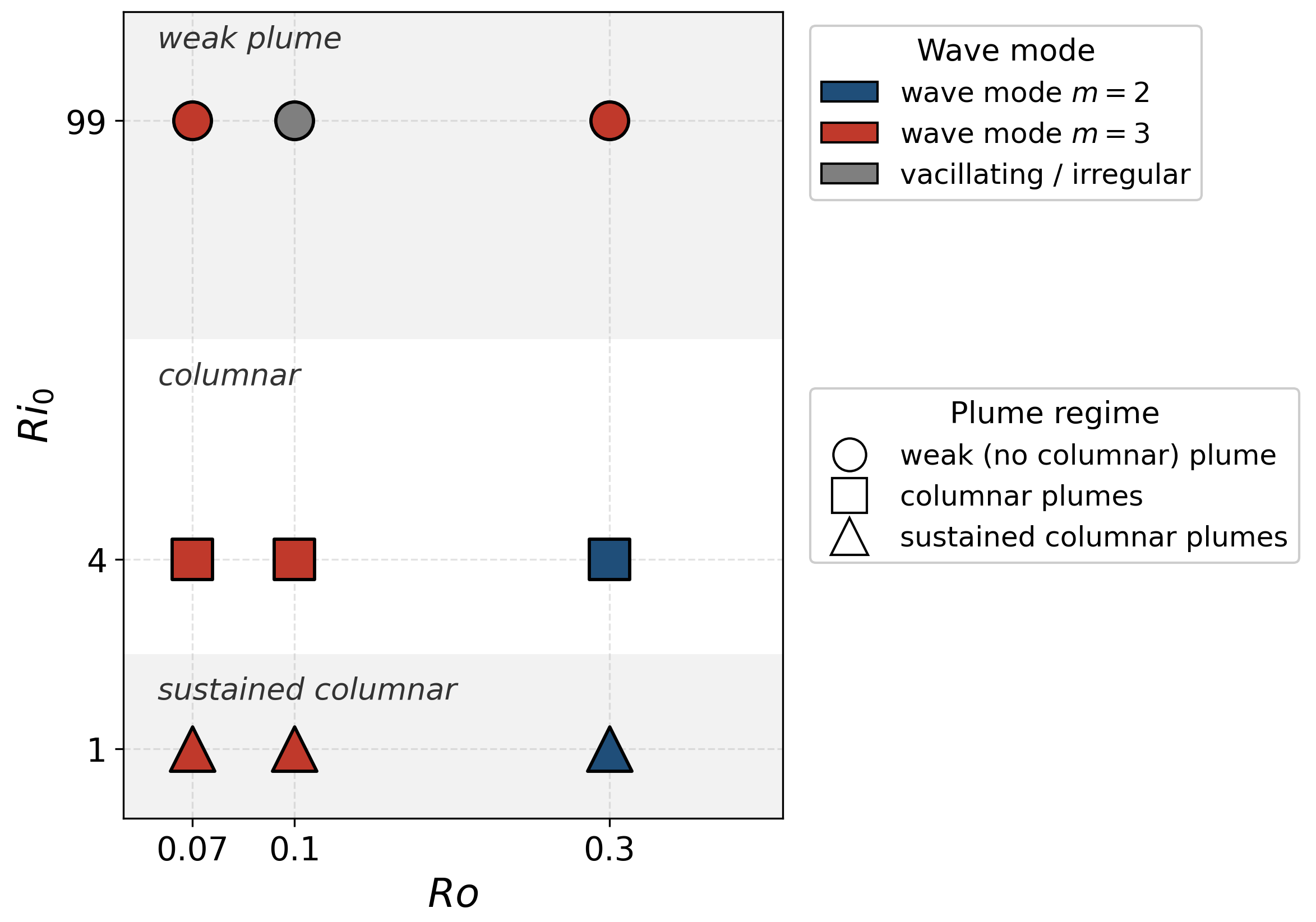}
  \caption{Regime map of the rectangular rotating annulus with
  localized forced heated plume in the $(Ri_{0},Ro)$ parameter
  space. Marker colour encodes the dominant baroclinic-wave mode
  identified from the instantaneous temperature snapshots of
  Figure  \ref{fig:mean_plot_T}
 and the CEOF principal-component time series of
  Figure  \ref{fig:PC_plot}: blue, $m=2$; red, $m=3$; grey, vacillating /
  irregular. Marker shape encodes the plume regime inferred
  from the $y$--$z$ snapshots of Figure~\ref{fig:YZ_vector_plot}: circle, weak
  (non-columnar) plume; square, columnar plumes that traverse a
  substantial fraction of the baroclinic zone; triangle,
  sustained columnar plumes reaching the top of the baroclinic
  zone. The alternating grey/white horizontal stripes carry no
  physical meaning and serve only to visually group the three
  $Ri_{0}$ rows. The map exhibits a clean horizontal banding of
  plume regimes (set by $Ri_{0}$, via the inlet vertical-momentum
  flux $M_{0}\propto 1/Ri_{0}$) and an $m=2\!\to\!m=3$ wave
  transition along each row as $Ro$ decreases (set by the
  contraction of the baroclinic deformation radius
  $L_{\rho}=NH/f$), indicating that within the parameter range
  explored the plume-regime and wave-selection problems are
  approximately separable. The single off-diagonal entry --
  vacillation at $Ri_{0}=99$, $Ro=0.10$ -- corresponds to the
  weak-plume / intermediate-rotation corner of the map where
  neither the buoyancy forcing nor the rotation is strong enough
  to lock the system into a clean wave mode.}
  \label{fig:regime_map}
\end{figure}

From the parameter range used in this study, we observe that heat flux is mostly well distributed for low values of the source Richardson number and has a weak dependence on the Rossby number (Figure~\ref{fig:x_enrgyflux_plot}). This is primarily explained by the structure of waves and plumes and their interaction in modulating heat transport in this configuration. A more quantitative rationale for these trends follows from a mixing-length argument that links the bulk eddy heat flux to the buoyancy supply delivered by the forced plume. The inlet buoyancy flux is defined as $F_{0} = \alpha g\,w\,\Delta T_{p}\,b$. Combining the thermal-wind velocity $U_{y} = \alpha g\,l_{z}\,\Delta T_{x}/(2\Omega l_{x})$ with the entrainment coefficient $\Gamma(z)$ (Figure~\ref{fig:entrnmnt_coeff}), standard manipulation yields

\begin{equation}
    \langle u'T'\rangle
    \;\sim\;
    \frac{F_{0}}{\alpha g}\,\frac{l_{z}}{l_{x}}
    \,(Pr)^{-1/2}
    \;\propto\;
    w\,\Delta T_{p}
    \;\propto\;
    \Delta T_{p}^{3/2}\,Ri_{0}^{-1/2}.
    \label{eq:uTscale}
\end{equation}
At fixed $\Delta T_{p}$, Eq.~\eqref{eq:uTscale} predicts a monotone increase in 
$\langle u'T'\rangle$ with decreasing $Ri_{0}$. The traversal from $Ri_{0} = 99$ to 
$Ri_{0} = 1$ produces an expected enhancement of $Ri_{0}^{-1/2}\Big|_{99}^{1} \;=\; \sqrt{99} \;\approx\; 10,$ which matches the order-of-magnitude amplification of bulk heat transport observed in 
Figure~\ref{fig:x_enrgyflux_plot}. The results suggest that bulk heat transport is controlled mainly by the plume-driven buoyancy flux rather than by rotational effects. This is consistent with the near-invariance of $\langle u'T'\rangle$ along each row of Figure~\ref{fig:regime_map}, except in the $Ri_{0} = 99$, $Ro \in \{0.1,\,0.07\}$ corner where weak plume forcing combines with rapid rotation to produce only weak, noise-dominated $u'$-$T'$ correlations. 

\subsection{Regime map in the $(Ri_{0},\,Ro)$ parameter space}
\label{sec:regime_map}

Secs.~\ref{sec:Rec_wave} -~\ref{sec:Rec_plume} have
classified each of the nine simulations along two independent
qualitative axes: the dominant baroclinic-wave mode ($m=2$, $m=3$, or a
vacillating/irregular state) extracted from the instantaneous
temperature contours at $z=0.15\,\mathrm{m}$ and the first CEOF mode of
$u$, and the plume regime (weak, columnar, or sustained columnar)
inferred from the instantaneous $y$--$z$ temperature contours over the
heating zone.  These classifications are summarized as a regime map in
the $(Ri_{0},Ro)$ plane in Figure~\ref{fig:regime_map}, where the
marker shape encodes the plume regime and the marker colour encodes the
wave mode.

Two complementary observations follow from
Figure~\ref{fig:regime_map}.

First, the plume regime is controlled almost entirely by the source
Richardson number and is essentially insensitive to the frame rotation
rate in the parameter range considered.  This is reflected in the
clean horizontal banding of the regime map: $Ri_{0}=99$ yields weak,
non-columnar plumes (circles) for all three $Ro$; $Ri_{0}=4$ produces
columnar plumes (squares); and $Ri_{0}=1$ produces sustained columnar
plumes (triangles) that traverse the full baroclinic depth.  Because
the inlet vertical momentum flux scales as $M_{0}=w^{2}b\propto
1/Ri_{0}$ (at fixed $\Delta T_{p}$), the plume's ability to penetrate
the stably stratified bulk is set by the ratio of its inlet inertia to
the buoyancy contrast with the ambient: rotation, characterized by
$Ro$, modulates the detrainment height through
$\Gamma(z)$ (Figure~\ref{fig:entrnmnt_coeff}) but does not alter the existence
of the columnar core.

Second, the wave mode is primarily controlled by the Rossby number,
with $Ri_{0}$ acting as a secondary modulator.  At
$Ro=0.3$ -- the weakest rotation considered -- the $Ri_{0}=4$ and
$Ri_{0}=1$ cases select an $m=2$ wave, consistent with the larger
deformation radius $L_{\rho}=NH/f$ allowing a longer-wavelength
baroclinic instability to fit within the periodic zonal domain length
$l_{y}=0.80\,\mathrm{m}$.  As $Ro$ decreases to $0.1$ and $0.07$, the
deformation radius contracts and the dominant wave shifts to $m=3$ at
both moderate ($Ri_{0}=4$) and high ($Ri_{0}=1$) buoyancy forcing,
giving a clean $m{=}2 \rightarrow m{=}3$ transition along each row of
the map.  The $Ri_{0}=99$ row is the only exception: the weak inlet
buoyancy flux at $Ri_{0}=99$ produces a less robust thermal
stratification, so that the system remains in $m=3$ at the two end
$Ro$ values and falls into a vacillating regime at the intermediate
$Ro=0.1$ rather than executing a clean mode transition.  This is
consistent with the CEOF principal components in
Figure~\ref{fig:PC_w_1.2_Ta_low}--\ref{fig:PC_w_1.2_Ta_high}, where the $Ri_{0}=99$, $Ro=0.1$ time
series displays an irregular envelope while the two flanking cases are
quasi-periodic.

Taken together, the map indicates that, within the parameter range
explored, the plume and wave selection problems are approximately
\emph{separable}: $Ri_{0}$ controls the plume regime and $Ro$ controls
the wave mode, with weak cross-coupling.  This separation is
consistent with the heat-flux behaviour reported in
Sec.~\ref{sec:Rec_heat}: the bulk turbulent flux $\langle u'T'\rangle$
strengthens monotonically as $Ri_{0}$ decreases (vertical traversal of
the map) but is largely invariant with respect to $Ro$ at fixed
$Ri_{0}$ (horizontal traversal), except in the $Ri_{0}=99$,
$Ro\in\{0.1,0.07\}$ corner of the map, where weak plume forcing
combines with rapid rotation to produce only weak, noise-dominated
$u$--$T$ correlations.  \textcolor{blue}{The approximate separability is expected to persist only while two conditions hold. First, the plume-regime bands must remain ordered by the source momentum flux $M_0\propto1/Ri_0$, which is the case while the columnar-onset boundary is monotone in $Ri_0$. Second, the wave selection must remain plume-independent, which requires $\mathrm{Ro}_{p}\lesssim \mathcal{O}(1)$ so that the plume does not inject momentum directly into the wave field. We therefore anticipate that separability breaks down (i) at weak rotation, $\Ro\gtrsim \mathcal{O}(1)$, where the deformation radius $L_\rho=NH/f$ exceeds $l_y$ and the wave is geometrically quenched, and (ii) at very low $Ri_0\ll1$, where a momentum-dominated plume ($\mathrm{Ro}_{p}>1$) forces the wave directly; mapping these boundaries requires a denser parameter sweep.} We emphasise that the present nine simulations
provide a coarse sampling of the $(Ri_{0},Ro)$ plane; a denser
parameter sweep would be required to map the precise location of the
$m=2/m=3$ neutral curve and of the columnar-plume onset boundary, both
of which lie between the simulated grid points and have not been
resolved here.

\section{\label{sec:con}Conclusion}

We have performed numerical simulations of a rotating rectangular annulus 
with bi-directionally forced temperature gradient, configured to focus on the interior bulk dynamics through the removal of the upper and lower Ekman layers. The temperature gradients were imposed via a localized forced heated plume at the outer bottom and a uniformly cooled inner wall, 
and the parameter space spanned $Ri_0 = 99, 4, 1$ and $Ro = 0.3, 0.1, 0.07$.

A non-dimensional scaling of the governing equations established that the 
leading-order bulk state is geostrophic-hydrostatic, and the resulting 
closed-form thermal-wind solution for $\avg{v}(x,z)$ reproduced the 
simulated zonal jet to within $\sim 20\%$, using only the plume detrainment 
height as a physical input. The slanted bulk isotherms, the prograde-upper / 
retrograde-lower jet asymmetry, and the $\mathcal{O}(Ro)$ weakness of the secondary 
circulation in the bulk all followed directly from the same thermal-wind 
argument. The validity of these scalings was confirmed a posteriori by the 
$x-$momentum budget, which exhibits a clean Coriolis-pressure-gradient 
balance throughout the bulk, and by the $z-$momentum budget, in which the 
buoyancy and vertical pressure-gradient terms dominate above the inlet, 
with the advection term comparable only in the near-source region. 

Instantaneous temperature contours and CEOF analysis revealed a clean 
$m = 2 \to m = 3$ baroclinic wave transition as $Ro$ decreased, consistent 
with the contraction of the Eady deformation radius $L_\rho = NH/f$. The 
$Ri_0 = 99,~Ro = 0.1$ case exhibited an amplitude-modulated, quasi-periodic 
CEOF principal component diagnostic of a Hopf-bifurcated vacillating regime, 
while the remaining cases were quasi-steady propagating waves whose phase speeds were a small fraction ($\sim$1--9\%) of the bulk thermal-wind velocity and varied systematically with $Ri_0$ and $Ro$.

Plume morphology was organized by the Morton length scale 
$L_M = b\, Ri_0^{-1/2}$ and the source flux-balance parameter. The 
$Ri_0 = 99$ runs produced weak, laterally-swept plumes; $Ri_0 = 4$ produced 
columnar plumes of intermediate vertical extent; and $Ri_0 = 1$ produced 
sustained columnar plumes traversing the full baroclinic depth. At low 
$Ri_0$ the plumes organized into clusters whose number locks to the 
baroclinic wave mode -- a purely kinematic coupling in which the wave, generated independently of the plume, organises the columnar plumes into its $m$ convergence zones; this clustering appears wherever a coherent columnar plume exists ($Ri_0 = 4$ and $1$) and is absent at $Ri_0 = 99$, where no columnar plume forms, with the sharpness of the imprint set by the transit-time ratio $\tau_p/T_w$. The plume entrainment coefficient $\Gamma(z)$ displayed 
opposite rotational sensitivities at low and high $Ri_0$, which we 
rationalized through a local plume Rossby number $Ro_p = w/(2\Omega b)$: 
at $Ro_p \ll 1$ rotation suppresses lateral ambient inflow and hinders 
detrainment, whereas at $Ro_p = \mathcal{O}(1)$ the Coriolis force deflects the 
rising plume and aids detrainment, with the crossover sitting at the 
$Ri_0 = 4$ cases that are correspondingly the least sensitive to $Ro$.

The turbulent heat flux $\avg{u'T'}$ in the bulk followed a mixing-length 
scaling $\avg{u'T'} \propto \Delta T_p^{3/2} Ri_0^{-1/2}$, predicting an 
order-of-magnitude enhancement of bulk transport as $Ri_0$ traverses from 
$99$ to $1$, in close agreement with the contours of Figure~\ref{fig:x_enrgyflux_plot}. 
The flux remained nearly invariant with respect to $Ro$ at fixed $Ri_0$ 
across most of the parameter map. These trends were synthesized in the 
$(Ri_0, Ro)$ regime map of Figure~\ref{fig:regime_map}, which exhibits a clean horizontal banding of plume regimes (controlled by $Ri_0$) and a clean 
$m = 2 \to m = 3$ wave transition along each row (controlled by $Ro$), 
indicating that within the parameter range explored the plume-regime and 
wave-selection problems are approximately separable.

Taken together, the results demonstrate that an Ekman-free rectangular 
section reproduces the essential mean-flow, wave, and heat-transport 
signatures of the cylindrical baroclinic annulus, and additionally permits 
an independent assessment of the role of plume forcing $(Ri_0)$ and frame 
rotation $(Ro)$. Several lines of investigation remain open. The kinematic 
plume-wave locking advanced in Sec.~\ref{sec:Rec_plume} can be developed into a closed-form theory for the plume tracer distribution as a function of 
$\tau_p/T_w$, including the dependence of the clustering contrast on this 
ratio. The integral plume framework can likewise be extended to a 
quantitative entrainment law $\Gamma(z; Ro_p)$ with a closed-form 
detrainment height $h_{tr}(Ro_p)$, calibrated against the $\Gamma(z)$ 
profiles of Figure~\ref{fig:entrnmnt_coeff}. A denser sweep of the $(Ri_0, Ro)$ plane 
is needed to map the precise location of the $m = 2 / m = 3$ neutral 
curve and the columnar-plume onset boundary, both of which lie between 
the simulated grid points and have not been resolved here. Finally, the 
sensitivity of the present results to finite curvature ($\beta_{cyl}$ 
corrections), to water-like Prandtl number $Pr \approx 7$, and to topographic forcing all merit dedicated study, and would 
clarify the extent to which the separability observed in 
Figure~\ref{fig:regime_map} survives outside the parameter range considered 
in this work. \textcolor{blue}{In
particular, although the present regime map is expected to remain
qualitatively unchanged, increasing $Pr$ would reduce the thermal
boundary-layer thickness ($\delta_T\sim Pr^{-1/2}$), modify the
convective plume structure \cite{zhong2009prandtl}, and shift the location of
baroclinic regime transitions \cite{fowlis1965thermal}. Consequently, the
absolute entrainment coefficient $\Gamma(z)$ and heat-transport
magnitudes are expected to change quantitatively at $Pr\approx7$, making
a dedicated water-Prandtl-number study the natural next step.}

\begin{acknowledgments}
The authors are grateful for funding support from Ministry of Earth Sciences (MoES) in form of a research grant.

\end{acknowledgments}

\section*{AUTHOR DECLARATIONS}
\subsection*{Conflict of Interest}
The authors have no conflicts to disclose.

\section*{DATA AVAILABILITY}
The data that support the findings of this study are available
from the corresponding author upon reasonable request.

\begin{figure}
\centering
\subfigure[]{%
\includegraphics[scale = 0.39]{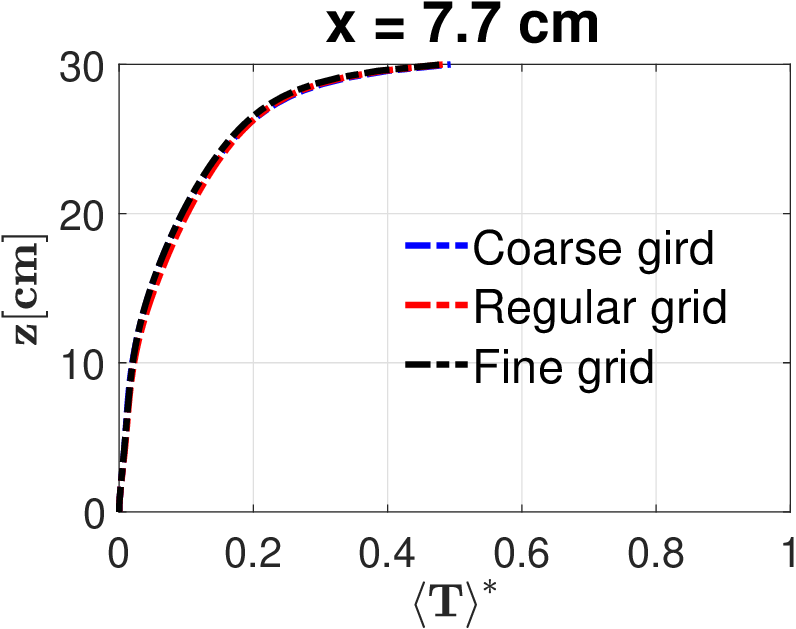}
\label{fig:grid_T_a}}
\subfigure[]{%
\includegraphics[scale = 0.39]{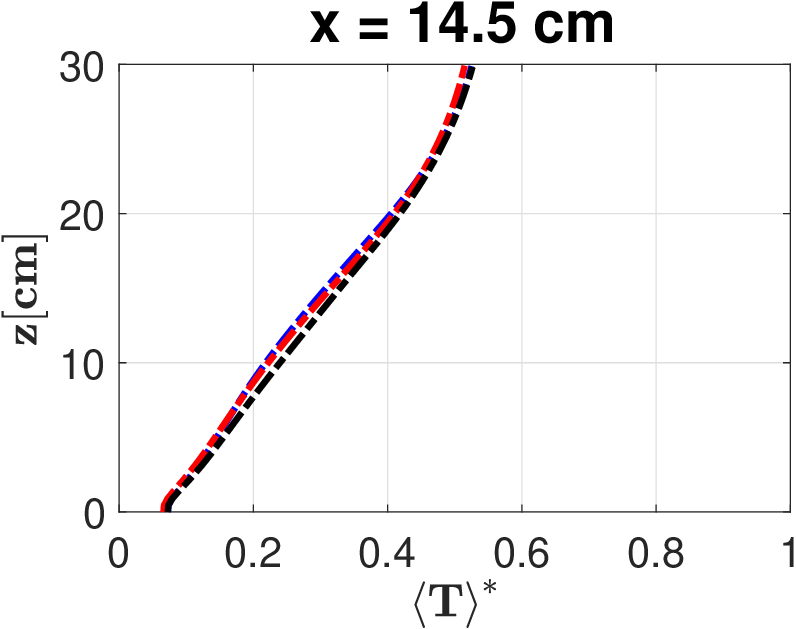}
\label{fig:grid_T_b}}
\subfigure[]{%
\includegraphics[scale = 0.39]{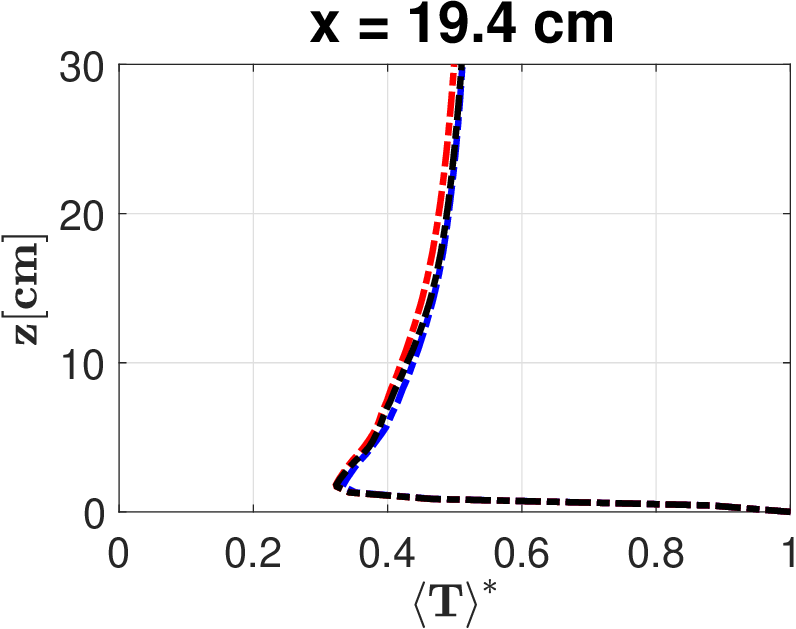}
\label{fig:grid_T_c}}\\
\subfigure[]{%
\includegraphics[scale = 0.155]{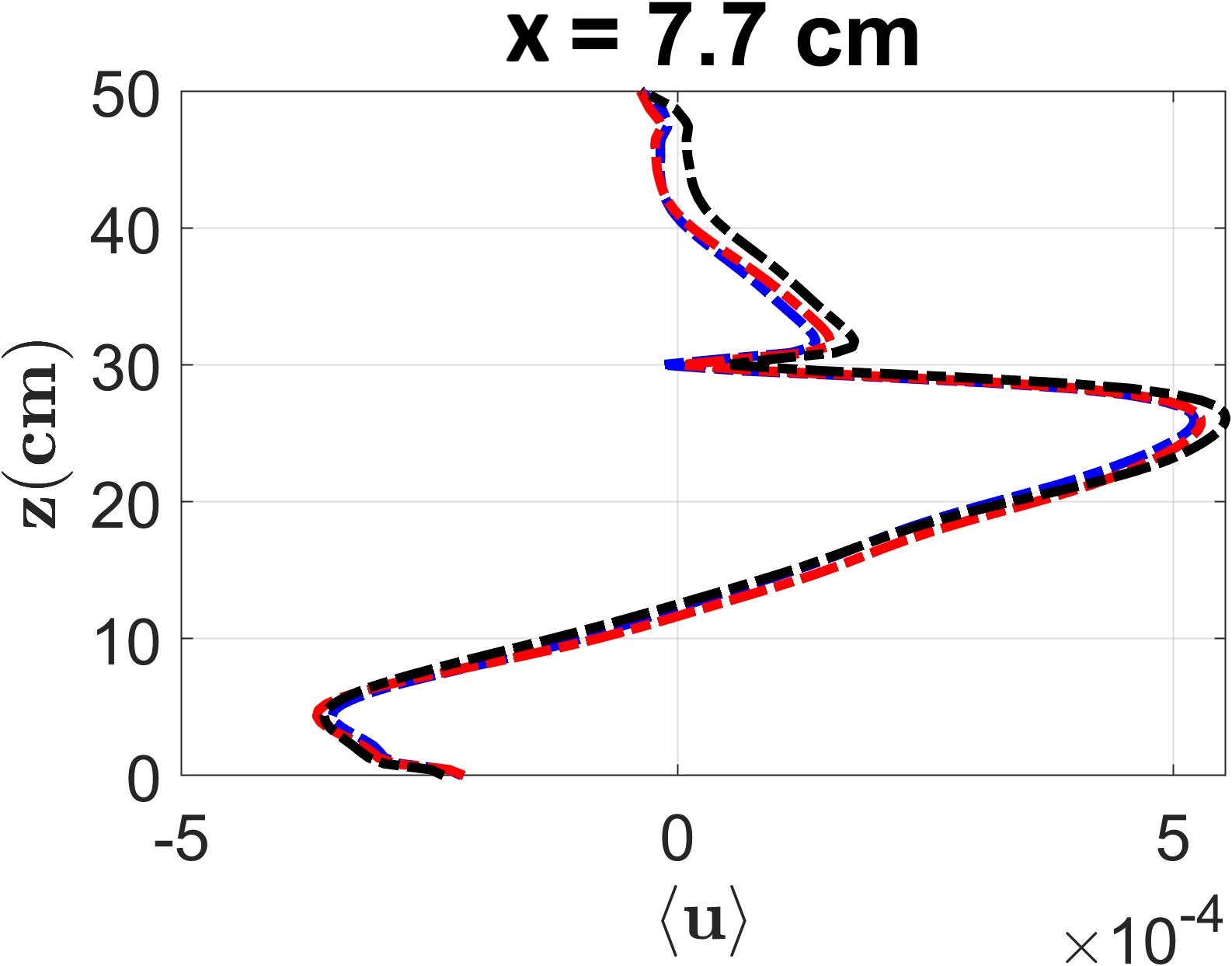}
\label{fig:grid_u_a}}
\subfigure[]{%
\includegraphics[scale = 0.155]{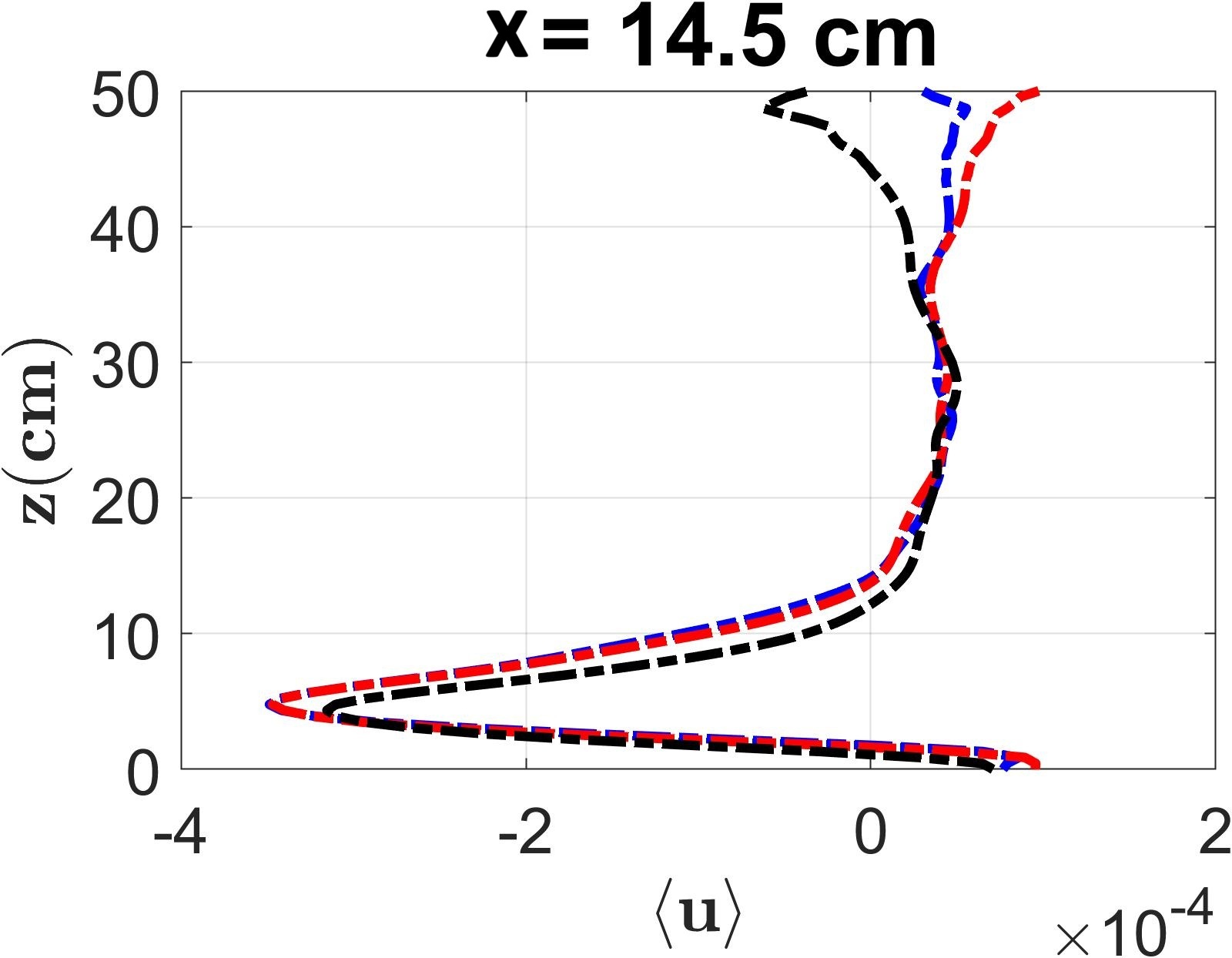}
\label{fig:grid_u_b}}
\subfigure[]{%
\includegraphics[scale = 0.155]{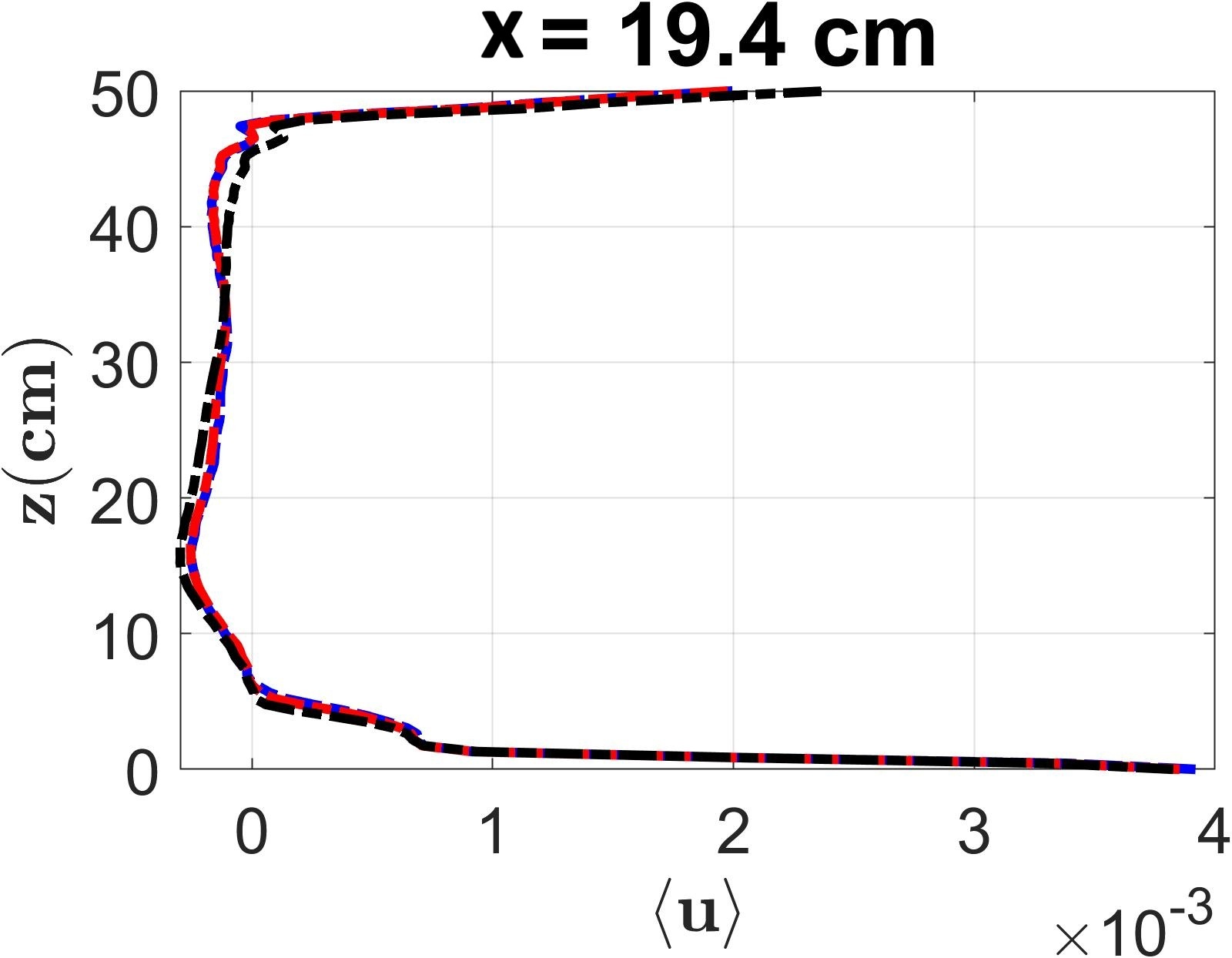}
\label{fig:grid_u_c}}\\
\subfigure[]{%
\includegraphics[scale = 0.152]{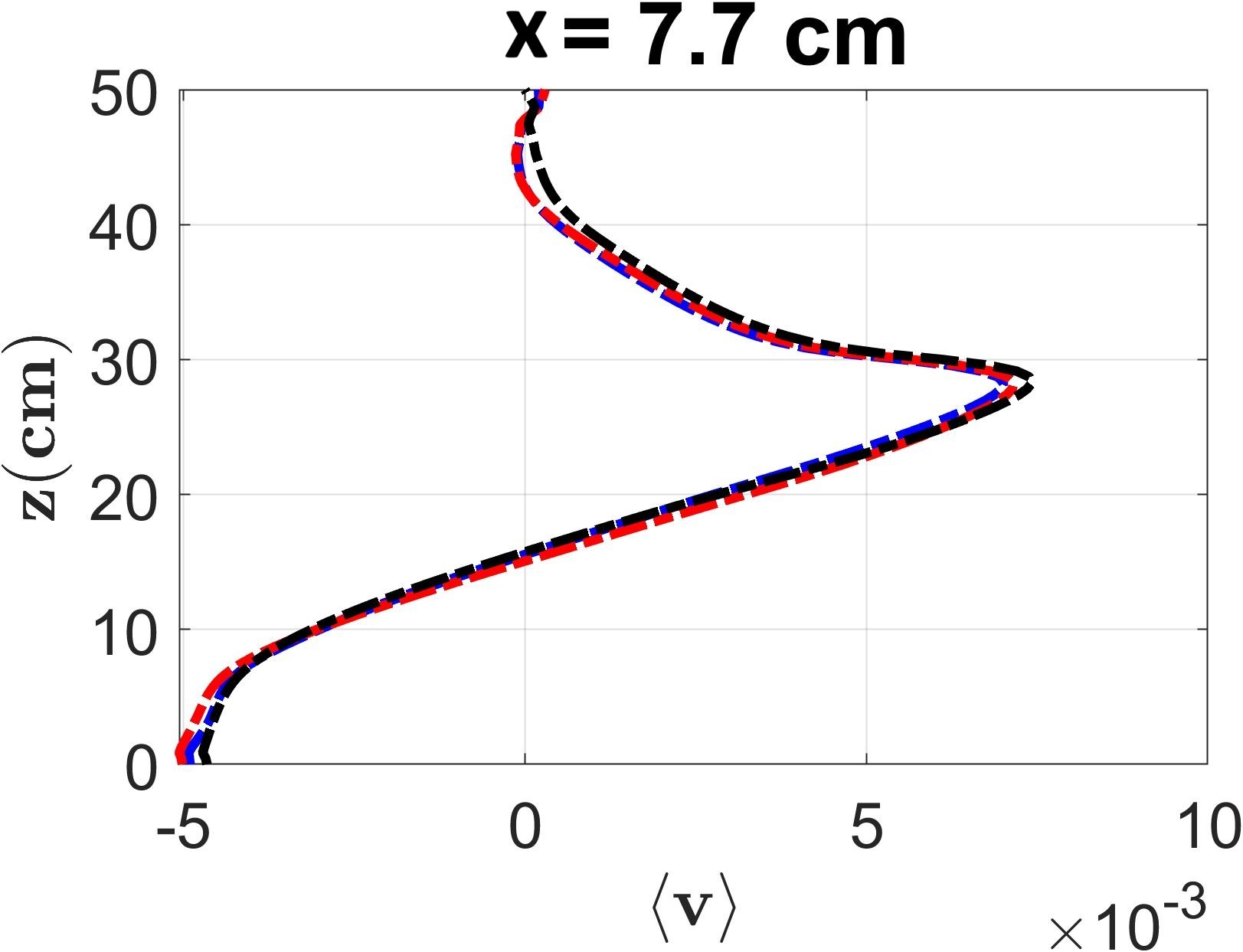}
\label{fig:grid_v_a}}
\subfigure[]{%
\includegraphics[scale = 0.152]{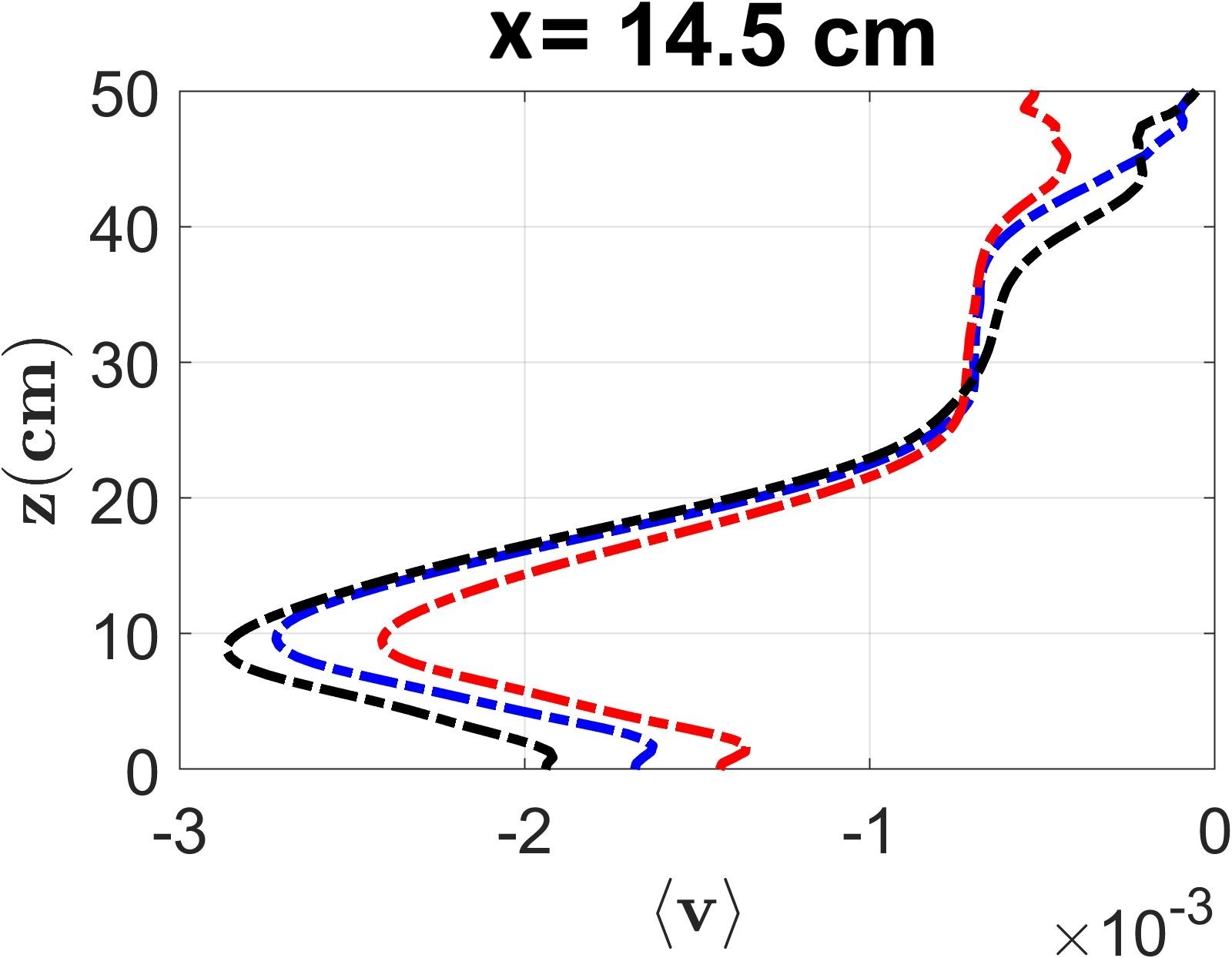}
\label{fig:grid_v_b}}
\subfigure[]{%
\includegraphics[scale = 0.152]{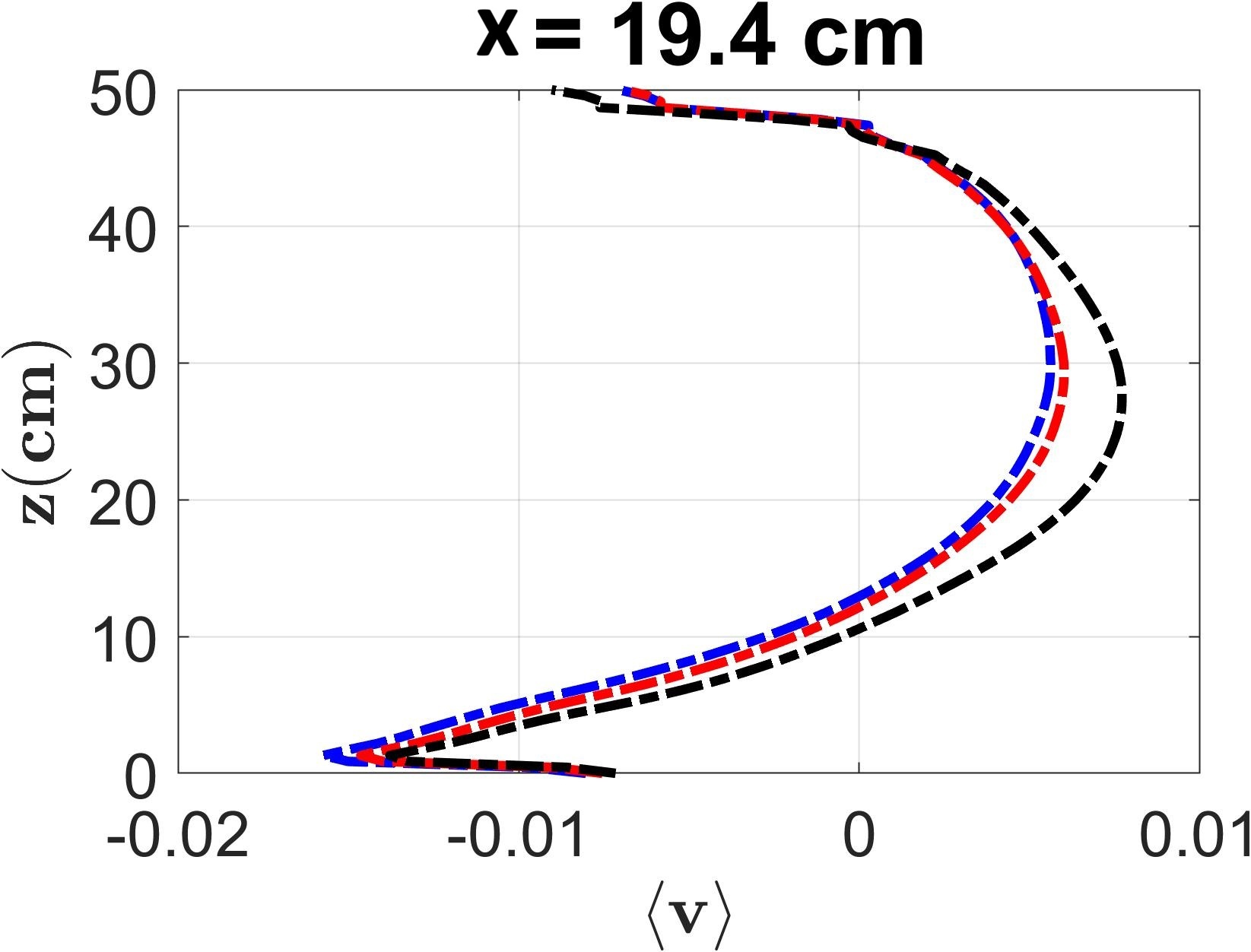}
\label{fig:grid_v_c}}\\
\subfigure[]{%
\includegraphics[scale = 0.155]{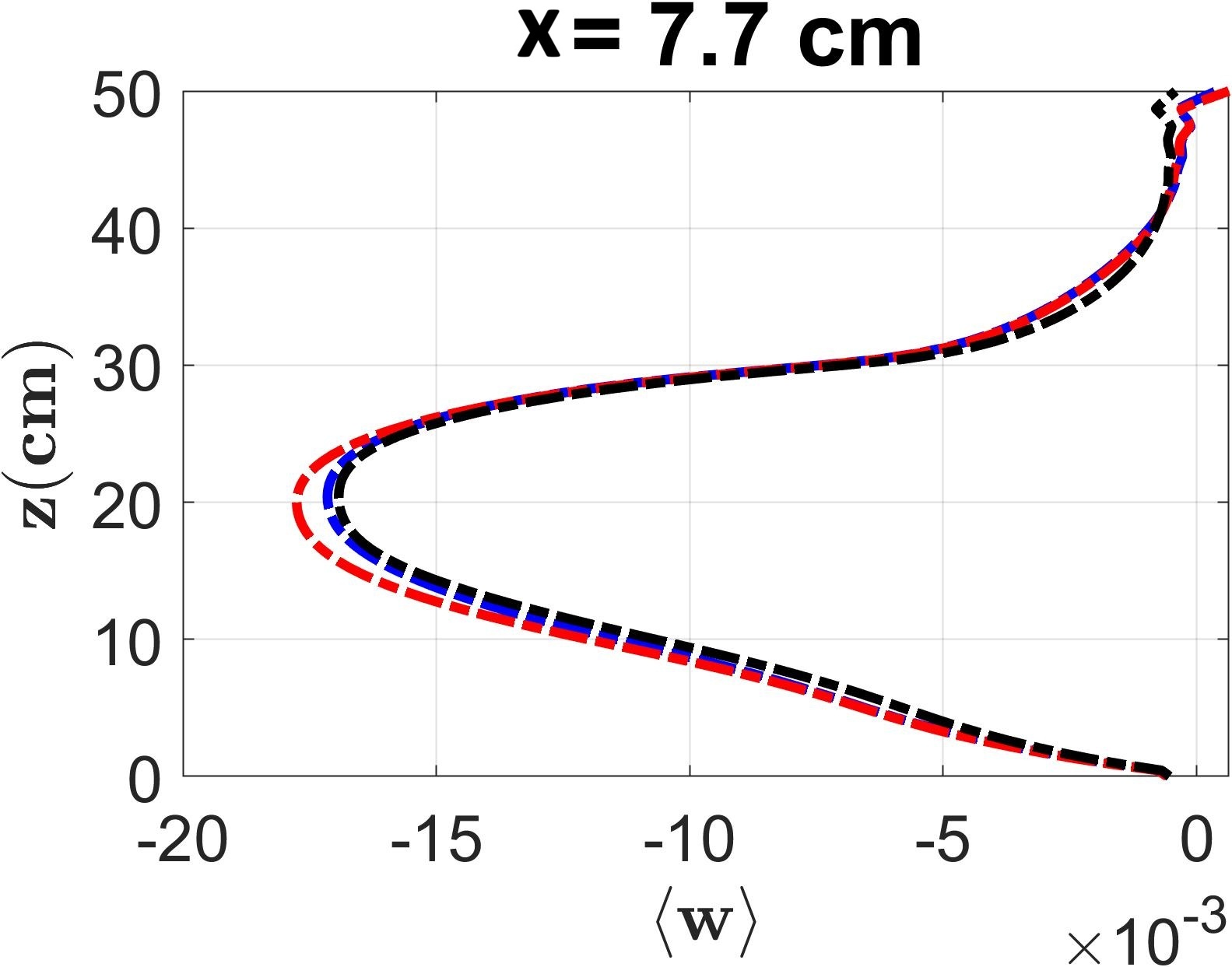}
\label{fig:grid_w_a}}
\subfigure[]{%
\includegraphics[scale = 0.155]{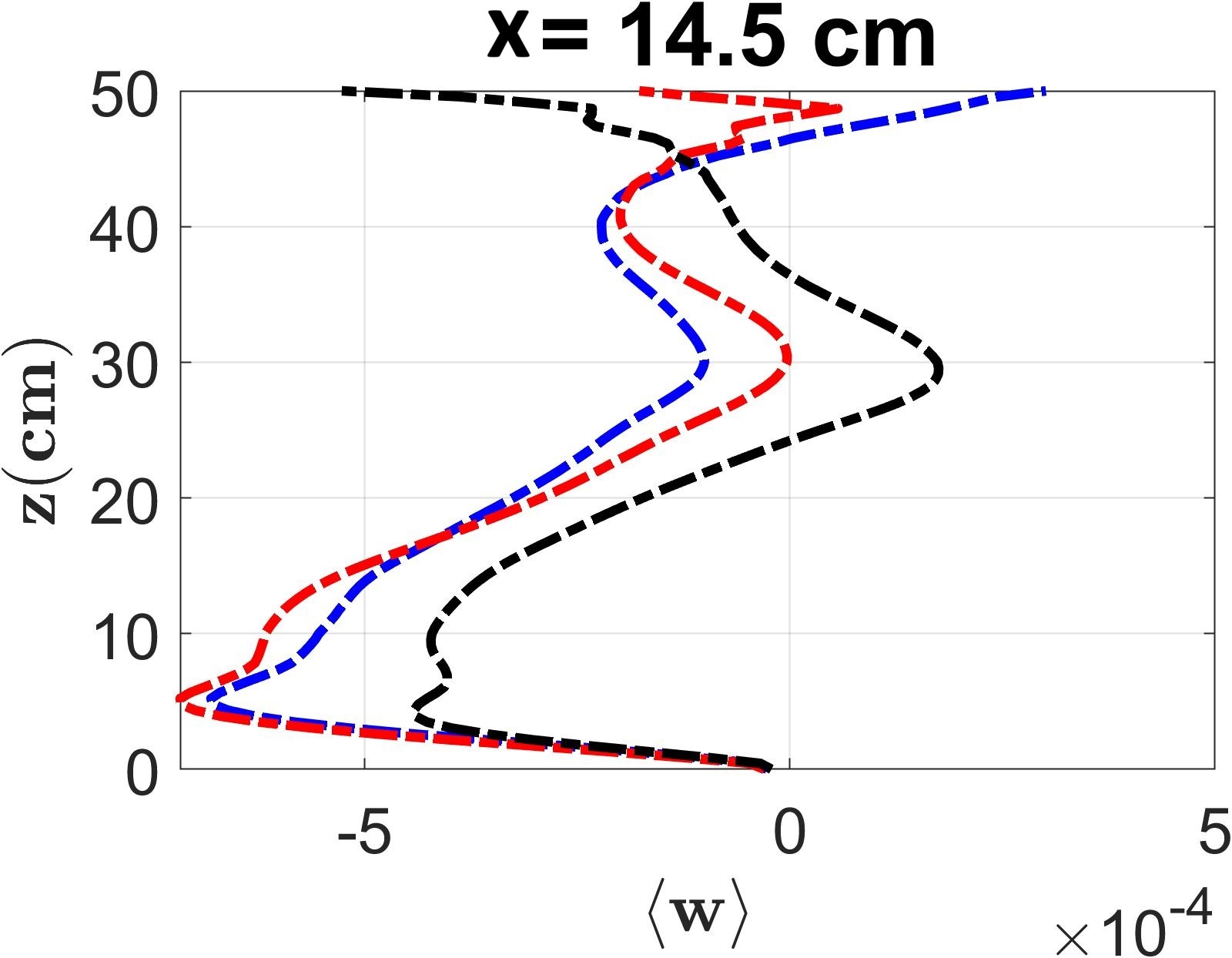}
\label{fig:grid_w_b}}
\subfigure[]{%
\includegraphics[scale = 0.155]{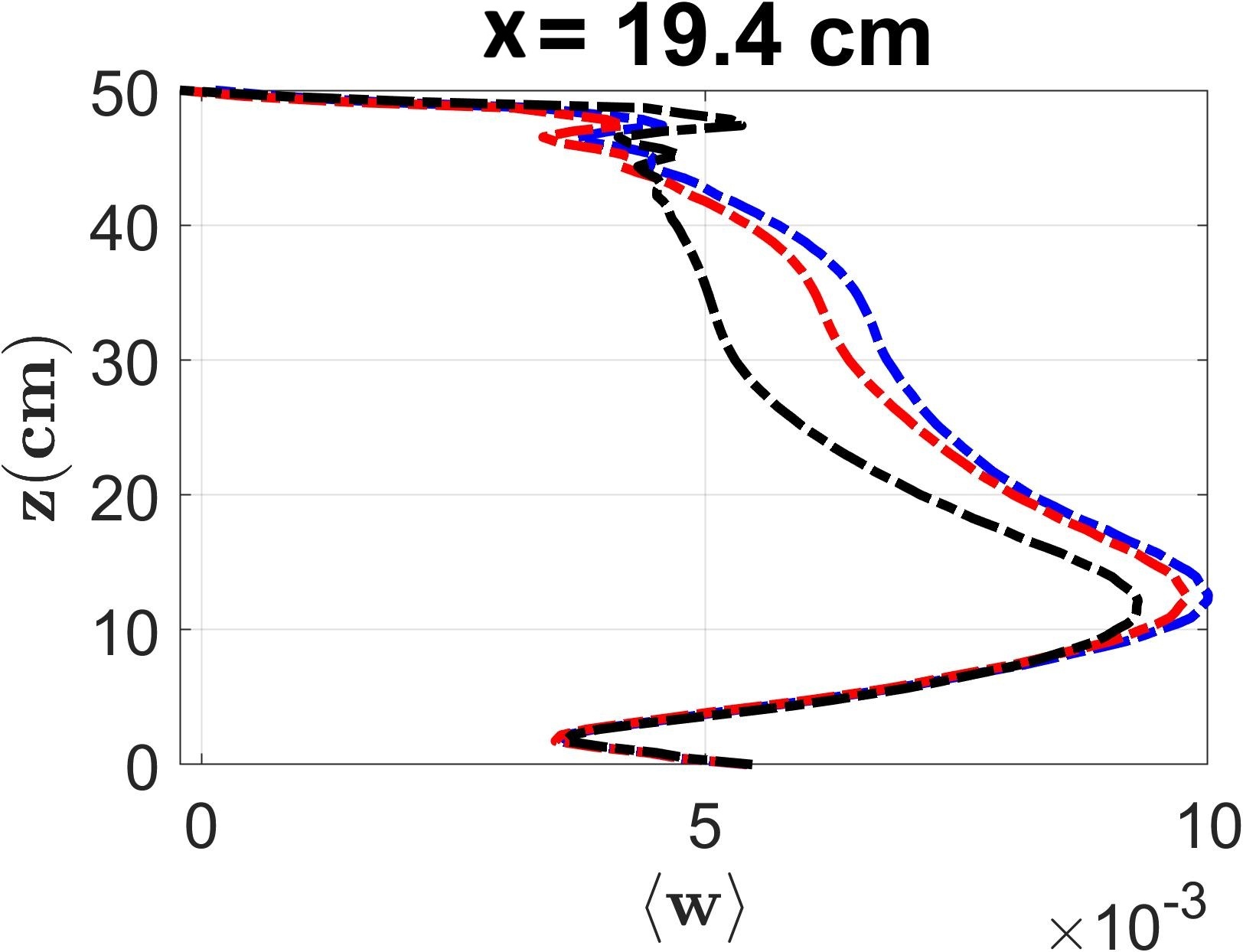}
\label{fig:grid_w_c}}

\caption{Grid-sensitivity study for $Ri_0=4,\Ro=0.1$. Mean non-dimensional
temperature and velocity
components $\avg{T}^{*}$, $\avg{u}$, $\avg{v}$, $\avg{w}$ as a function of
height $z$ at three radial stations $x=7.7$, $14.5$ and $19.4$~cm, for the Coarse, Regular ($51\times96\times116$) and Fine
meshes.}
\label{fig:grid_sensitivity}
\end{figure}

\section*{APPENDIX}
\label{appendix}
\noindent{\bf Grid-sensitivity study\\[0.3em]}

\textcolor{blue}{A grid independence study was conducted to determine the optimal mesh resolution for the numerical simulations. To assess the sensitivity of the results to mesh refinement, the representative case with $Ri_0 = 4$ and $\Ro = 0.1$ was simulated using three different grids: a coarse mesh (51 × 80 × 116), a regular mesh (51 × 96 × 116), and a fine mesh (51 × 140 × 116). The resulting time-averaged non-dimensional mean temperature ($\avg{T}^{*}$) and mean velocity components ($\avg{u}$, $\avg{v}$, and $\avg{w}$) were then compared directly. The time-averaged non-dimensional mean temperature $\avg{T}^{*}$ was calculated as,}

\begin{equation}
\avg{T}^{*}=\frac{\avg{T}-\avg{T}_{\min}}{\avg{T}_{\max}-\avg{T}_{\min}} 
\end{equation}
\textcolor{blue}{where $\avg{T}_{\max}$ and $\avg{T}_{\min}$ denote the maximum and minimum time-averaged, y-directional mean temperatures respectively.}

\textcolor{blue}{The temperature profiles (Figures~\ref{fig:grid_T_a}--\ref{fig:grid_T_c})
are essentially grid-independent. At the interior and the near-outer-wall stations at $x=7.7, 14.5,19.4$~cm, the three curves are indistinguishable over the entire
depth. The height profile of $\avg{v}$ is the wave-carrying component and shows
the tightest collapse: at $x=7.7$~cm and $x=19.4$~cm the three profiles
overlap along their full extent, and even at the mid-gap station
$x=14.5$~cm the shape, sign and location of the extrema are preserved
across meshes, with amplitude differences below $\approx 15\%$. The height profile of $\avg{u}$ is one order of magnitude smaller
than $\avg{v}$ and shows the same qualitative structure on all three
meshes; residual amplitude differences at $x=7.7$ and $14.5$~cm are on
the same order as the ageostrophic signal itself. The vertical component
$\avg{w}$ is the most sensitive of the three, as expected: it
is set locally by the forced plume, and the plume source is spanned by
only $\sim 2$ cells in $x$. At $x=7.7$~cm the strong descending lobe
around $z\approx 20$~cm is captured on all three meshes to within
$\sim 5\%$. Further into the plume-influenced region, at $x=14.5$~cm and
$x=19.4$~cm, the meshes differ visibly in the upper part of the
baroclinic zone ($z\gtrsim 25$~cm) and in the height at which
$\avg{w}$ peaks, but the amplitude of the near-source vertical velocity
at $x=19.4$~cm (the plume outlet, $z\lesssim 15$~cm) collapses across
meshes to within $\sim 10\%$. }

 \textcolor{blue}{Overall, the mean-velocity components profiles (Figure~\ref{fig:grid_u_a}--\ref{fig:grid_w_c}) obtained using the 51 × 96 × 116 grid were found to be in close agreement with those computed using the finer meshes of 51 × 140 × 116, indicating that further mesh refinement results in negligible changes in the solution. Considering both solution accuracy and computational cost, the 51 × 96 × 116 grid provides an optimal balance between numerical precision and efficiency. Therefore, all simulations presented in this study were performed using the 51 × 96 × 116 mesh.}

\bibliography{refs}

\appendix 

\end{document}